\documentclass[floatfix, reprint, nofootinbib, amsmath, amssymb, prd, aps]{revtex4-2}

\usepackage{bm} 
\usepackage{amsmath, amssymb}
\usepackage{hyperref} 

\usepackage{silence}
\usepackage[T1]{fontenc}

\edef\restoreparindent{\parindent=\the\parindent\relax}
\usepackage{parskip} 
\restoreparindent 

\usepackage{amsfonts}
\usepackage{booktabs}
\usepackage{siunitx}
\usepackage{dcolumn} 

\usepackage{lipsum}
\usepackage{comment}

\usepackage{algpseudocode}
\usepackage{algorithm}

\usepackage{graphicx}
\usepackage{fontawesome}

\usepackage{multirow}
\usepackage{chngcntr}
\usepackage{xcolor}

\usepackage[nolist]{acronym}
\newacro{GW}{gravitational-wave}
\newacro{ML}{machine-learning}
\newacro{PSD}{power spectral density}
\newacro{FAR}{false-alarm rate}
\newacro{SNR}{signal-to-noise ratio}
\newacro{ROC}{receiver operating characteristic}
\newacro{BBH}{binary black-hole}
\newacro{IID}{independent and identically distributed}
\newacro{CBC}{compact binary coalescences}
\newacro{OOD}{out-of-distribution}
\newacro{ISCO}{innermost stable circular orbit}
\newacro{CBAM}{Convolutional Block Attention Module}
\newacro{MLP}{multi-layered perceptron}

\makeatletter
    \newcommand{\obast}{\mathbin{\mathpalette\make@circled\ast}}
    \newcommand{\make@circled}[2]{%
        \ooalign{$\m@th#1\smallbigcirc{#1}$\cr\hidewidth$\m@th#1#2$\hidewidth\cr}%
    }
    \newcommand{\smallbigcirc}[1]{%
        \vcenter{\hbox{\scalebox{1.075}{$\m@th#1\bigcirc$}}}%
    }
\makeatother

\usepackage{algorithm} 
\usepackage{algpseudocode} 

\begin{document}

\preprint{APS/123-QED}

\title{Identifying and Mitigating Machine Learning Biases for the Gravitational Wave Detection Problem}

\author{Narenraju Nagarajan}
    \email{n.nagarajan.1@research.gla.ac.uk}
    \email{nnarenraju@gmail.com}
    \affiliation{SUPA, School of Physics and Astronomy, University of Glasgow, Glasgow G12 8QQ, United Kingdom}

\author{Christopher Messenger}
    \email{christopher.messenger@glasgow.ac.uk}
    \affiliation{SUPA, School of Physics and Astronomy, University of Glasgow, Glasgow G12 8QQ, United Kingdom}

\date{\today}

\begin{abstract}
Matched filtering is a long-standing technique for the optimal detection of known signals in stationary Gaussian noise. However, it has known departures from optimality when operating on unknown signals in real noise and suffers from computational inefficiencies in its pursuit of near-optimality. A compelling alternative that has emerged in recent years to address this problem is deep learning. Although it has shown significant promise when applied to the search for gravitational waves in detector noise, we demonstrate the existence of a multitude of learning biases that hinder generalisation and lead to significant loss in detection sensitivity. Our work identifies the sources of a set of 11 interconnected biases present in the supervised learning of the gravitational-wave detection problem and contributes mitigation tactics and training strategies to concurrently address them. In light of the identified biases, we demonstrate that existing detection sensitivity metrics are not reliable for machine-learning pipelines and discuss the trustworthiness of previous results. We use gravitational-wave domain knowledge to build a bespoke machine-learning based binary black hole search pipeline called \texttt{Sage} that addresses these biases. Via the injection study presented in the Machine Learning Gravitational-Wave Search Challenge, we show that \texttt{Sage} detects $\approx11.2\%$ more signals than the benchmark \texttt{PyCBC} analysis at a false alarm rate of one per month in O3a noise. Moreover, we also show that it can detect $\approx48.29\%$ more signals than the previous best-performing machine-learning pipeline on the same dataset. We empirically prove that \texttt{Sage} can: [i] effectively handle out-of-distribution noise power spectral densities, [ii] strongly reject non-Gaussian transient noise artefacts, and [iii] achieve higher detection sensitivities using less data than network architectures of a similar size. By studying machine-learning biases and conducting empirical investigations to understand the reasons for performance improvement/degradation, we aim to address the need for interpretability of machine-learning methods for gravitational-wave science. All code and implementations are available \href{https://github.com/nnarenraju/sage}{here} \href{https://github.com/nnarenraju/sage}{\faGithubSquare}.
\end{abstract}

\maketitle

\section{\label{sec:Introduction}Introduction}
    
    The promise of deep learning in the field of applied data science has been to develop fast approximation machines capable of emulating physical principles. Although it has been successful in seemingly approaching this ideal, it falls short due to its inability to truly generalise or adapt outside its training distribution (otherwise called brittleness) \cite{measure_of_intelligence, OOD_learnability}, being biased \cite{ISSUES_bias_and_fairness} and data-intensive \cite{data_intensive_ml}. We hope that by acknowledging the issues with \ac{ML}, we can be better informed of their capabilities and more importantly, their deficits. Whilst these deficits exist, it is possible to mitigate them and produce reliable pipelines via careful consideration of domain-specific knowledge. In this paper, we focus on identifying and addressing some of the biases that prevail in the application of \ac{ML} to the \ac{GW} detection of \ac{BBH} mergers. Although we focus on the \ac{BBH} detection problem and provide explanations from this perspective, the salient ideas can be extrapolated and adapted to suit other sources of \ac{GW}s and domains of \ac{GW} data analysis.

    Matched filtering \cite{matched_filtering_dhurandhar_sathyaprakash, matched_filtering_owen_sathyaprakash, matched_filtering_babak, pycbc} is an optimal linear filter that maximises the \ac{SNR} in additive white Gaussian noise for a known signal \cite{Whalen1984}. With careful considerations for the detection of unknown \ac{GW} signals in coloured Gaussian noise and non-Gaussian transient noise artefacts \cite{lvk_detnoise}, it has aided in the discovery of 90 \ac{GW} events as of the third observation run of the advanced \ac{GW} detector network~\cite{GWTC-3, lvk_ref_1, lvk_ref_2, lvk_ref_3, lvk_ref_4, lvk_ref_5, lvk_ref_6, lvk_ref_7, lvk_ref_8, lvk_ref_9}. To better understand the necessity for looking at alternate ways of detecting \ac{GW}s, we first build a case against matched filtering in section~\ref{subsec:Matched filtering}. We then highlight the advantages that an alternate \ac{ML}-based method has over matched-filtering in section~\ref{subsec:Machine learning}, but also consider its current issues in section~\ref{subsec:Issues}. We then provide effective mitigation tactics, in section~\ref{sec:Methods}, that aim to address these issues concurrently. To validate our methods, we conduct an injection study with 1 month of O3a noise from the Hanford (H1) and Livingston (L1) detectors (see section~\ref{sec:Results}), as described in the Machine Learning Gravitational-Wave Search Mock Data Challenge (MLGWSC-1)~\cite{DP5_MLGWSC1_2023}. This includes a comparative investigation of the detection performance of our pipeline, \texttt{Sage}, with the production level matched-filtering pipeline, \texttt{PyCBC} \cite{findchirp, pycbc, PyCBC_reference_Nitz_2017, pycbc_live}, and previous works in \ac{ML}-based \ac{GW} detection. Finally, we systematically remove important components of \texttt{Sage} and investigate their contribution to the overall detection performance, thus justifying our method choices in section \ref{sec:Methods}.

    Although investigations into these biases have existed in the machine-learning literature in some form, we translate it to the \ac{GW} detection problem and provide methods to mitigate them. We show that addressing these biases leads to a significant boost in detection performance compared to previous works, and argue that the simultaneous mitigation of these issues is a necessary condition for \ac{ML} to match matched-filtering. As a consequence, we show that \texttt{Sage} can detect more signals than matched-filtering while operating on the same parameter space and is capable of effective glitch mitigation.
    
    
    \subsection{\label{subsec:Matched filtering}Matched Filtering for GW Detection}
    The definition of an optimal receiver is often muddied when dealing with real-world data \cite{Whalen1984}, such as for the detection of unknown \ac{GW}s in detector noise, making the matched-filter \ac{SNR} a non-optimal detection statistic. We can deconstruct the major deviations from the optimal detection scenario into the following: unknown signal parameters, non-stationary noise, uncertainty in the estimates of noise \ac{PSD} and non-Gaussian noise artefacts. All of these factors contribute to the departure of matched-filtering from Neyman-Pearson optimality \cite{neyman_pearson_older_ref_1, neyman_pearson_older_ref_2, bbh_tuning_mf_1, bbh_tuning_mf_2, DP3_Yan_2022} to varying degrees \cite{Whalen1984, lvk_detnoise}.

    Say we have the time series data $d=n+h$ collected from a \ac{GW} detector, where $n$ is the combination of all possible noise sources and $h$ is the sum of the \ac{GW} response of the detector. In the case of an unknown signal, the optimal detection statistic would be obtained by marginalising the likelihood ratio $\Lambda(d|\theta)$ over all unknown signal parameters, $\theta$, by integrating $\Lambda(d|\theta)$ over $\theta$. The likelihood ratio will generally be sharply peaked about its maximum, given that the log-likelihood ratio is a linear function of the signal model. Thus, maximising $\Lambda(d|\theta)$ over $\theta$ is expected to be a good approximation of the marginalised likelihood ratio, up to a constant rescaling factor \cite{lvk_detnoise}. However, this approximation is a non-optimal treatment of the parameter space. 

    The template bank used for matched filtering over the multidimensional \ac{GW} signal parameter space is constructed such that at least one template in the bank is close enough to a potential astrophysical signal, while not losing more than a predetermined fraction of the \ac{SNR} under ideal conditions \cite{making_template_bank, matched_filtering_dhurandhar_sathyaprakash, matched_filtering_owen_sathyaprakash}. Although this condition leads to a reasonably sized template bank for \ac{CBC} in the current detector era, it is well known that template bank based approaches will become too computationally expensive in the future \cite{lisa_large_emri_template_bank, cw_number_of_templates}.

    Deviations of the detector noise from \ac{IID} Gaussian noise, in the time domain, include coloured noise, non-stationary noise and the presence of non-Gaussian transient instrumental artefacts arising from known and unknown sources \cite{noise_source_1, noise_source_2, noise_sources_3}. For a known noise covariance matrix, we can reduce the problem of detecting a signal in coloured noise to the problem of detecting a signal in \ac{IID} Gaussian noise. In practice, this is typically performed by whitening the signal in the frequency domain by estimating the \ac{PSD} using the off-source or on-source methods. Both methods assume that the estimated \ac{PSD} is exact and that the noise is stationary and Gaussian. These assumptions do not hold in either case, but more so for the off-source method, since it requires a longer duration of data than the on-source method \cite{psd_uncertainty_and_pe}.

    The presence of non-Gaussian transient instrumental artefacts or glitches can induce certain templates to produce high \ac{SNR} outputs, making the matched-filter \ac{SNR} highly non-optimal as a detection statistic \cite{lvk_detnoise}. This is currently mitigated using witness sensors~\cite{data_quality_vetoes}, waveform consistency checks~\cite{pycbc, waveform_consistency_1, findchirp, chi_squared_test} and time coincidence tests~\cite{GWTC-1}. However, artefacts of unknown origin that are similar in time and frequency evolution to \ac{GW} signals are difficult to veto out. This is especially true in the case of short-duration \ac{GW} signals, where the background might be contaminated by numerous short-duration transient noise \cite{GW150914}. Although there exist waveform consistency tests to identify these types of noise, they are based on assumptions that might not always hold true and would thereby only target a small subset of all possible glitches \cite{blip_glitch_veto}. These assumptions include modelling short-duration noise transients as sine-Gaussian, targeting a particular frequency range for discriminative features and using parameter distributions for the assumed noise model that may not completely coincide with the truth \cite{blip_glitch_veto}. Such assumptions have to be made since it is impossible to accurately model noise of unknown origin or create a waveform consistency test that is reliable for all future observed glitches. The utmost that we can do to mitigate the effects of unknown noise on detection efficiency, is to incorporate all prior non-Gaussian noise data into a generalised waveform consistency test that is capable of being continually updated. Deep learning is a promising framework that is capable of handling problems of this nature.

    
    \subsection{\label{subsec:Machine learning}Machine Learning for GW Detection}
    The subtle but fundamental difference that allows for the increased efficiency of machine-learning based searches when compared to matched-filtering is the use of a non-parametric hypothesis instead of its parametric counterpart - ``is this a \ac{GW}?" instead of ``is this a \ac{GW} with parameters $\theta$?". A non-parametric approach allows for a template-free search throughout the considered parameter space.

    \textit{Problem Setup}: The search for \ac{GW}s, in the context of supervised machine-learning, can be interpreted as a binary classification problem between a pure noise and noisy signal class. Typically, we are given a collection of $n$ training data pairs, ${(\textbf{x}_i, \textbf{y}_i)}_{i=1}^n$, where $\textbf{x}_i$ is some input data sample and $\textbf{y}_i$ is its corresponding class label. In our case, the noise class contains detector noise taken from the 3$^{\text{rd}}$ advanced detector observing run (O3) \cite{GWTC-3}. The signal class contains simulated \ac{GW} signals with different realisations of additive detector noise. The signals are generated by randomising on the standard parameter distributions of all intrinsic and extrinsic \ac{GW} parameters. We seek an \ac{ML} \textit{model} $\Phi$ with two primary properties. First, the model should be able to \textit{interpolate} the training data such that $\Phi(\textbf{x}_i)\approx \textbf{y}_i$. Second, the model should learn a \textit{generalised} representation of the training data, such that when it is shown an unseen testing data pair $(\textbf{x}^{'}_i, \textbf{y}^{'}_i)$, $\Phi(\textbf{x}^{'}_i)\approx \textbf{y}^{'}_i$. If the model has interpolated the training data but does not generalise, the model is said to have \textit{overfit}. The output of the model $\Phi(\textbf{x}^{'}_i)$ can be interpreted as a ranking statistic of the provided input sample $\textbf{x}^{'}_i$ and if $\Phi(\textbf{x}^{'}_i)$ is normalised to lie in $[0, 1]$, it can be interpreted as a prediction probability. During the \textit{training} or \textit{learning} phase, the trainable parameters $\Theta$ of the model $\Phi$ is optimised to minimise a \textit{loss function} that reduces some distance measure between $\Phi(\textbf{x}_i)$ and $\textbf{y}_i$ \cite{geometry_of_deep_learning}.

    Machine learning is a promising alternative to tackle the detection problem in non-optimal settings. This is corroborated by previous machine learning literature, some of which are chosen here: \cite{ML_GOOD_noisy_data_1, ML_GOOD_noisy_data_2} exhibit the capability of \ac{ML} in detecting low \ac{SNR} signals, \cite{ML_GOOD_complex_1} classifies accurately on highly complex and large datasets, and \cite{ML_GOOD_generalisation, ML_GOOD_biased_simple_func} provide theoretical insights into why \ac{ML} generalises well. It has also been shown that neural networks have the capability to approach Neyman-Pearson optimality for the noisy signal detection problem when trained with a sufficient amount of data on a large enough network under appropriate training strategies \cite{neyman_pearson_ml_1, neyman_pearson_ml_2}. Thus, the \ac{ML} literature has more than sufficient evidence for its capability to handle \ac{GW}s in detector noise when used appropriately.

    The usage of \ac{ML} for \ac{GW} detection is strongly supported, given that \ac{GW}s from \ac{BBH} mergers were successfully distinguished from noise samples using machine learning, as shown in the progenitor works \cite{DP1_George_Huerta_2018, DP2_Gabbard_Messenger_2018}. The Neyman-Pearson optimality of these search methods was later proven in \cite{DP3_Yan_2022}. Although these studies have demonstrated that they match or surpass matched filtering, it is albeit under restricted conditions. Under more realistic conditions, \cite{DP5_MLGWSC1_2023} showed that machine-learning achieves at most $\approx70\%$ of the detection sensitivity when compared to \texttt{PyCBC} at an \ac{FAR} of one per month in O3a noise. \texttt{AresGW}, one of the teams that participated in the mock data challenge \cite{DP5_MLGWSC1_2023}, later showed that they were able to achieve detection sensitivities higher than \texttt{PyCBC} at all false alarm rates above one per month \cite{DP6_Nousi_2023}\footnote{The \texttt{AresGW} pipeline was later enhanced in \cite{Koloniari_gw_events}. We discuss more about this in section \ref{subsec:Comparison with ML}. Any mention of \texttt{AresGW} throughout this paper refers to \cite{DP6_Nousi_2023} and the term enhanced-\texttt{AresGW} refers to \cite{Koloniari_gw_events}.}. At face value, achieving a higher sensitive distance, as defined in \cite{DP5_MLGWSC1_2023}, than \texttt{PyCBC} suggests that machine learning has surpassed matched-filtering at the \ac{GW} detection problem. In the next section, we will argue against this statement and discuss the necessary conditions that need to be satisfied for \ac{ML} to match matched-filtering in detection performance.

    \subsection{\label{subsec:Issues} Acknowledging Potential Problems of Machine Learning Models}
    In this section, we discuss a comprehensive set of issues that systematically bias \ac{ML} models to disproportionately favour the learning of certain parts of the input parameter space more than others. Say we have two \ac{GW} signals with parameters $\theta$ and $\theta'$ in some noise realisation. There is a maximum amount of \textit{information} that can be extracted from these noisy signals, for the purpose of classification. The supervised training of an \ac{ML} model on these noisy signals is not guaranteed to extract the maximum possible information for every possible \ac{GW} signal. If the amount of information learned from a \ac{GW} signal with parameters $\theta$ differs from that learned from a signal with parameters $\theta'$, then the \ac{ML} model is said to exhibit a learning bias. In this section, we identify and describe several possible sources of bias that could affect the effective learning of the \ac{GW} parameter space.

    It is important to note that the issues discussed in this section correspond to \ac{ML} models trained using a \textit{supervised approach} on a labelled dataset, as typically seen in previous works \cite{DP1_George_Huerta_2018, DP2_Gabbard_Messenger_2018, DP3_Yan_2022, DP4_Schafer_Nitz_2022, DP5_MLGWSC1_2023, DP6_Nousi_2023, aframe}. These issues have generally been well-documented in the machine learning literature but have not become standard in the context of \ac{GW} detection using \ac{ML}. A subset of the following issues were discussed in~\cite{magic_bullet}, where they show different modes of failure for \ac{ML} that hinder detection performance. In this work, we try to provide a more comprehensive set of issues and investigate effective mitigation strategies that should serve as a guide to developing better \ac{ML} pipelines.

        \subsubsection*{\label{definitions}Definitions}
        Let us consider the inner product (.|.) \cite{Cutler_Flanagan_1994} between two real functions $a$ and $b$ expressed in the frequency domain as
        \begin{equation}
            (a|b) = 2\int_0^\infty\frac{\tilde{a}(f)\tilde{b}^*(f)+\tilde{a}^*(f)\tilde{b}(f)}{S(f)}df,
        \end{equation}
        where $S(f)$ is the one-sided power spectral density and $*$ represents the complex conjugate operation. The optimal \ac{SNR} is defined as the square root of the norm of $h$,
        \begin{equation}
            (h|h)=4\int_0^\infty\frac{|\tilde{h}(f)|^2}{S(f)}df,
        \end{equation}
        where $h$ is the Fourier-transformed waveform. The matched-filter \ac{SNR} $\rho$ is defined as
        \begin{equation}
            \rho  = \frac{(s|h)}{\sqrt{(h|h)}},
        \end{equation}
        where $s$ is the Fourier transformed detector data. The term integrated \ac{SNR} is used interchangeably with the matched-filter \ac{SNR} within the text.

        \textit{Inductive Bias}: Almost all learning biases can be directly or indirectly related to \textit{inductive bias}. Inductive bias is the biased preference in machine learning models to constrain the hypothesis space, given the training data. A simple example is the difference between a $100^{\text{th}}$ order polynomial fit and a $100$-parameter neural network trying to fit discrete data points generated by a $5^{\text{th}}$ order polynomial. The polynomial fit will have large errors, but the over-parameterised neural network is capable of fitting the relatively simple function easily. This is due to the innate bias of \ac{ML} models toward learning simpler functions and hypothesis \cite{biased_toward_simpler_hypothesis_1, biased_toward_simpler_hypothesis_2}. The architecture of the model can be altered to point the inductive bias toward the correct hypotheses using domain knowledge. For example, the detection problem of a set of objects in an image should not depend on the order or arrangement of the objects in that image, so one can use a 2D permutation-invariant architecture that treats all arrangements equally \cite{permutation_invariant_architecture}. Unwanted learning biases typically arise from a lack of or improper use of inductive bias.
        
        \subsubsection{\label{subsubsec:Learns Biases}Biases in the Training Data}
        \textit{Bias due to insufficient information}: Insufficient information can be interpreted in a multitude of ways. Say we generate a training dataset using some chirp mass distribution for our \ac{ML} model. This will determine the chirp times (also referred to as signal duration and defined as the time spent within the detection band prior to the merger) of the signals up to the first order, with lower chirp masses generating signals with longer chirp times. We define the chirp time as a function of the signal low-frequency cut-off $f_l$ as
        \begin{equation}
        \label{eqn:t_from_mc_Cutler_Flanagan}
            \tau_0(f_l)=t_c-5\left(8\pi f_l\right)^{-8/3}\mathcal{M}_c^{-5/3},
        \end{equation}
        where $t_c$ is the time of coalescence and $\mathcal{M}_c=\frac{(m_1m_2)^\frac{3}{5}}{(m_1+m_2)^\frac{1}{5}}$ is the chirp mass \cite{Cutler_Flanagan_1994}. The length of the samples used for the training dataset must be long enough to contain most of the integrated \ac{SNR} of all possible signal durations dictated by the chirp mass distribution. Using an insufficient sample length leads to reduced integrated \ac{SNR} for signals that are longer than that sample length in the training dataset. This biases the \ac{ML} model, proportionate to the amount of lost \ac{SNR}, against low chirp mass signals. Some previous works, such as the \ac{ML} pipelines in MLGWSC-1 \cite{DP5_MLGWSC1_2023}, \texttt{AresGW} \cite{DP6_Nousi_2023} and \texttt{Aframe} \cite{aframe}, are affected by this bias. This bias negates the need for the \ac{ML} model to learn discriminative features from lower chirp mass signals and narrows the operating parameter space. This situation is similar to running a matched-filter search on a narrower parameter space. Assuming Gaussian noise, the false alarm threshold on a detection statistic at a given false alarm rate will be lowered compared to a matched-filter search on the full parameter space. This allows \texttt{AresGW} to discover more high chirp mass signals than \texttt{PyCBC} that is running on the entire parameter space, at a given \ac{FAR}, in the mock data challenge dataset \cite{DP6_Nousi_2023}.

        Other examples include using an insufficient sampling frequency to represent merger and ringdown or using a filter to discard important discriminative features present in non-Gaussian noise transients. Although removing glitch features might seem intuitively correct, we found that keeping them within the training dataset aided in rejecting unseen glitches and allowed the \ac{ML} model to be more confident in separating noise from signals. We will explore this further in section \ref{subsubsec:Effects of PSD}.
        
        \textit{Bias due to class overlap}: The boundary between a detectable \ac{GW} and pure noise is soft and broad. In the context of generating a set of noisy signals for the training data, we have freedom over the choice of optimal \ac{SNR}. When injected into arbitrary samples of detector noise, the resulting distribution of matched-filter \ac{SNR} will have an innate variance around the optimal \ac{SNR}. Moreover, in non-Gaussian noise, the matched-filter \ac{SNR} threshold above which a trigger can be considered an event at a reasonable \ac{FAR} is not well defined and can vary within the parameter space. Our choice of training data \ac{GW} parameter distributions, alongside the considered detector noise realisation, dictates the extent of overlap between the signal and noise classes. This bias also ties back to the \textit{bias due to insufficient information} - all signals with an integrated \ac{SNR} lower than the minimum \ac{SNR} threshold are essentially noise and are thus wrongly labelled. Wrong or noisy labelling in the class overlap problem is associated with brittle generalisation \cite{ISSUES_noisy_labels, overlap_and_imbalance}.
        
        \textit{Bias due to imbalance between classes}: Supervised machine learning is known to learn the distribution of the data, including any innate biases between classes, regardless of the relative importance of each sample \cite{ISSUES_bias_and_fairness}. Counteracting the noise-dominance of the training dataset by using more samples in the signal class compared to the noise class has been seen in previous works \cite{DP7_Zelenka_2024, magic_bullet}. Introducing a bias toward the signal class will lead to worse noise rejection capabilities at lower \ac{FAR}s. Since reducing the loss function only amounts to placing more samples closer to their corresponding class label, the imbalance leads to an increase in false positives for the sake of increasing true positives. The joint effect of class overlap and imbalance is still poorly understood, but it is widely agreed that both phenomena impede unbiased model learning.

        \textit{Bias due to lack of variation}:
        Variation within the signal class is defined by the changes in morphology produced by differences in the intrinsic and extrinsic \ac{GW} parameters. It is typical in supervised learning settings to generate a limited training dataset and to augment the samples to introduce more variation. This is required across dimensions in the parameter space, along which the classifier cannot interpolate trivially. The inability to interpolate trivially is defined as the inability of the model to comprehend a continuous transformation space from the given finite set of discrete transformed instances. It is difficult to determine the number of discrete transformations or its bounds in the continuous space, required for a given model to understand the overall transformation. This will vary depending on the \ac{GW} parameter in question. For example, although a simple scaling transformation as seen in \ac{SNR} can be understood easily with a relatively small number of discrete \ac{SNR} instances within a sufficient upper and lower bound, the same might not be true for chirp mass or mass ratio, which require more complex transformations.

        Similarly, for the noise class, variation amounts to different estimated \ac{PSD}s, different noise realisations, and the presence of distinct glitches. Assuming that we use real detector noise for the training dataset, it is safe to make some assumptions. First, the different noise realisations are uncorrelated. Second, the different instances within a glitch type are not reliably present within a continuous transformation space. One of the easiest ways to perform good noise rejection is to provide the model with more training data. However, due to limitations in the amount of available detector noise, there is a possibility for the network to become biased toward the available data.

\begin{figure}[htbp]
            \centering
            \includegraphics[width=1.0\linewidth]{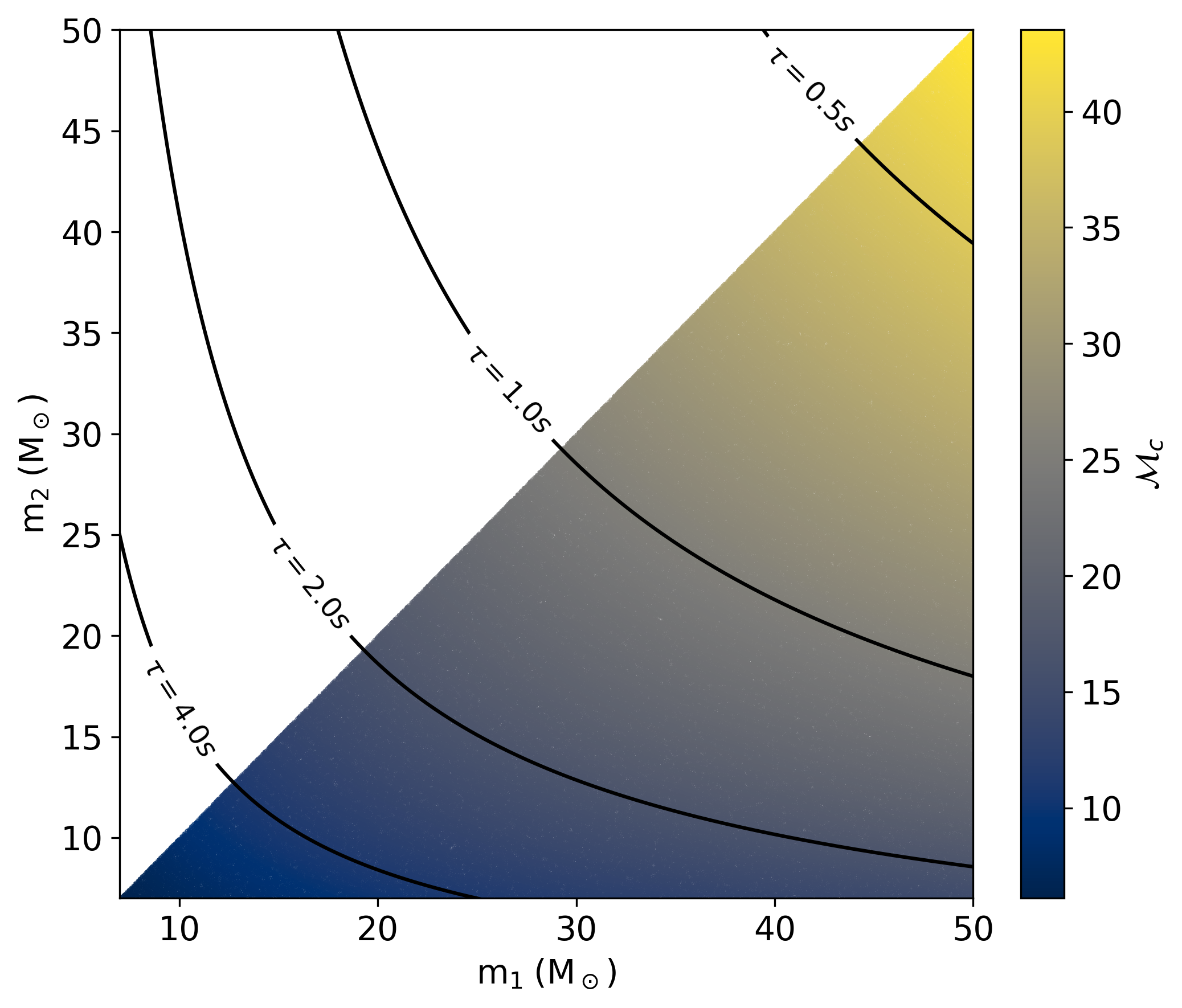}
            \caption{Diagram showing the boundaries of the $(m_1, m_2)$ space coloured by chirp mass, $\mathcal{M}_c$. The solid lines show curves of constant chirp time calculated using equation \ref{eqn:t_from_mc_Cutler_Flanagan}.}
            \label{fig:testing_dataset_priors}
        \end{figure}

        \textit{Bias due to limited sample representation}: The choice of the training distribution on intrinsic and extrinsic \ac{GW} parameters is one of the primary sources of biased learning. Say we sample uniform in the $(m_1, m_2)=U[7, 50]\text{M}_\odot$ space subject to the constraint $m_1>m_2$, where $m_1$ and $m_2$ are the binary source masses. Although this is a very common choice and a uniform sampling for $(m_1, m_2)$ might seem akin to an unbiased representation, evidence of the contrary can be appreciated by considering the associated chirp time $\tau_0$. Figure \ref{fig:testing_dataset_priors} shows lines of constant $\tau_0$ in the $(m_1, m_2)$ space. $\approx25.1\%$ of the samples are above a duration of $2$ seconds, $\approx4.6\%$ above $4$ seconds and only $\approx0.22\%$ above $8$ seconds. Given that the longest $\tau_0$, obtained via equation \ref{eqn:t_from_mc_Cutler_Flanagan} for this parameter space, should be around $11$ seconds, a training dataset built in this framework will lead to heavily biased learning against longer $\tau_0$. Using an input sample length that is longer than the longest duration signal (addressing \textit{bias due to insufficient information}) is not solely sufficient to claim to have built a model that is operating equally on the full parameter space. This bias can also be related to the \textit{bias due to lack of sufficient variation} in the $\tau_0$ dimension. A change in $\tau_0$, with all other parameters fixed, is a linear time dilation or contraction transformation. Even if we had a distribution that is uniform on $\tau_0$, the difficulty of learning different signal durations is directly proportional to $\tau_0$ (see \textit{bias due to sample difficulty}), meaning uniform on $\tau_0$ is still not sufficient to be unbiased.

        Previous works like \cite{DP5_MLGWSC1_2023}, \cite{aframe} and \cite{Koloniari_gw_events} acknowledged the existence of this bias for chirp mass due to lower detection sensitivities observed for low chirp mass signals. 
        
        \textit{Bias due to limited feature representation}: A feature is defined as an individual measurable property or attribute of data in \ac{ML} and pattern recognition according to \cite{bishop_2006}. In our problem, this can include frequency, amplitude, and phase evolution profiles of \ac{GW}s, properties of noise \ac{PSD}s and glitches, possible correlations between detector data in a coherent search or a lack of discriminative features in white Gaussian noise.

        Say we choose a parameter distribution for the training dataset that is uniform on $\tau_0$. We could define a feature in the frequency evolution profile of the \ac{GW}s to be the instantaneous frequency and \ac{SNR} pair along this profile. The instantaneous \ac{SNR} of \ac{BBH} chirps will typically be higher at higher frequencies till merger. The \ac{ML} model could thus disproportionately place a lower weight toward learning early inspiral features. 

        Similarly, with detector noise, there could be realisations of \ac{PSD} and glitches, with outlier features that are under-represented in the training data. A lower weight toward learning these realisations would lead to the model not being able to confidently reject similar realisations of noise in the unseen testing dataset, leading to worse performance at lower \ac{FAR}s.

        \textit{Bias due to train-test mismatch}: The typical desire in machine learning is consistency between training and testing feature distributions, and a mismatch might lead to degraded generalisation and overly optimistic measures of performance during training \cite{train_test_distribution_mismatch}. In the context of \ac{GW} detection, this is particularly important, as models trained on idealised injections may underperform when exposed to real detector noise with unmodeled non-stationarities.
        
        \subsubsection{\label{subsubsec:Biases in the Model}Biases in the Learning Model}
        
        \textit{Bias due to sample difficulty}: Not all samples are equally learnable for a given \ac{ML} model if the proper inductive biases are not incorporated into its architecture~\cite{ISSUES_example_difficulty}. Learning long-term dependencies has been a persistent issue in \ac{ML} \cite{long_term_dependencies_difficult} and many architectural advances have been made to tackle it \cite{rnn_long_term_dependencies, lstm, attention_is_all_you_need}. Solving this issue hinges on two factors: the largest region of the input sample that the model can use to create a feature, also known as the \textit{maximum effective receptive field} \cite{effective_receptive_field}, and the ability to treat dependencies over different distances (near-)equally. For our problem, both factors depend on proper inductive biases of the architecture, and the former also depends on the maximum extent to which the input can be compressed. If the receptive field is not large enough to handle the longest possible signal in the input data, providing the model with $100\%$ of the \ac{SNR} (addressing \textit{bias due to insufficient information}) is not sufficient to ensure unbiased learning. Since standard \ac{ML} architectures must be deeper and take longer to learn long-term dependencies \cite{effective_receptive_field} and this is the hallmark of example difficulty for a given architecture according to \cite{ISSUES_example_difficulty}, we can conclude that longer signals are more difficult to learn. Even with a sufficient receptive field and $100\%$ of the \ac{SNR}, a standard architecture will be biased against longer-duration signals due to relative example difficulty. 
        
        \textit{Spectral bias}: Work on spectral bias \cite{ISSUES_spectral_bias, spectral_bias} has shown that neural networks are biased toward learning lower frequency features first, and their generalisation capability degrades with an increase in the frequency content of the features. In the context of \ac{GW} chirps, this amounts to generalising differently to different \ac{GW} frequency profiles and even parts of a given chirp signal. The bias due to the skewed concentration of integrated \ac{SNR} toward higher frequencies will amalgamate with spectral bias and result in some uncertain learning dynamics. With regard to detector noise, neural networks should be more prone to higher frequency noise than lower frequency. Models may struggle to consistently differentiate between sharp glitch features and the late inspiral or merger phases of true \ac{GW} signals, potentially leading to degraded detection sensitivity or misclassification in these regions.
        
        \subsubsection{\label{subsubsec:User-introduced Biases}Experimenter Bias}
        
        \textit{Bias due to disproportionate evaluation}: The mock data challenge, MLGWSC-1 \cite{DP5_MLGWSC1_2023}, describes the sensitive volume as
        \begin{equation}
            \label{eqn:sensitive_volume}
            \mathcal{V}(\mathcal{F})\approx\frac{V(d_{\text{max}})}{N_I}\sum_{i=1}^{N_{I,\mathcal{F}}}\left(\frac{\mathcal{M}_{c,i}}{\mathcal{M}_{c,\text{max}}}\right)^{5/2},
        \end{equation}
        where $V(d_{\text{max}})$ is the volume enclosed within a radius $d_{\text{max}}$, $\mathcal{F}$ is the \ac{FAR}, $d_{\text{max}}$ is the maximum luminosity distance of all injected sources, $N_I$ is the number of simulated signals present in the data, $N_{I,\mathcal{F}}$ is the number of found signals at a given \ac{FAR} and $\mathcal{M}_c$ is the chirp mass. The sensitive volume metric is evaluated at the required false alarm rates, and the sensitive distance is defined as $(3\times\mathcal{V}(\mathcal{F})/4\pi)^{1/3}$. The metric places an increased weighting to higher chirp mass detections given by $(\mathcal{M}_{c,i}/\mathcal{M}_{c,max})^{5/2}$ on each detection indexed by $i$. An \ac{ML} model that is biased against low chirp mass signals and operating on a subset of the parameter space considered by matched filtering will find more high chirp mass signals comparatively at a given \ac{FAR}. Given the imbalanced chirp mass weighting, the biased model will achieve a higher sensitive distance but detect a lower number of signals overall compared to matched-filtering at all \ac{FAR}s.
        
        \textit{Bias due to train-test overlap}: Train-test overlap is measured as the extent to which the testing data features are present in the training data. Significant overlap will lead to overly optimistic evaluation results and will not reflect the true generalisation performance of the model on unseen data.
        
        For the signal class, the features are defined in a continuous parameter space with well-defined bounds given by the chosen training distributions, which are expected to be the same between the training and testing datasets. In this case, we seek an \ac{ML} model that interpolates the training data. Detrimental overlap for the signal class amounts to having signals produced using the same combination of \ac{GW} parameters in both datasets. Signs of model overfitting will be masked due to this overlap. However, approximate overlap, where the parameters are similar but not the same (from the same distribution), will always exist between the datasets and is not detrimental.
        
        For the noise class, no well-defined definition exists for the features, and there is often a mismatch between training and testing datasets. In this case, we seek an \ac{ML} model that \textit{extrapolates} the training data. Detrimental overlap for the noise class amounts to using the same realisations of noise \ac{PSD} or glitches in both datasets. Using a testing dataset that does not have appropriate \ac{OOD} noise realisations is detrimental, as it does not reflect the model's extrapolation capabilities.
    

    \subsection{\label{subsec:Key results} Key Contributions to the Field}

    \noindent \textbf{Identified sources of learning bias}: We provide explanations for several sources of \ac{ML} learning bias for the \ac{BBH} \ac{GW} detection problem. However, they are general enough to be easily extrapolated to other domains of \ac{GW} data analysis.
    
    \noindent \textbf{Provided bias mitigation tactics}: In section \ref{sec:Methods}, we will provide translatable mitigation tactics and training strategies to \textit{concurrently} address almost all the aforementioned biases.
    
    \noindent \textbf{Improvements to detection performance}: In section \ref{subsec:Comparison with matched-filtering}, we evaluate our pipeline on the injection study presented in MLGWSC-1 and show that our detection pipeline, \texttt{Sage}, is capable of detecting $\approx11.2\%$ and $\approx48.29\%$ more signals than the \texttt{PyCBC} submission in \cite{DP5_MLGWSC1_2023} and the \texttt{AresGW} results in \cite{DP6_Nousi_2023}, at an \ac{FAR} of one per month in O3a noise. 

    \noindent \textbf{Negating the need for \ac{PSD} estimation}: Contrary to previous \ac{ML}-based detection pipelines operating on real detector noise \cite{DP6_Nousi_2023, aframe, DP7_Zelenka_2024, Koloniari_gw_events}, we show that \texttt{Sage} does not need to use a \ac{PSD} estimate for whitening any input sample (see section \ref{sec:Methods}, \textit{Data Transformation}). We use a single approximate \ac{PSD} instead for all samples. The primary benefit of this feature is not that it reduces computational cost during data preprocessing, but the idea of having ``less moving parts" in a low-latency search setting, thus reducing the maintenance needs and improving the efficiency of troubleshooting issues. 
    
    \noindent \textbf{Mitigating biases might aid in glitch rejection}: In section \ref{subsubsec:Glitch rejection capability}, we will obtain all the triggers generated by \texttt{Sage} and \texttt{PyCBC} during glitch events in one month of O3a noise and prove that \texttt{Sage} is capable of effectively rejecting glitches. We also argue about the potential benefits of mitigating biases for glitch rejection.
    
    \noindent \textbf{\ac{OOD} \ac{PSD}s are good for classification}: In section \ref{subsubsec:Effects of PSD}, we will show that using a broader range of noise \ac{PSD}s for the training data, which could even be \ac{OOD} to the testing dataset, greatly aids in detection performance.

    \noindent \textbf{Glitches make classification easier}: We show in section \ref{subsubsec:Effects of PSD} that models trained with real detector noise produce more confident classifiers than models trained with simulated coloured Gaussian noise.

    \noindent \textbf{Parameter efficiency via inductive biases}: Via an ablation study (see section \ref{subsec:Ablation study}) we show that a model that is larger than \texttt{Sage} in terms of number of trainable parameters is not able to achieve nearly the same detection sensitivity as \texttt{Sage}, \textit{if} it does not have the proper inductive biases.
    

\begin{figure}[b!]
    \centering
    \includegraphics[width=1.0\linewidth]{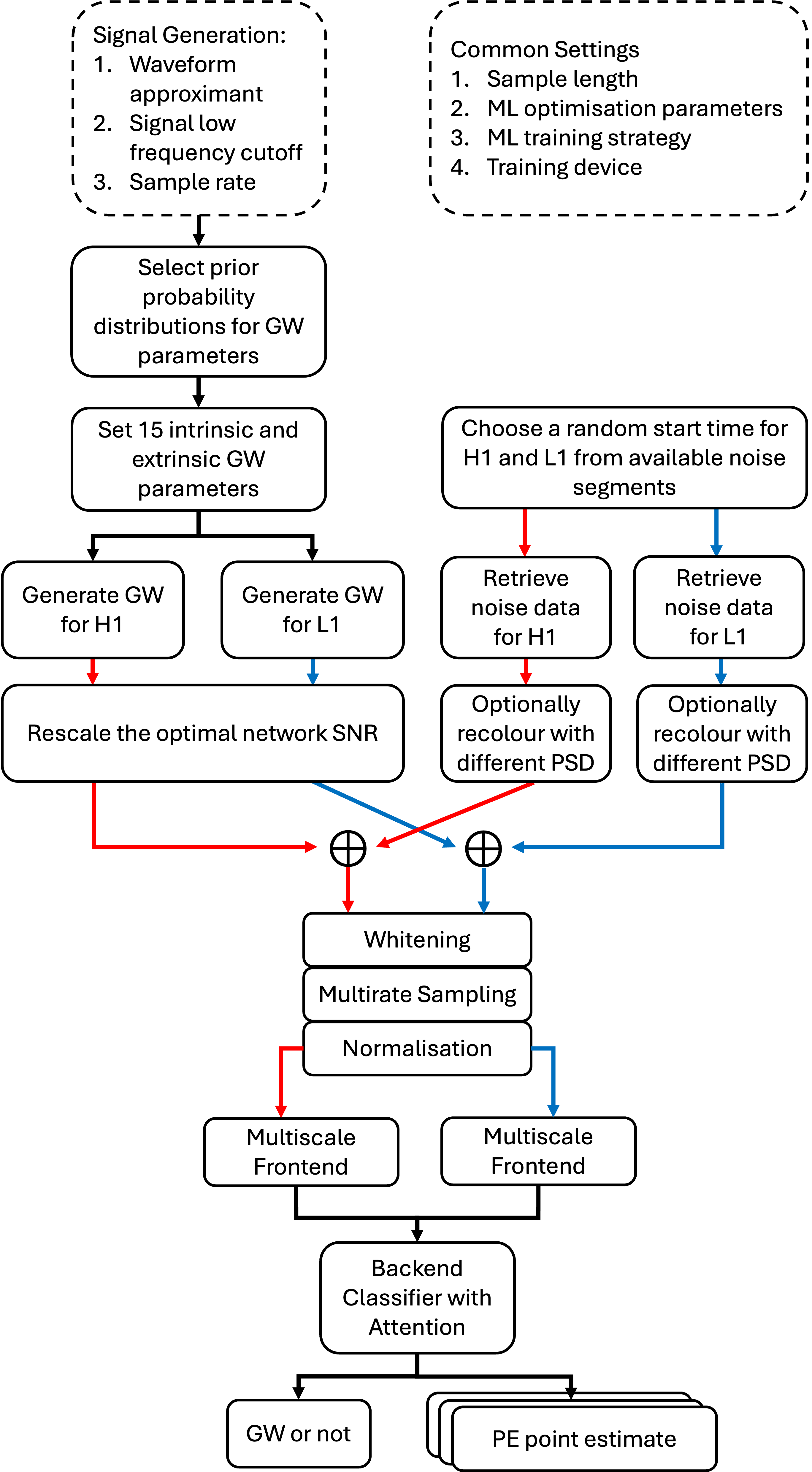}
    \caption{Flowchart of \texttt{Sage} methodology. The details in the dotted box are set once per run; the signal generation box is specific for generating the signal class; the common settings apply for both the signal and noise class. All boxes with solid lines are iterated over during the training process.}
    \label{fig:methodology_flowchart}
\end{figure}

\begin{table*}[t]
\setlength{\tabcolsep}{12pt}
\begin{tabular}{c c c c} 
    \toprule \toprule
    {Notation} & {Parameter} & {Parameter Distribution} & {Constraint}\\ \midrule
    $m_1, m_2$ & Masses & $m_1, m_2\in U(7.0, 50.0) \, \textup{M}_\odot$ & $m_1>m_2$ \\
    $\chi$ & Spins & $\;\;\;\chi\in U(0.0, 0.99)$ & distributed isotropically \\ \midrule
    $\iota$ & Inclination & $\cos{\iota}\in U(-1, 1)\;\;\;\;\;\;$ & - \\ 
    $\Phi_0$ & Coalescence Phase & $\Phi_0\in U(0, 2\pi)\;\;\;\;$ & - \\
    $\Psi$ & Polarisation & $\Psi_0\in U(0, 2\pi)\;\;\;\,$ & - \\
    $\alpha$ & Right ascension & $\alpha\in U(0, 2\pi)\,\,\,$ & - \\
    $\delta$ & Declination & $\sin{\delta}\in U(-1, 1)\;\;\;\;\,\,\,$ & - \\ 
    $d_c$ & Chirp Distance & $\;\;\;\;\;\;\;\;\;\;\;\;\;\;\,d_c^2\in U(130^2, 350^2) \, \textup{Mpc}^2$ & - \\ 
    \bottomrule \bottomrule
\end{tabular}
\caption{Summary of testing dataset \ac{GW} parameter distributions used for intrinsic and extrinsic \ac{GW} parameters. This corresponds to the D4 realistic dataset of the MLGWSC-1\cite{DP5_MLGWSC1_2023}.}
\label{tab:testing_dataset_priors}
\end{table*}

\section{\label{sec:Methods}Methods}
In this section, we elaborate on our \texttt{Sage} pipeline, which enables the detection of \ac{GW}s at an efficiency that rivals the production-level search algorithm \texttt{PyCBC} in real detector noise. We provide effective mitigation tactics and training strategies that aid in mitigating the biases mentioned in section \ref{subsec:Issues} concurrently. Mechanisms that are requisite for full reproducibility are explained in detail. All code and implementations are available at: \url{https://github.com/nnarenraju/sage} \href{https://github.com/nnarenraju/sage}{\faGithubSquare}.

Figure \ref{fig:methodology_flowchart} shows a flowchart of the entire methodology for \texttt{Sage}. The following sections have been arranged in the order of the decision-making process. The entire methodology is fine-tuned based on prior knowledge about the testing dataset.

\begin{figure*}[htbp]
    \centering
    \includegraphics[width=1.0\linewidth]{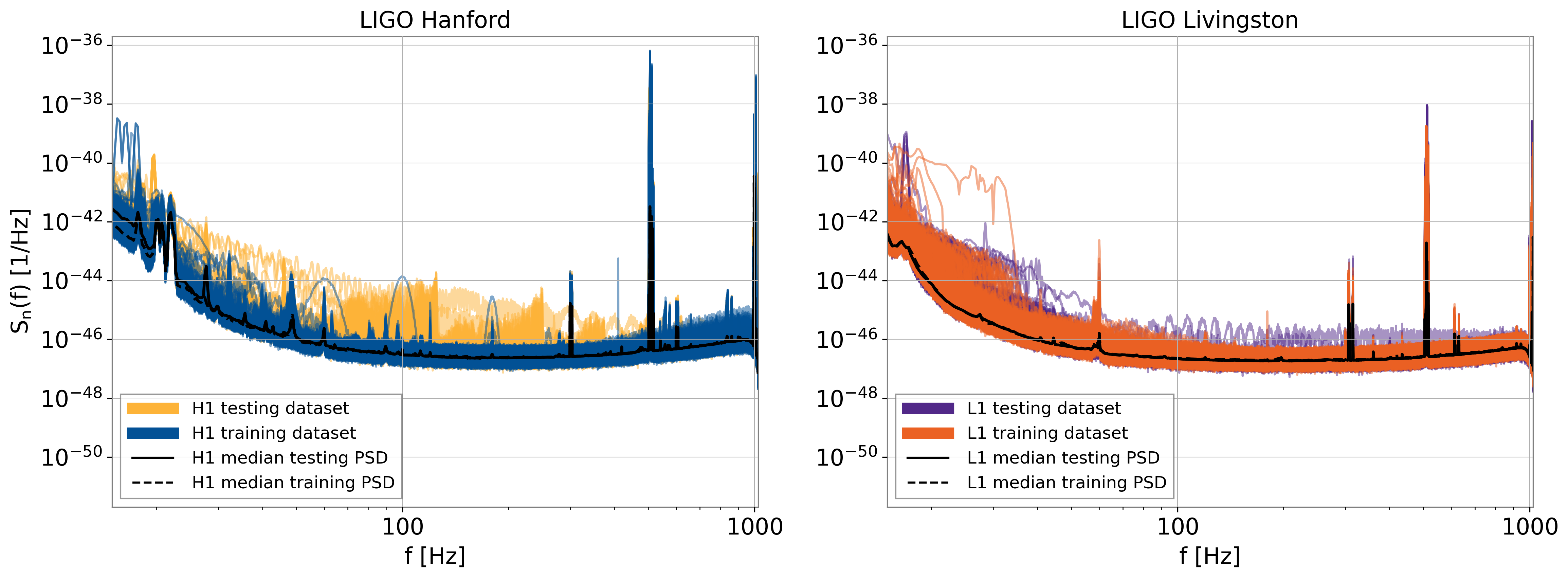}
    \caption{Amplitude of estimated strain power as a function of frequency for different segments of O3a noise. We compare the estimated \ac{PSD}s in the training (blue and orange curves) and testing (yellow and purple) datasets for the LIGO Hanford and Livingston detectors. Both plots have an $f_{\text{low}}$ of $15$ Hz at the noise low-frequency cut-off and an $f_{\text{high}}$ of $1024$ Hz at the Nyquist limit of the sampling rate. The solid and dashed black lines show the median \ac{PSD} of all estimated \ac{PSD}s in the testing and training dataset, respectively.}
    \label{fig:o3a_psds}
\end{figure*}

    \subsection{\label{subsec:Testing Dataset Priors}Testing Dataset Parameter Distributions}
    
    Any and all prior knowledge about the testing data is crucial in determining the \ac{GW} parameter distributions for the training dataset, the network architecture and the training strategies required for efficient learning. 
    
    \textit{Testing dataset configuration}: A summary of the testing data parameter distributions used for all \ac{GW} parameters is provided in table \ref{tab:testing_dataset_priors}. The signals are generated as described in the realistic ``dataset 4" (D4) of the MLGWSC-1 \cite{DP5_MLGWSC1_2023}. We use the \verb|IMRPhenomXPHM| waveform approximant \cite{IMRPhenomXPHM} that includes the effect of precession and higher-order modes. All higher-order $(l, m)$ modes available in \verb|IMRPhenomXPHM| are included, namely $(2, 2)$, $(2, -2)$, $(2, 1)$, $(2, -1)$, $(3, 3)$, $(3, -3)$, $(3, 2)$, $(3, -2)$, $(4, 4)$ and $(4, -4)$. Signals have a low-frequency cut-off of $20$ Hz, and at least 33 per cent of them should have an optimal network \ac{SNR} below the threshold of 4 \cite{DP5_MLGWSC1_2023}. Figure \ref{fig:testing_mock_datset_all_priors} in appendix \ref{app:testing dataset priors} shows the 1D histograms of all relevant intrinsic and extrinsic parameters of $10^6$ signals, using table \ref{tab:testing_dataset_priors} for the parameter distributions. A \textit{sample} in the testing dataset is defined as the time series data from a two-detector network (H1 and L1).

    \texttt{PyCBC} only uses the fundamental mode and does not consider spin-induced precession effects in MLGWSC-1 \cite{DP5_MLGWSC1_2023}. If we were to include these effects in the training dataset configuration, gains in detection efficiency compared to \texttt{PyCBC} seen in our evaluation of the testing dataset could be attributed to this disparity. However, this is \textit{contingent} on the detectability of the two effects in the testing dataset.
    
    \textit{Detectability of higher-order modes}: The relative power of higher-order modes is primarily dependent on, and proportional to, mass-ratio \mbox{($q=m_1/m_2$)} and total mass ($M$). Ignoring spins and the dependency on orientation, the detectability of higher-order modes relative to the \mbox{$l=2,\;|m|=2$} fundamental mode for an \ac{SNR}$=20$ signal is provided in \cite{higher_order_multipoles_detectability}. All measures were estimated for a single LIGO detector with a sensitivity comparable to that achieved during the third observing run. The parameter space in table \ref{tab:testing_dataset_priors} dictate that generating a signal with an optimal network \ac{SNR} of $10$ is in the $\approx99^{\text{th}}$ percentile and a mass-ratio of $2$ is in the $\approx65^{\text{th}}$. Given the statistical improbability of generating a high \ac{SNR}, high mass ratio signal with sufficient total mass, we can expect negligible influence due to higher-order modes on the detection efficiency of our model.
    

    \textit{Detectability of precession}: The impact of varying different \ac{GW} parameters on the detectability of precession is presented in \cite{precession_detectability}. They conduct their analyses assuming O2 detector sensitivities for LIGO  but show that their results are unlikely to be affected when considering O3 sensitivities. Their results indicate that the recovered posterior distribution of precessing spin for a low \ac{SNR}($=10$) signal resembles the prior, indicating the non-informativeness of the noisy data. Moreover, a signal \ac{SNR} of $20$-$30$ is shown to be required for a $5^{\text{th}}$ percentile of the precession \ac{SNR} distribution to be marginally detectable. This shows that the testing dataset is unlikely to have any signals with detectable precession effects.
    
    The noise characteristics, although not analytic in the same sense as signals, can be summarised using the behaviour of the PSDs and the non-Gaussian transient noise artefacts. MLGWSC-1 provides approximately 81 days of coincident real noise for the LIGO Hanford (H1) and LIGO Livingston (L1) detectors from the O3a observing run \cite{GWTC-2}. This consists of several noise segments that coincide in GPS time between H1 and L1 (referred to as time-correlated hereafter), with a duration of at least 2 hours. All segments are $10$ seconds away from any detection listed in GWTC-2 \cite{GWTC-2}. They use coincident noise segments that span a total of 30 days (day 1 to day 30 inclusive), between GPS time 1238205077 (2$^{\text{nd}}$ of April 2019, 01:50:59) to 1244605921 (15$^{\text{th}}$ of June 2019, 03:51:43) to create the ``background" testing dataset. Signals were injected every $24$ to $30$ seconds into the background to form the ``foreground" testing dataset. The parameters of the injected signals and their times of coalescence within the testing dataset noise are fixed using the seed provided by MLGWSC-1 for D4. All noise segments were ensured to have the \verb|data| flag\footnote{Definition of the flag provided at \href{https://gwosc.org/datadef/}{this https url}. Accessed on Nov 2024.} and to not have any of the \verb|CBC_CAT1|, \verb|CBC_CAT2|, \verb|CBC_HW_INJ|, or \verb|BURST_HW_INJ| flags. The noise has a low-frequency cut-off of $15$ Hz and is downsampled to a sampling rate of $2048$ Hz. As a result of choosing noise segments over a long period of time, the \ac{PSD} should vary by a non-negligible amount in the testing dataset. Figure \ref{fig:o3a_psds} shows estimated \ac{PSD}s of every $15$ seconds of the 30 days of the background testing data using the Welch method \cite{welch_method}, alongside the median of all estimated \ac{PSD}s. Compared to the median, we see some outlier \ac{PSD} realisations that are likely due to transient non-Gaussian noise artefacts but could also be caused by other known or unknown factors affecting the detector behaviour.
    
    The frequency of non-Gaussian noise artefacts was obtained by comparing the glitch event times reported in Gravity Spy \cite{gravity_spy_o3ab} to the GPS times of the noise segments used for testing. A total of 16,531 and 28,739 glitches were present in the H1 and L1 O3a testing data respectively. We can validate the glitch rejection capabilities of \texttt{Sage} compared to \texttt{PyCBC} using this information. Any conclusions obtained from our analysis are strictly under the caveat that the evaluation was performed on a limited testing dataset. We leave an in-depth investigation into the glitch rejection and \ac{PSD} handling capabilities of \texttt{Sage} to the future.

    We primarily evaluate \texttt{Sage} on the realistic D4 dataset described above. However, to experiment on a testing dataset without glitches and significantly varying noise \ac{PSD}s, we also use the ``dataset 3" (D3) testing dataset from the MLGWSC-1. All aspects of the methodology are tailored to D4, and any changes made to accommodate D3 will be explicitly mentioned in the relevant sections.
    
    \textit{Modifications for D3}: This dataset uses the same \ac{GW} parameter distributions for generating the signals but uses 30 days of simulated coloured Gaussian noise instead of real O3a noise. The 30 days of noise is split into several segments that are at least 2 hours in length. For a given detector, each noise segment is coloured using one of 20 noise \ac{PSD}s estimated from real O3a noise for that detector \footnote{\ac{PSD}s provided in the MLGWSC-1 GitHub repository at \href{https://github.com/gwastro/ml-mock-data-challenge-1}{this https url}}. We use the seed provided in the challenge paper to produce the testing dataset for a fair comparison to the D3 results submitted by \texttt{PyCBC} and ensure that we use a different seed for training \texttt{Sage}.

    \subsection{\label{subsec:Training Dataset Priors}Training Dataset Parameter Distributions}
    It is typical and necessary in supervised machine learning to choose a feature distribution during training that closely matches the unseen testing features, for optimal generalisation. For the most part, we abide by the waveform generation parameters and testing dataset parameter distributions mentioned in section \ref{subsec:Testing Dataset Priors}. We will use this section to provide justification for any deviations. Although we argued in section \ref{subsec:Testing Dataset Priors} that neither precession nor higher-order modes are detectable in the testing dataset, it does not warrant the usage of a waveform model like \verb|IMRPhenomD|, which does not account for those effects, for training. Although the usage of a more accurate waveform model like \verb|IMRPhenomPv2| or \verb|IMRPhenomXPHM|, will take some unnecessary learning capacity from the model, when evaluated on the proposed testing dataset, the additional \ac{GW} domain knowledge allows the model to more effectively reject samples from the realistic noise class. In the interest of preserving computational resources, we use \verb|IMRPhenomPv2|, which is less expensive, instead of \verb|IMRPhenomXPHM|. In order to facilitate this usage, it is necessary to employ an \ac{SNR} distribution that operates in the regime of generating signals where precession is detectable. We use this necessity as a gateway argument to further investigate the requirements of the training dataset parameter distributions. 
    
    As mentioned in sections \ref{subsubsec:Learns Biases} and \ref{subsec:Testing Dataset Priors}, using the testing dataset parameter distributions for training is not a sufficient condition to guarantee that the model learns an unbiased representation of the required parameter space. For example, the sample distribution of a given parameter and the relative sample difficulty within the parameter bounds will heavily affect the generalisation outcome. However, given the number of intrinsic and extrinsic \ac{GW} parameters, it is not trivial to determine the joint distribution that ensures the fastest possible convergence of a training model to an unbiased representation.
    
    As an empirical alternative, we study the model's testing performance by training on the joint distribution of the testing dataset described in MLGWSC-1 (reproduced in table \ref{tab:testing_dataset_priors}) and alter the probability distribution function of the parameters over which the model exhibits learning bias. We define the joint parameter distribution of \ac{GW} parameters that leads to the fastest convergence of a given \ac{ML} model to its global minimum, without the biases due to limited sample and limited feature representation, as the optimal joint parameter distribution. Through experimentation, we determined chirp mass, mass ratio and \ac{SNR} as being the central source of such biases and altered their sample distributions in multiple intuitive ways to try to move closer to the optimal joint parameter distribution. The upper and lower bounds for the chirp mass and mass ratio distributions remain unchanged between the training and testing datasets, but we do change the bounds on network optimal \ac{SNR} for the training distribution, which is reflected on the distance parameter.
    
\begin{figure}[htbp]
        \centering
        \includegraphics[width=1.0\linewidth]{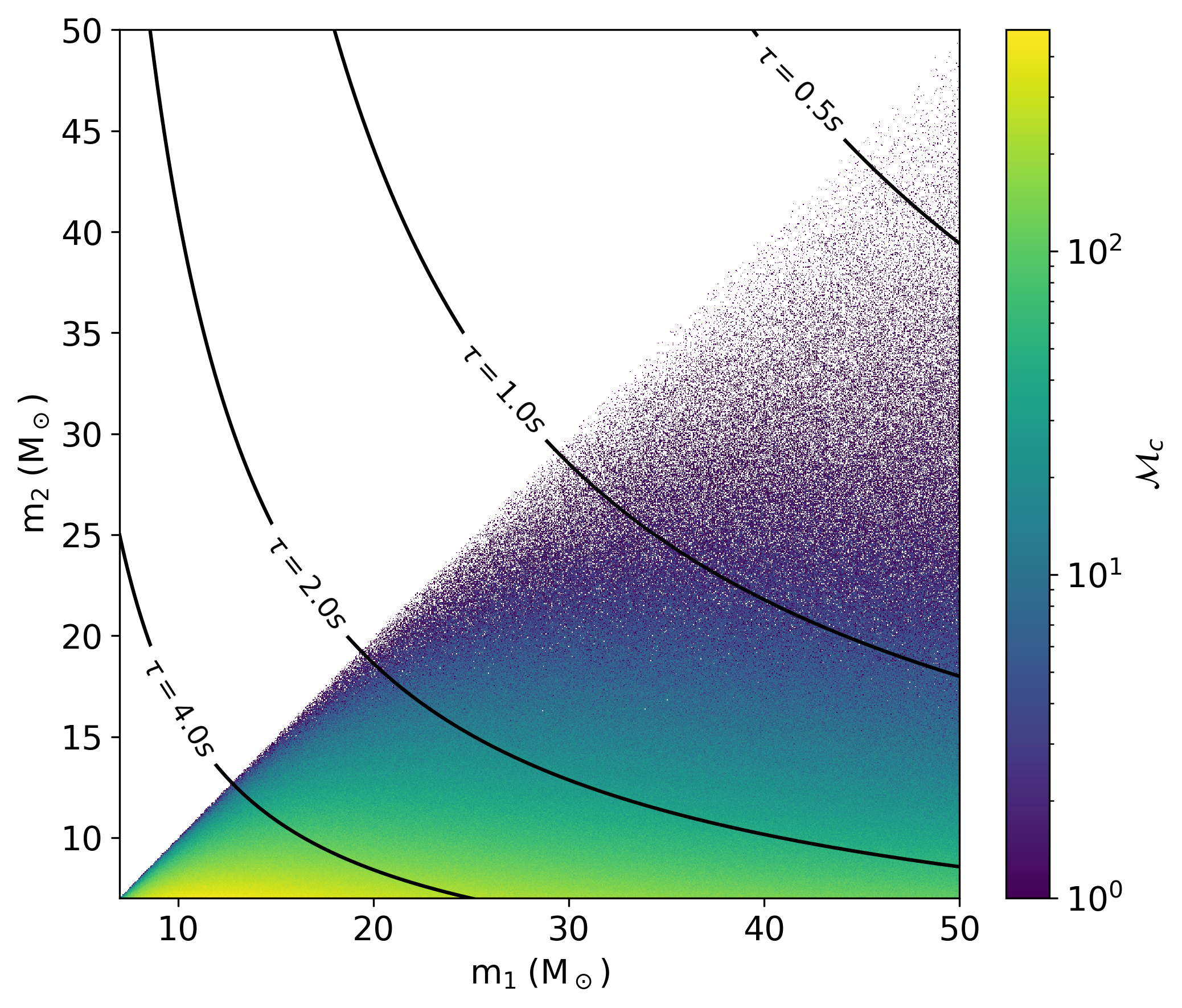}
        \caption{2D histogram of the joint distribution of component masses, $\textup{P}(m_1, m_2)$ used to generate the training dataset. $10^8$ points were uniformly sampled from $U(\tau_0, \tau_3)$; points that lie within the mass bounds were accepted ($\approx13.4\%$) and the rest were rejected.}
        \label{fig:training_dataset_priors}
    \end{figure}

    Chirp mass $\mathcal{M}_c$ and mass ratio $q$ are two independent mass coordinates, which we can analytically invert to obtain the component masses. This allows us to choose any arbitrary probability density $p(\mathcal{M}_c)$ and $p(q)$, and populate the $(\mathcal{M}_c, q)$ space within the required bounds (analytically derived from the bounds on $m_1$ and $m_2$). We can then analytically invert from $(\mathcal{M}_c, q)$ to $(m_1, m_2)$, and apply rejection sampling based on the constraints: $m_1>m_2$, $m_{\text{min}}<m_1<m_{\text{max}}$ and $m_{\text{min}}<m_2<m_{\text{max}}$. This procedure can be exploited to experiment with any chosen mass coordinate pair. For experimentation, we chose various realisations of the power-law and beta distributions alongside the uniform distribution for some chosen mass coordinate pair. To address the arguments in section \ref{subsubsec:Learns Biases} about a potential bias against signals with longer signal duration, we experimented with the $(\tau_0, q)$ and $(\tau_0, \tau_3)$ pairs as well. Here, we redefine $\tau_0$ from equation \ref{eqn:t_from_mc_Cutler_Flanagan} in terms of $M$ and $\eta$, and define $\tau_3$ as
    \begin{align}
        \tau_0 &= \frac{5}{256 \pi f_L \eta} (\pi M f_L)^{-5/3} \label{eqn:tau0},\\
        \tau_3 &= \frac{1}{8 f_L \eta} (\pi M f_L)^{-2/3} \label{eqn:tau3}.
    \end{align}
    We chose the $(\tau_0, \tau_3)$ pair to investigate the use of the template density distribution inspired by the metric based on the SPA model \cite{SPA_model_metric}. The choice of chirp times $\tau_0$ and $\tau_3$ as coordinates on the parameter space comes with the benefit of the metric being constant at 1PN and essentially near-constant for higher PN-orders in the local vicinity of every point in the parameter space, and almost constant throughout \cite{tau0_tau3_cartesian, hexagonal_template_placement}. Moreover, using the ($\tau_0, \tau_3$) pair comes with the added benefit of simple analytical inversion with the ($M, \eta$) pair \cite{hexagonal_template_placement}. If $p(\tau_0, \tau_3)$ were uniform within the bounds imposed by table \ref{tab:testing_dataset_priors}, it would translate to an $(m_1, m_2)$ space as shown in Fig. \ref{fig:training_dataset_priors}. 
    
    \begin{figure*}[htbp]
        \centering
        \includegraphics[width=1.0\linewidth]{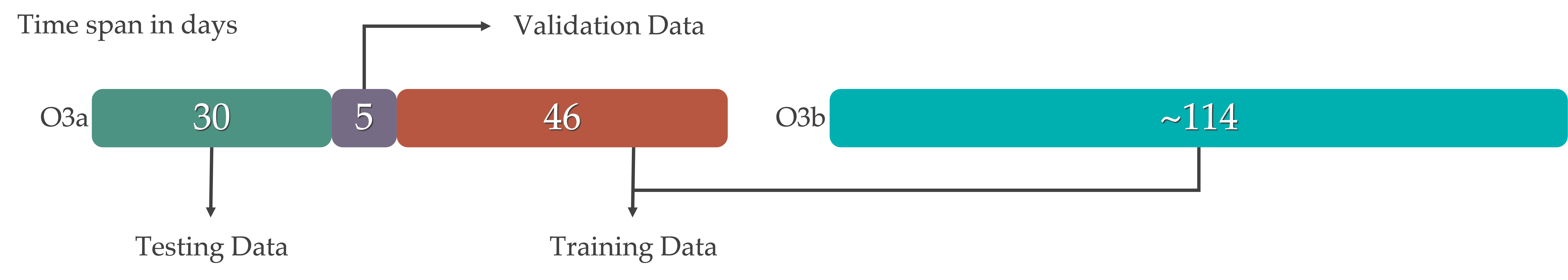}
        \caption{Time spans (in days) for the training, validation, and testing datasets taken from the O3a and O3b observing runs for the H1 and L1 detectors. The O3a noise is provided by the MLGWSC-1 \cite{DP5_MLGWSC1_2023} and is time-correlated between the H1 and L1. We downloaded the O3b noise from GWOSC, and it is not time-correlated.}
        \label{fig:datasets}
    \end{figure*}
    
    Finally, the network optimal \ac{SNR} distribution is varied to conform to a uniform distribution, a half-normal distribution, or different variations of a beta distribution. The lower bound on \ac{SNR} is set to 5, and the upper bound is allowed to differ with experiments but is always $\geq15$. Change to the \ac{SNR} is assumed to reflect only on the distance parameter. Using a lower bound of 5 ensures a lower overlap with the noise class and a significant reduction in noisy labels. The upper bound must be high enough for the network to learn the scaling transformation of a signal resulting from a change in \ac{SNR}. However, we also found that the network does not require as many high \ac{SNR} signals in the training dataset as low \ac{SNR}. A uniform distribution for \ac{SNR}, for example, biased the network against learning low \ac{SNR} samples. For our experiments, we typically choose a parameter distribution that is skewed toward low \ac{SNR}s. 
    
    It is essential to mention that any significant results obtained from the above experimentation is only the best result from a limited set of joint distribution realisations, and it may or may not be close to the optimal joint parameter distribution.

    The testing dataset parameter distributions dictate that the longest possible waveform that could be present in the unseen data will be $\approx 12.74$ seconds in length for a low-frequency cut-off of $20$ Hz, ignoring the effects due to higher-order modes and precession. This was calculated using \verb|SimIMRPhenomDChirpTime| in \texttt{lalsimulation} with a $10\%$ margin of error. Our primary results in section \ref{sec:Results} all use a $12$-second sample length for the training dataset, which should contain the majority of the signal \ac{SNR} for all considered signals. Since the signal duration could be longer when considering more realistic estimates of chirp time, we conducted an experiment with a $20$-second sample length as well. Using a $20$-second sample length would encompass $100\%$ of the signal \ac{SNR}, but would require longer training and evaluation times, while also increasing the problem difficulty. We found that using a $12$-second was sufficient to learn the parameter space well.

    With regard to noise, we have the latter 51 days (day 31 to day 81 inclusive) of the provided 81 days of O3a noise via MLGWSC-1 that is not used for the testing dataset, GPS time 1244608795 (15$^{\text{th}}$ of June 2019, 04:39:37) to GPS time 1253966055 (1$^{\text{st}}$ of October 2019, 11:53:57). We use the first 5 days of the 51 days as the validation dataset. The rest of the data (day 36 to day 81 inclusive), alongside $\approx114$ days and $\approx113$ days of H1 and L1 O3b noise, is used for training the model (refer to Fig. \ref{fig:datasets}). The downloaded O3b noise segments span between GPS time 1256655618 (1$^{\text{st}}$ of November 2019, 15:00:00) to GPS time 1269363618 (27$^{\text{th}}$ of March 2020, 17:00:00). They have a duration of at least 1 hour and are 30 seconds away from any detection listed in \texttt{GWOSC}\footnote{Event list obtained from \href{https://gwosc.org/eventapi/html/allevents/}{this https url}, and noise segment metadata obtained from \href{https://gwosc.org/timeline/show/O3b_4KHZ_R1/H1_DATA*L1_DATA*V1_DATA/1256655618/12708000/}{this https url} (accessed May 2024)}. All noise segments were ensured to have the \verb|data| flag and to not have any of the \verb|CBC_CAT1|, \verb|CBC_CAT2|, \verb|CBC_HW_INJ|, or \verb|BURST_HW_INJ| flags. The noise segments provided by the MLGWSC-1 for O3a noise were constrained to coincide in GPS time between the two detectors, which limits the total amount of data used for training. Given that time correlation is not a requirement for training, the O3b noise segments that we downloaded are not constrained in this manner. More information on how this data is used to create the training samples is provided in section \ref{subsec:Data Generation}.
    
    Similar to the testing dataset, we estimated the PSDs for the entirety of the training data, and the results are compared in Fig. \ref{fig:o3a_psds} for both detectors. It was a cause for concern to notice several outlier PSDs in the H1 testing dataset that had higher energies in the relevant frequency range when compared to the training dataset. These PSDs in the testing dataset can be considered out-of-distribution (OOD) and the model might struggle during testing when encountering them. There is ample precedence for this expectation in the machine learning literature \cite{ml_deployment_problems, google_retinopathy_issues}. Since the noise \ac{PSD}s and glitches are different between the training and testing datasets, validating the model on a fixed set of noise realisations will bias our choice of the best-performing network instance. A significant overlap between the training and validation set will hinder the ability to detect the onset of overfitting, and using the testing dataset to validate will again bias our choice of the best performing network. The five-day validation dataset is not the best choice given these biases, but will still be useful in detecting signs of overfitting.

    \textit{Modifications for D3}: To evaluate \texttt{Sage} on the D3 testing dataset, we train a separate \ac{ML} model using simulated coloured Gaussian noise instead of real O3a noise.

    \textbf{Biases Addressed}: We address the \textit{bias due to limited sample/feature representation} by choosing an appropriate training dataset parameter distribution for chirp mass, mass ratio and \ac{SNR}. For example, signals with longer duration or higher mass ratio are represented much better in $U(\tau_0, \tau_3)$ than $U(m_1, m_2)$ mass distribution. \textit{Bias due to sample difficulty} can be addressed by choosing the optimal \ac{SNR} distribution to be a half-normal distribution, which places a higher weight on the harder-to-classify low \ac{SNR} signals. Placing the lower bound for optimal \ac{SNR} at 5 reduces the \textit{bias due to class overlap}. Using a $12$-second sample length addresses the \textit{bias due to insufficient information} and using a broad enough parameter space for \ac{SNR} addresses \textit{bias due to lack of variation}. Finally, using a larger amount of noise leads to a broader training distribution for noise \ac{PSD}s and glitches, and addresses the \textit{bias due to lack of variation}.

    \subsection{\label{subsec:Data Generation}Data Generation}
    
    \textit{Generation of signal samples}: All previous studies on machine-learning based \ac{GW} detection have used a fixed dataset size to train their model. Our investigations on mitigating learning bias showed that \ac{IID} training dataset size plays an equal role alongside the joint sample distribution of the \ac{GW} parameters. We use \texttt{PyCBC} to generate frequency domain \verb|IMRPhenomPv2| waveforms on-the-fly (OTF) during the training phase of our model. This allows for unlimited signal data, constrained only by the number of training epochs, and is equivalent to augmentation on all \ac{GW} parameters. Each sample is injected into a noise realisation, the generation of which is described in the remainder of this section. 
    
    \textit{Generation of noise samples}: Unlike signals, we do not have the liberty to produce unlimited \ac{IID} samples from $\approx$160 days of coincident detector noise. However, we can obtain unlimited non-\ac{IID} noise realisations as a near-ideal proxy. Inspired by the time-slides method for background estimation in a network of \ac{GW} detectors \cite{time_slides}, we select start times independently for each detector from a uniform distribution over the entire $\approx$160-day span of available data. These samples may or may not be time correlated and result in a large number of non-\ac{IID} noise realisations. However, due to the limited number of non-Gaussian noise transients and varying \ac{PSD} realisations, the advantage that the model gains from this method will eventually saturate. An estimate of the number of \ac{IID} samples will provide us with some insight on when this will happen. At a sampling frequency of $2048$ Hz, 160 days of noise gives us access to $\approx1.15\times10^{6}$ \ac{IID} samples. Between 2 detectors, there are $\approx1.33\times10^{12}$ noise realisations, which is the Cartesian product of two sets of $\approx1.15\times10^{6}$ samples in H1 and L1 (essentially time-shifting the detector data by more than the sample length of $12$ seconds). The final number depends on noise augmentation and is provided in section \ref{subsec:Data Augmentation}.

    \textit{Modifications for D3}: Signal generation is exactly the same for D3 and D4. For noise generation, we first produce white Gaussian noise with a different seed for each sample. Each noise sample of the noise class and additive noise in the signal class is coloured using one of the 20 estimated noise \ac{PSD}s that is used for creating the D3 testing dataset. This nullifies the difference between the training and testing noise \ac{PSD} distribution, meaning the trained model will not encounter any \ac{OOD} noise samples during testing.

    \textbf{Biases Addressed}: Using on-the-fly signal generation is equivalent to augmenting on all \ac{GW} parameters and addresses the \textit{bias due to lack of variation} and \textit{bias due to limited feature representation}. This increases the number of discrete points within the continuous transformation space of all \ac{GW} parameters and allows the network to learn the transformation well. This bias is also addressed for noise by using random start times for each detector.

    \subsection{\label{subsec:Data Augmentation}Data Augmentation}
    \textit{Signal Augmentation}: Since we use on-the-fly signal generation, which essentially augments on all \ac{GW} parameters, it is not necessary to include any explicit form of signal augmentation. The optimal network \ac{SNR} is, however, shifted and re-scaled in accordance with probability distributions mentioned in section \ref{subsec:Training Dataset Priors}. This action effectively alters the distance and, thereby, the chirp distance distribution. Any correlation that the original optimal network \ac{SNR} had with the other \ac{GW} parameters is lost. Although there is some loss of astrophysical prior knowledge, the null correlation should help to simplify the data manifold of the signal class. This reduces the \textit{bias due to sample difficulty} affecting the learning of low \ac{SNR} regions of the parameter space. For example, binary systems with edge-on inclinations will typically have lower \ac{SNR}s than face-on systems. Without removing this correlation, the network will be biased against learning edge-on systems. To show the benefit of on-the-fly signal generation, \texttt{Sage} also includes limited signal augmentation procedures for the sake of comparison - after generating a fixed size training dataset of $h_+$ and $h_{\times}$ signal components, we randomly choose the polarisation and sky-position based on the distributions provided in table \ref{tab:testing_dataset_priors}, before projecting the componenets into $h(t)$.

    \textit{Noise Augmentation}: The noise sample generation procedure explained in section \ref{subsec:Data Generation} innately includes noise augmentation via the choice of random start times and the non-necessity of time correlation between detectors. Given that not all \ac{IID} noise samples are significantly different, the potency of this augmentation to lower the training loss eventually saturates. However, we postpone this saturation effect by optionally recolouring (with a $40\%$ probability of being recoloured) the noise realisation of each detector with a pre-computed set of noise PSDs from the $\approx160$ days of O3 training noise. We compute a total of 250,000 \ac{IID} PSDs using $15$-second non-overlapping noise segments for each detector using the Welch method.
    
    To investigate any changes in detection efficiency during the testing phase due to the lack of training on OOD PSDs seen in Fig. \ref{fig:o3a_psds}, we also precompute a 172,800 set of PSDs from the 30 days of O3a testing data (again using $15$-second non-overlapping noise segments). Any increase in detection efficiency observed during testing due to augmentation using these PSDs during training could indicate the lack of proper generalisation to OOD PSDs. During the recolouring stage, we also have the option to shift the entire \ac{PSD} up or down to either increase or decrease the overall noise power.

    Previously, in section \ref{subsec:Data Generation}, we showed that our noise generation procedure has access to $\approx1.33\times10^{12}$ \ac{IID} noise realisations. Given that the recolouring procedure with $\approx160$ days of O3 training noise has 250,000 \ac{IID} \ac{PSD} realisations in each detector, this number increases to $\approx8.26\times10^{22}$. With 172,800 \ac{IID} PSDs, this number will be $\approx3.95\times10^{22}$. The noise samples will be exactly the same as seen in O3 when they are time-correlated between H1 and L1, and no recolouring is applied. Under this condition, we must include the number of non-\ac{IID} noise realisations to calculate the probability that the network sees such samples (following the same procedure as for \ac{IID} samples, the number of non-\ac{IID} samples is $\approx6\times10^{31}$). The probability of seeing an exact O3 noise realisation during training is, thus, $\approx1.4\times10^{-22}$ for recoloured noise samples and $\approx8.6\times10^{-11}$ for non-recoloured samples.

    \textit{Modifications for D3}: Signal augmentation is consistent between the models trained for D3 and D4. None of the above noise augmentation procedures apply to D3. 

    \textbf{Biases Addressed}: Using the recolouring augmentation further increases the amount of variation in the training noise features and addresses the \textit{bias due to lack of variation} by broadening the training noise feature distribution. Using an estimate of the testing dataset \ac{PSD}s to recolour the training dataset introduces a mild train-test overlap, but addresses the \textit{bias due to train-test mismatch}.

    \subsection{\label{subsec:Transforms}Data Transformation}
    \textit{Whitening}: Once the class sample (either signal or noise) is generated and augmented, we whiten it using a single, precomputed \ac{PSD} for each detector, which remains constant across all samples. This \ac{PSD} is estimated by taking the median over several noise PSDs from O3 training data and smoothed via inverse spectrum truncation using a Hanning window. To whiten the data, we divide the sample in the Fourier domain by the smoothed noise \ac{PSD}, which standardizes the amplitude distribution across frequencies, ensuring that the transformed data has approximately equal variance across the spectrum. Using a fixed \ac{PSD} for whitening all samples instead of estimating the \ac{PSD} for every sample results in mildly non-uniform power between different frequencies. We observed that the presence of this non-uniformity in the training data did not contribute to any noticeable degradation of the network when evaluated on the testing data, thus voiding the need to estimate the \ac{PSD} for every input sample. For D3, we compute the median \ac{PSD} from the 20 provided \ac{PSD}s for each detector, which is then used to whiten every input sample.

\begin{figure}[htbp]
        \centering
        \includegraphics[width=1.0\linewidth]{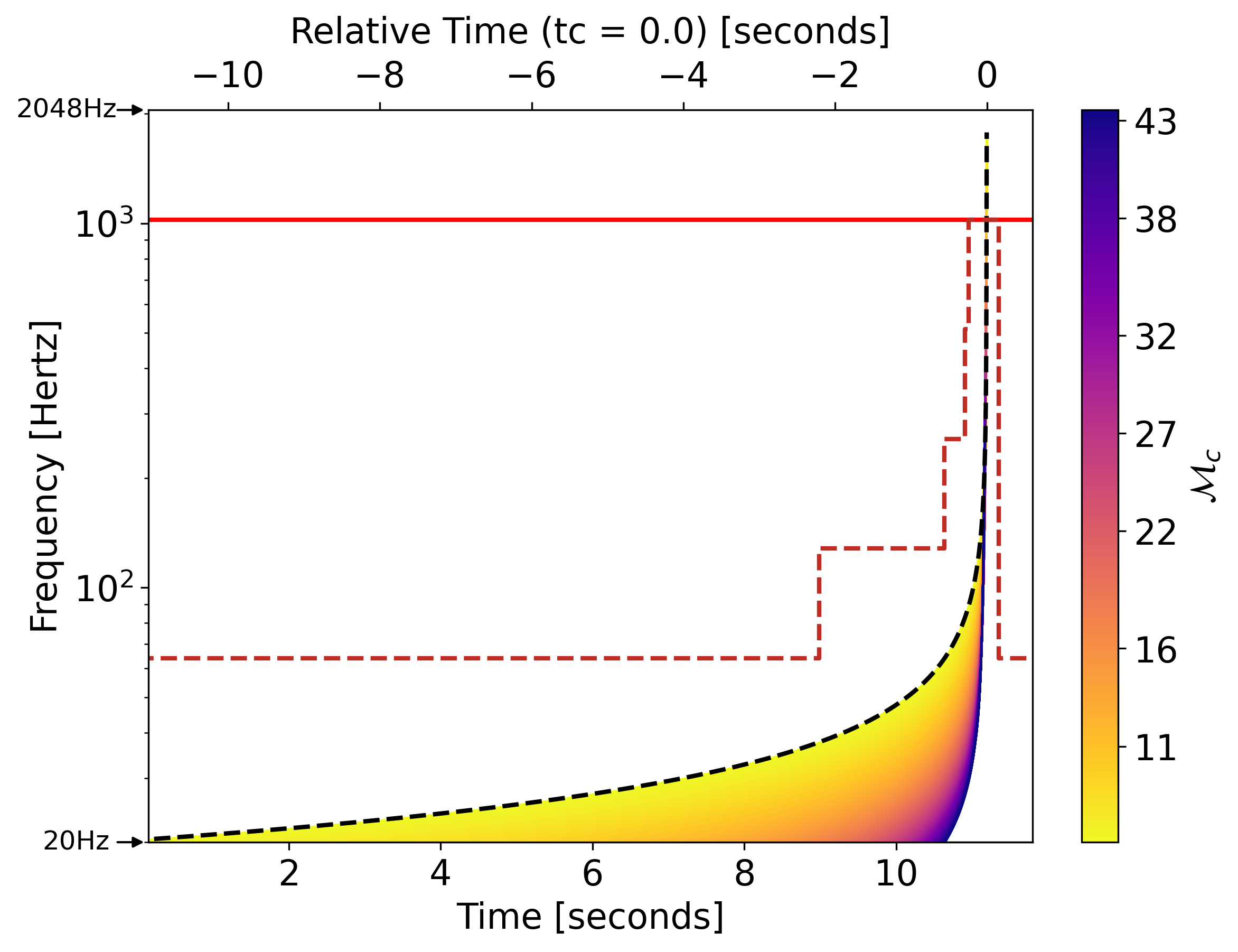}
        \caption{Frequency vs time curves of \ac{GW}s generated from the testing dataset parameter distributions described in table \ref{tab:testing_dataset_priors} coloured by chirp mass. The dashed black line shows the frequency profile of a \ac{GW} with the lowest possible chirp mass or longest duration. The solid red line shows the Nyquist limit of the generated data at $1024$ Hz given that the sampling rate is $2048$ Hz. The dashed red line shows the modified Nyquist limit after applying the multi-rate sampling algorithm.}
        \label{fig:multirate_sampling_template}
    \end{figure}

    \textit{Multi-rate Sampling}: Each input, being $12$ seconds in duration at a sampling frequency of $f_\text{s}=2048$Hz, has 24576 points. Training a neural network on such a long sample has two main disadvantages: [i] it demands a substantial computational cost, especially for complex problems requiring large datasets [ii] most standard architectures will struggle to formulate long-range dependencies due to an insufficient effective maximum receptive field~\cite{effective_receptive_field}. It is, therefore, crucial to compress the data to its permissible limits. The solid red line on Fig. \ref{fig:multirate_sampling_template} indicates the Nyquist limit of the input data, $0.5f_s$, alongside the frequency evolution curves of the chirp mass distribution for the training data. Given that the time of coalescence ($t_c$) of any signal is placed within a predefined time range, it is straightforward to see the inefficiency in using an unnecessarily high $f_s$ to represent regions in time where we are \textit{apriori} aware of the maximum possible \ac{GW} frequency.

    \begin{figure*}[htbp]
        \centering
        \includegraphics[width=1.0\linewidth]{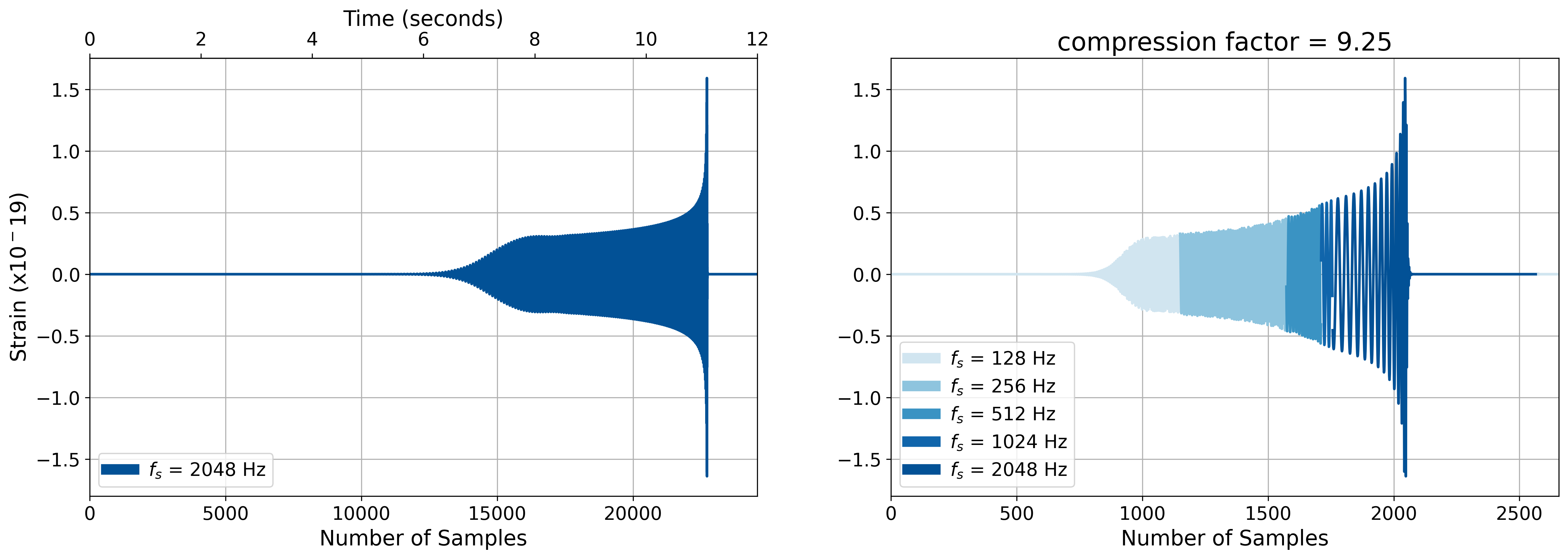}
        \caption{Example application of multirate sampling on an IMRPhenomXPHM BBH template with the following parameters: $m_1=20\;\mathrm{M}_\odot$, $m_2=10\;\mathrm{M}_\odot$, zero spins and a time of coalescence of 11.1 s. The colours depict the sampling frequency ($f_s$) of different sections of the compressed signal. $f_s=2048\;\textup{Hz}$ corresponds to an unaltered section of the generated waveform.}
        \label{fig:multirate_sampling_example}
    \end{figure*}
    
    We take inspiration from \cite{CONTRIB_multirate_sampling} that implements multi-rate sampling for the detection of binary neutron star mergers. The goal is to modify the Nyquist limit at different points along the given time series according to the maximum possible instantaneous \ac{GW} frequency dictated by the training \ac{GW} parameter distributions. This can be seen as a form of physics-informed time series compression. Consistency for all training and testing samples can be guaranteed with the following bottleneck criterion:
    
    Bottleneck 1. For a $12$-second sample length, we choose the training distribution for $t_c$ to be $U[11.0, 11.2]$ s. Let us assume that the frequency evolution profile of the signals in the training distribution are well described by equation \ref{eqn:t_from_mc_Cutler_Flanagan}. At any given time instant till the merger, the instantaneous frequency of the longest possible signal placed at $t_c=11.0$ s will be higher than that of any other signal in the training distribution. Thus, if we modify the Nyquist limit to lie just above the frequency profile of this signal, it will ensure that \textit{all} possible \ac{GW} signals in the training distribution are captured completely. In the time domain, we can split the signal into time bands of constant sampling frequency above the minimum allowed value.
    
    Say we place the longest possible signal in the training distribution, described by the lowest possible chirp mass (to first order), at $t_c=11.0$ s. We move back in time along the frequency profile of this signal, starting from $f_{\text{gw}}^{0}=f_{\text{ISCO}}$, where $f_{\text{ISCO}}$ is the instantaneous frequency at the \ac{ISCO} of a single black hole whose mass is a sum of the source components, and $f_{\text{gw}}^{i}$ is the maximum \ac{GW} frequency in time band $i$. We use $f_{\text{ISCO}}$ as a proxy for the frequency at merger. We start at a sampling frequency $f_s^{0}=1024$ Hz at $f_{\text{ISCO}}=314.1$ Hz, where $f_s^{i}$ is the constant sampling frequency within time band $i$. The sampling frequency is halved $f_s^{i}=f_s^{i-1}/2$ for every corresponding halving of the \ac{GW} instantaneous frequency $(f_{\text{gw}}^{i-1}/f_{\text{gw}}^{i})-2\approx0$. This process is repeated until we reach the lowest allowed sampling frequency at $128$ Hz. The region beyond $f_{\text{ISCO}}$ can contain \ac{GW} frequency components that are high enough to require an $f_s$ of $2048$ Hz. Thus, the time instant at $f_{\text{gw}}^{0}$, defines the lower boundary of the $f_s=2048$ Hz time band.
    
    Bottleneck 2. The second bottleneck is defined by the ringdown of the \ac{GW} signal, with the highest total mass $M=m_{\text{max}}+m_{\text{max}}=100\,M_\odot$ with the maximum allowed spin placed at the upper bound of $t_c$ at $11.2$ s. This configuration gives us the longest possible ringdown signal dictated by the quasi-normal mode damping time at a given spin and mode (we use only $l=2$, $m=2$) and typically consists of high-frequency components. Thus, we set the highest available $f_s$ of $2048$ Hz for this region. This defines the upper boundary of the $f_s=2048$ Hz time band.

    Bottleneck 3. Additionally, we impose an offset to each boundary due to $t_c$ being measured in geocentric time. This accounts for the $t_c$ in detector time being outside the two boundaries. The final boundaries for the $f_s=2048$ Hz time band are $\approx[10.95, 11.35]$ s. 
    
    The region beyond 11.35 s is assumed to be pure noise and is sampled at $128$ Hz. Figure \ref{fig:multirate_sampling_template} shows the frequency profile of the longest possible \ac{GW} as a dashed black line. The modified Nyquist limit is shown as a dashed red line. Some high-frequency features close to $t_c$ will not be registered due to the maximum possible $f_s$, as seen in Fig. \ref{fig:multirate_sampling_template}, but the loss in \ac{SNR} should be negligible. Figure \ref{fig:multirate_sampling_example} shows an example of the procedure applied to a \ac{GW} within the considered parameter space.

    \textit{Normalisation}: Each multirate sampled time series is treated independently and normalised to lie within the range of $[-1, +1]$. We also use an input standardisation layer within the network architecture, that re-scales the features to ensure a zero mean and unit standard deviation. Say we have a batch with dimensions $(N, C, L)$ where $N$ is the batch size, $C$ is the number of channels/detectors, and $L$ is the length of the input sequence. \verb|BatchNorm| uses all samples in $(N, L)$ and standardises each detector data $C_i$ separately. This would work well in simulated Gaussian noise, but the presence of an outlier, like a loud glitch, in one of the batch elements for a particular detector will heavily influence the normalisation for that detector in all batch elements. \verb|LayerNorm| uses all samples in $(C, L)$ and standardises each batch element $N_i$ separately. This is equivalent to standardising the time-correlated strains from the two detectors together. \verb|LayerNorm| will bias the network against detectors that typically have lower strain amplitudes. \verb|InstanceNorm| standardises each individual strain data from each detector independently. This is the most suited layer for our application, and we use it to transform each time series before providing it to the main model architecture.

    \textbf{Biases Addressed}: Whitening reduces the \textit{bias due to limited feature representation} and \textit{spectral bias} by normalising the power due to different frequencies. Multirate sampling reduces the sample length and negates the need for a larger receptive field to learn low chirp mass signals (aids in learning long-term dependencies). This addresses the \textit{bias due to insufficient information} and \textit{bias due to sample difficulty}. Normalising the amplitude to lie within $[-1, +1]$ also reduces to \textit{bias due to limited feature representation}, similar to whitening. 

    \subsection{\label{subsec:Train optimisation}Optimisation}
    \textit{Multi-objective Loss Function}: We use the standard Binary Cross Entropy (BCE) loss for the purpose of classifying a sample as either a noisy signal or pure noise. The unreduced form of the BCE loss for a batch of network outputs $\mathbf{x}$ and targets $\mathbf{y}$ can be written as
    \begin{align*}
    \label{eqn:bce_loss}
            \mathcal{L}^{\text{gw}}_{bce}(\mathbf{x},\mathbf{y}) &= \{l_1,...,l_n\}^\top, \\
            l_i &= -w_i[y_i\cdot \log x_i + (1-y_i)\cdot \log(1-x_i)], \\
            \text{where}\;i&=1,...,n
    \end{align*}
    and for each $i^{\text{th}}$ batch element (for a batch size of $n$), $x_i$ and $y_i$ are the network output and target, $l_i$ is the loss, $w_i$ is a manual re-scaling weight. The unreduced loss $\mathcal{L}^{\text{gw}}_{bce}(\mathbf{x},\mathbf{y})$ is reduced as $\overline{\mathcal{L}^{\text{gw}}_{bce}(\mathbf{x},\mathbf{y})}$ for every batch (the bar represents sample mean). There is a possibility for one of the log terms to be mathematically undefined or for the network to tend to an infinite gradient. \texttt{PyTorch} deals with this possibility by clamping its log function outputs to be greater than or equal to -100. Thus, the loss function is always finite, and backpropagation is linear. Alongside the BCE loss, we also optimise on a Mean-Squared Error (MSE) loss function for each \ac{GW} parameter $\phi$ in ${t_c, \mathcal{M}_c}$. The unreduced form of this loss is given by
    \begin{align*}
        \mathcal{L}_{mse}^{\phi}(\mathbf{x^{\phi}}, \mathbf{y^{\phi}})&=\{l^{\phi}_1,...,l^{\phi}_n\}^\top, \\
        l^{\phi}_i&=(x^{\phi}_i-y^{\phi}_i)^2, \\
        \text{where}\;i&=1,...,n
    \end{align*}
    and noise samples that do not have a target $y^{\phi}_i$ and are ignored. For each $i^{\text{th}}$ batch element that is not a noise sample (for a batch size of $n$), $x^{\phi}_i$ and $y^{\phi}_i$ are the network output and normalised target for the given \ac{GW} parameter and $l^{\phi}_i$ is the loss. Each network output $x^{\phi}_i$ is limited within the range $[0, 1]$ using a Sigmoid function, and the unreduced loss $\mathcal{L}^{\phi}_{mse}(\mathbf{x^{\phi}},\mathbf{y^{\phi}})$ is reduced as $\overline{\mathcal{L}^{\phi}_{mse}(\mathbf{x^{\phi}},\mathbf{y^{\phi}})}$ for every batch. The primary purpose of these loss functions is \textit{not} to obtain a point estimate of the parameters, but to instead regularise the classification problem by acting as a soft boundary condition. The final batch loss is given by
    \begin{equation*}
        \label{eqn:final_loss_function}
        \mathcal{L} = \overline{\mathcal{L}^{\text{gw}}_{bce}(\mathbf{x},\mathbf{y})} + \sum_{\phi\in\{t_c,\, \mathcal{M}_c\}}\overline{\mathcal{L}_{mse}^{\phi}(\mathbf{x}^{\phi},\mathbf{y}^{\phi})}.
    \end{equation*}
    
    \textit{Optimisation Parameters}: The learning dynamics is optimised using an \verb|Adam| optimiser \cite{Adam_optimiser} with an initial learning rate of $2\times10^{-4}$ and a weight decay of $10^{-6}$. The learning rate is then scheduled using \verb|CosineAnnealingWarmRestarts| that cycles every 5 epochs, with a minimum learning rate of $10^{-6}$. We use gradient clipping set to a maximum allowed gradient norm of 1 and a batch size of 64 for all our runs.

    \textbf{Biases Addressed}: The MSE loss term, $\overline{\mathcal{L}^{\phi}_{mse}(\mathbf{x^{\phi}},\mathbf{y^{\phi}})}$, provides an inductive bias to learn the \ac{GW} signals properly. In other words, \ac{GW} signals become easier to learn, thus addressing the \textit{bias due to sample difficulty}. 

    \subsection{\label{subsec:Architecture}Network Architecture}
    \texttt{Sage} uses a bottom-up hierarchical architecture with two main stages. The first stage is a frontend multiscale feature extractor for each detector that highlights the essential regions and, more importantly, de-emphasises unwanted attributes at various scales in the time domain. The second stage is a backend classifier powered by attention to form temporal relationships within each detector and to emphasise any correlation, or lack thereof, between detectors. There is ample precedence in the machine learning literature to suggest that a multi-stage detection process that fragments the primary problem into multiple sub-problems will lead to better generalisation and faster convergence over single-stage approaches \cite{multi_stage_detection_1, multi_stage_detection_2, multi_stage_detection_3}. The rest of this section is dedicated to explaining parts of the two-stage architecture, their intended purposes, advantages, and necessity. 
    
    \textit{Frontend Feature Extractor}: The frontend architecture is inspired by two main works\footnote{We also took inspiration from and had useful discussions with the G2Net 2021 Kaggle competition winners via private communications\cite{g2net_2021}}: [i] \cite{large_kernel_matters} that argues for the need for large kernels and provides design principles for learning contradictory tasks such as classification and localisation of features simultaneously [ii] \cite{OSNet} that designs a multiscale residual block called the \texttt{OSNet}. 
    
    The key feature of the \texttt{OSNet} \cite{OSNet} is to simultaneously analyse an input over multiple scales to create a comprehensive representation. The scale, dictated by the effective receptive field, can be varied by stacking $n_c$ convolutional layers, with a higher value of $n_c$ resulting in larger receptive fields. Stacking small kernel convolutional layers was initially introduced as an efficient alternative to using a single large kernel, as it results in a similar effective receptive field with a lower computational cost. However, \cite{large_kernel_matters} argue and empirically prove that large kernels lead to better generalisation in per-pixel classification tasks, compared to an equivalent stacking of small kernels. They conclude that the denser connection between large kernels and the output enables the network to handle input transformations more resiliently. Additionally, according to \cite{large_kernel_matters}, to retain the ability to localise important features, the architecture must be fully convolutional and not contain any fully connected layers or global pooling layers. 
    
\begin{figure}[htbp]
        \centering
        \includegraphics[width=1.0\linewidth]{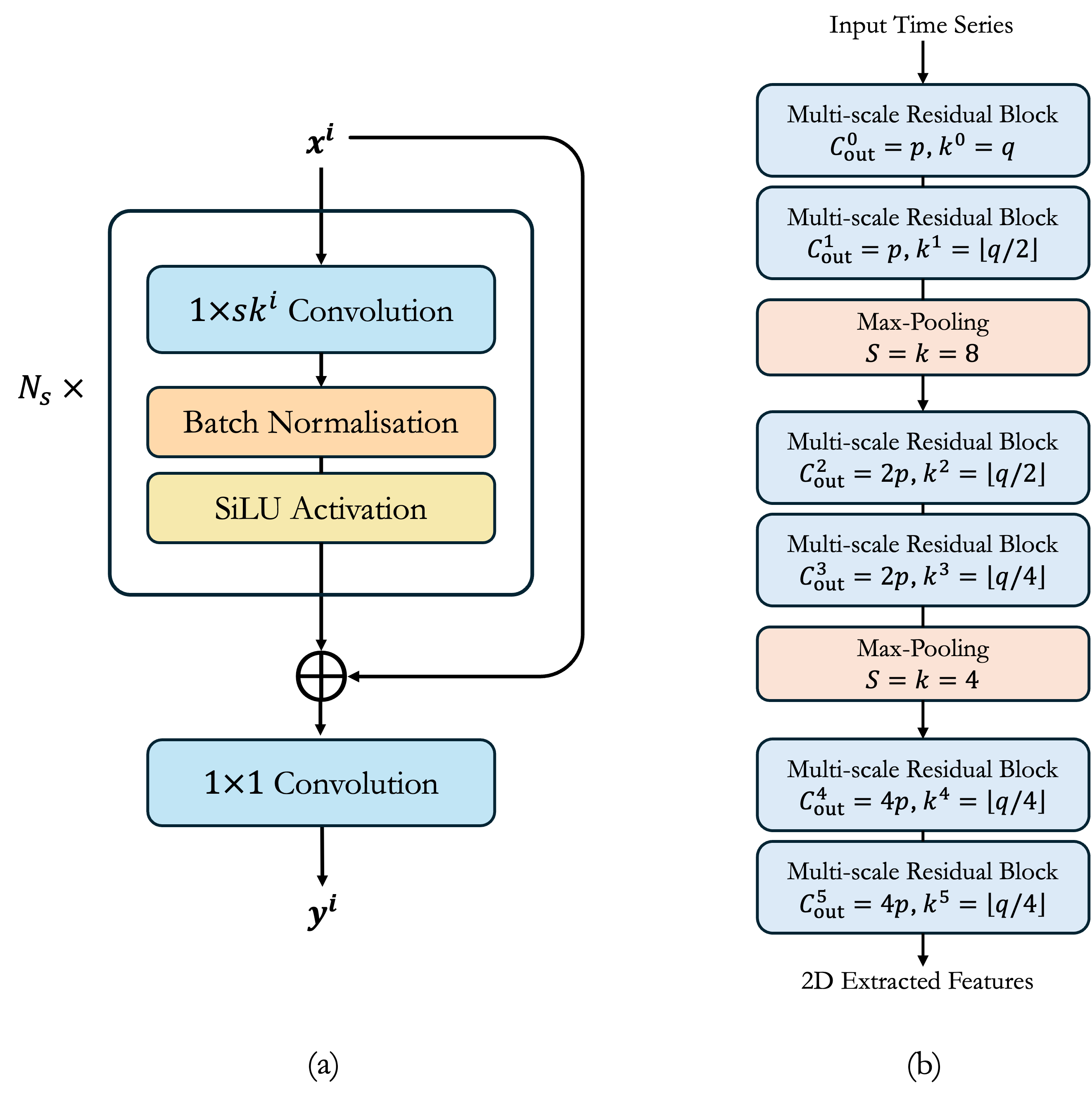}
        \caption{A diagrammatic depiction of the multiscale frontend feature extractor. (a) shows a single multiscale residual block. The input $x^i$ is analysed separately and simultaneously by $N_s$ different 1D large convolutional kernels with kernel size $sk^i$, where $N_s$ is the number of scales, and $s$ is a set of constant prefactors applied to the base kernel size $k^i$. The $\bigoplus$ is an element-wise addition. (b) shows the full frontend architecture.}
        \label{fig:frontend}
    \end{figure}

    Based on these key insights, we build our frontend architecture. A diagram of our large kernel multiscale residual block is shown in Fig. \ref{fig:frontend} (a). In Fig. \ref{fig:frontend} (b), these blocks are then stacked with intermittent downsampling via 1D max pooling to create our full feature extractor. Although we downsample the features along the temporal dimension using a stride $S$, this is performed after generating $C_{\text{out}}$ multiscale output channels, where $S$ and $C_{\text{out}}$ are chosen such that the 2D output feature map is larger overall than the input. Thus, we retain the dense connections between kernels and the extractor feature output. 
    
    Reproducibility. With reference to Fig. \ref{fig:frontend}(a), the large 1D kernels used for the multiscale feature extractor uses $N_s=5$ scales with prefactors $s=\{0.25, 0.5, 1, 2, 4\}$. In Fig. \ref{fig:frontend}(b), the initial kernel size $k^0$ is set to $q=64$ and the initial number of output channels $C_{out}^0$ is set to $p=32$. These values were fine-tuned such that they minimise the loss during the training phase.

    The input time series for each detector is provided to a separate frontend feature extractor (Fig. \ref{fig:frontend}) and there is no weight sharing at this point between detector data. This is done to ensure that subtle differences within the data collected from each detector can be learnt without any interference. The small and long-range features extracted with the multiscale architecture should encode rich spectro-temporal information from the GWs, non-Gaussian noise artefacts and varying noise PSDs.
    
    \textit{Backend Classifier}: We use a variation of the ResNet architecture \cite{resnet} introduced in \cite{CBAM} which integrates the \ac{CBAM} into the ResBlock of a ResNet. The ResNet typically comprises a stack of ResBlocks, with the output feature map of one block feeding into the next. The residual nature of each ResBlock refers to the architectural motif $y=\Phi_s(x)+x$, where $\Phi_s$ is some arbitrary neural network module acting on an input feature map $x$. The recasting of the input feature map $x$ into $\Phi_s(x)$ is called a residual connection, and it allows for the model to mitigate the vanishing gradient problem \cite{resnet}. 
    
    The salient design feature of CBAM is to sequentially infer attention maps along the channel and spatial dimensions of $\Phi_s(x)$. Say we define the output of the arbitrary neural network module within the ResBlock $F=\Phi_s(x)$ as an intermediate feature map. CBAM applies the spatial and channel-wise attention maps ($M_s, M_c$) to produce a final output $F''=M_s(F')\otimes F'$ where $F'=M_c(F)\otimes F$ and $\otimes$ denotes element-wise multiplication. So $y$ will be $F''+x$ instead of $\Phi_s(x)+x$ (in reference to ResBlock motif). We run \texttt{Sage} on a CBAM-ResNet50 backend variant for all experiments. The channel-wise attention $M_c(F)$ is computed as,
    \begin{equation}
        \label{eqn:channel_attention}
        {M_c}(F) = \sigma\left[\Phi_c(P_{\text{avg}}(F)) + \Phi_c(P_{\text{max}}(F))\right]
    \end{equation}
    where $\Phi$ is some \ac{MLP}, ($P_{\text{avg}}$, $P_{\text{max}}$) are average pooling and max pooling operations respectively, and $\sigma$ is the sigmoid activation function. The \ac{MLP} has a single hidden layer that is smaller than the input, thus acting as a bottleneck, similar to an autoencoder architecture \cite{autoencoder}. The spatial attention $M_s(F')$ is computed as,
    \begin{equation}
        \label{eqn:spatial_attention}
        M_s(F') = \sigma\left[P_{\text{avg}}(F') \circledast P_{\text{max}}(F')\right]
    \end{equation}
    where $\circledast$ is the convolution operation. The usage of pooling operations for the aggregation of information is well-founded \cite{avgpool_for_context_1, avgpool_for_context_2} and CBAM exploits this feature across the spatial and channel dimension and empirically prove that the usage of both pooling operations together greatly improves detection capability. The spatial attention will aid greatly in forming temporal relationships within the time series of each detector, and the channel-wise attention will construct correlations between detectors or indicate a lack thereof. 
    
    \textit{Point-estimate Outputs}: The CBAM-ResNet50 is made to output a 512-dimensional embedding, which is passed through three separate (i.e., no shared weights) fully connected layers with 512 neurons each. Each layer is then responsible for one unnormalised output. A sigmoid activation function is then applied to each output to restrict it between $[0,\,1]$. The three normalised outputs are point estimates on the probability of a \ac{GW} event, $p_{\text{GW}}$, and the normalised \ac{GW} parameters $\{t_c,\,\mathcal{M}_c\}$. The normalised values are used to compute the loss function and the unnormalised counterpart of $p_{\text{GW}}$ is used as the output ranking statistic (also referred to as the network output). Through backpropagation, the 512-dimensional embedding should encode information about all outputs, given the composite loss function in equation \ref{eqn:final_loss_function}. We choose chirp mass as it is the best-measured parameter for the most observed mergers, as it determines the leading-order \ac{GW} radiation during the inspiral phase \cite{chirp_mass_best_measured}. This acts as a soft constraint and provides an additional inductive bias to aid the network in distinguishing between the signal and noise classes. The time of coalescence estimate acts as a reference point and aids in feature localisation. Furthermore, we can also use the $t_c$ estimate as input to parameter estimation pipelines to investigate the generated triggers. An example of this estimate for parameters $\{t_c,\,\mathcal{M}_c\}$ is shown in Fig. \ref{fig:parameter_point_estimate} (see appendix \ref{app:parameter point estimate}) for two different \texttt{Sage} runs (see section \ref{subsubsec:Injection Study} for details about runs).

    \textbf{Biases Addressed}: The multiscale frontend with large kernels is useful when learning time series signals with a high dynamic range of frequencies \cite{multiscale_for_timeseries, large_kernel_matters}. Spatial attention aids the model in learning amplitude, frequency, and phase evolution profiles by forming temporal relationships using the multiscale features. The attention mechanism automatically increases the weight of under-represented scales, thus addressing \textit{spectral bias}. The attention mechanism and multiscale architectures have been used before to address this bias, as seen in \cite{attention_for_spectral_bias_1, multiscale_for_spectral_bias_1, multiscale_for_spectral_bias_2}. Channel-wise attention aids in forming a correlation between the temporal relationships formed for the signal in each detector. The two attention mechanisms aid significantly in learning long-term dependencies \cite{attention_is_all_you_need, attention_for_long_dependencies_1, attention_for_long_dependencies_2}, addressing the \textit{bias due to sample difficulty}. Moreover, training a deeper network for longer (without overfitting) increases the effective receptive field, which also aids in learning long sequences \cite{effective_receptive_field}.

\section{\label{sec:Results}Results and Discussion}

    We will use this section to further describe the injection study, elaborate on the various training configurations employed, and compare our results with previous works on the same datasets (D3 and D4) ~\cite{DP5_MLGWSC1_2023, DP6_Nousi_2023, DP7_Zelenka_2024}. Furthermore, we analyse the behaviour of \texttt{Sage} under different conditions and discuss the validity of juxtaposing different results on established evaluation metrics. For the sake of completeness, we also investigate the performance of \texttt{Sage} by removing certain important components from our pipeline to understand its contribution to the overall system (also known as an ablation study).
    
    We ran \texttt{Sage} on multiple training configurations to investigate the impact of different biases, time-varying \ac{PSD}s and glitches. Table \ref{tab:summary_of_runs} gives a brief overview of the runs that we will be discussing in this section. Any configuration that is not mentioned in this table is consistent between all runs and is as described in section \ref{sec:Methods}. The steps taken to evaluate \texttt{Sage} after the end of training are provided here in sequential order:

    \begin{table*}[htbp]
    \setlength{\tabcolsep}{12pt}
    \begin{tabular}{p{3.5cm} p{5cm} p{6.5cm}}
        \toprule \toprule
        {Run Name} & {Training Data - Mass Distribution} & {Noise Segment for Recolouring}\\ \midrule
        Sage — Broad & U$(m_1, m_2)$ & $\{\text{O3a}(\text{Day}\;1-30)\}$ \\
        Sage — Limited & U$(m_1, m_2)$ & $\{\text{O3a}(\text{Day}\;36-81),\,\text{O3b}(\text{Day}\;1-113)\}$ \\
        Sage — D4 Metric & U($\tau_0$, $\tau_3$) & $\{\text{O3a}(\text{Day}\;1-30)\}$ \\
        Sage — Annealed Train & U$(m_1, m_2)$ $\longrightarrow$ U($\tau_0$, $\tau_3$) & $\{\text{O3a}(\text{Day}\;1-30)\}$ \\
        Sage — D3 Metric & U($\tau_0$, $\tau_3$) & -- \\
        \bottomrule \bottomrule
    \end{tabular}
    \caption{Summary of all Sage runs discussed in the results section. \mbox{Column 1} gives a shorthand for the discussed runs and will be used to describe a singular training instance of that type that achieved the highest sensitive distance at an \ac{FAR} of one per month. \mbox{Column 2} describes the distribution of component masses \mbox{$(m_1, m_2)$}. The \mbox{$\tau_0$-$\tau_3$} coordinate pair can be analytically inverted to the \mbox{$M$-$\eta$} pair, as described in section \ref{subsec:Training Dataset Priors}, which can then be used to derive the distribution of component masses. The $\longrightarrow$ symbol refers to a particular training strategy of transitioning from one distribution to another (refer to the Sage — Annealed Train run in section \ref{subsubsec:Injection Study} for more details). Finally, \mbox{column 3} provides details about the noise segment used to obtain precomputed \ac{PSD}s that are meant to recolour the training noise realisations.}
    \label{tab:summary_of_runs}
    \end{table*}

    \textit{Common settings}: As mentioned in sections \ref{subsec:Training Dataset Priors} and \ref{subsec:Testing Dataset Priors}, the training noise is \mbox{O3a(Day 36-51)} and \mbox{O3b(Day 1-113)} and the testing noise is \mbox{O3a(Day 1-30)} for all our runs. All runs were trained with 80 million samples (1:1 ratio between the signal and noise class) split equally over 40 epochs. 

    \textit{Best trained model}: The validation noise dataset will be slightly different, compared to the training and testing datasets, in terms of noise \ac{PSD}s and glitches. Thus, we retrieve the network learnable parameters (i.e., weights and biases) that correspond to the 5 lowest epoch validation losses for testing. We then test the model on the ``background" and ``foreground" testing datasets discussed in section \ref{subsec:Testing Dataset Priors} for each set of network parameters, and pick the one that detects the most signals at an \ac{FAR} of 1 per month. This procedure is exactly what the Neyman-Pearson lemma prescribes, but to be clear we \textit{do not} advocate for the number of detections to be a less-biased or better metric compared to any other. 
    
    \textit{Testing iteration}: The 30-day testing data is split as sequential windows with a step size of $0.1$ seconds. Each window is whitened, multirate sampled and normalised, consistent with the training dataset, prior to being provided as input to the trained network. The network outputs a ranking statistic for each $12$-second window, which we assume originates due to a possible \ac{GW} event with a time of coalescence within $[11.0, 11.2]$ s (the bounds used during training). This is called a \textit{trigger}. Given that the step-size is 0.1 s, a \ac{GW} event can have at most 3 triggers.

    \textit{Trigger clustering}: To identify independent gravitational wave candidates, we cluster nearby triggers based on a predefined time threshold ($2$ seconds). We iterate through the list of triggers, grouping consecutive ones if their time difference is below the threshold; otherwise, a new cluster is initiated. Each cluster is then represented by its most significant trigger, determined by the maximum ranking statistic within the cluster. Any subsequent use of the term ``trigger" must be interpreted as this independent cluster maxima. This approach ensures that multiple responses to the same underlying signal, or to closely spaced noise fluctuations, are not overcounted.

    \textit{Labelling cluster maxima}: Only triggers above a ranking statistic of 0 are retained (to reduce compute cost). No prior knowledge about the location of the injected events is taken into account when assigning the trigger a ranking statistic. For the foreground dataset, triggers within $[t_c^{\text{inj}}-0.2,\,t_c^{\text{inj}}+0.2]$ seconds of the $12$-second window, where $t_c^{\text{inj}}$ is the time of coalescence of any one of the injected signals, are labelled as \textit{true positive} triggers for the corresponding injection (not to be confused with a binary \ac{ML} label). For each true positive trigger, we get a point estimate of the time of coalescence and chirp mass from \texttt{Sage} (see section \ref{subsec:Architecture}). All other foreground triggers that do not correspond to a known event are labelled as \textit{false positives}. For the background dataset, all triggers are labelled as false positives. The labels provided to the triggers are purely for the purpose of computing the following evaluation metrics. \footnote{For a fair comparison to the MLGWSC-1 benchmark results, we use the code provided at \url{https://github.com/gwastro/ml-mock-data-challenge-1/blob/master/evaluate.py} to evaluate our model.}
    
    \textit{Evaluation metrics}: The sensitive distance metric at different \ac{FAR}s is computed as described in section \ref{subsubsec:User-introduced Biases}. To compute \ac{ROC} curves, we use the true positive triggers for each signal from the foreground dataset and false positive triggers from arbitrary noise windows of the background dataset. Triggers in the foreground dataset marked as false positive can originate from windows where a \ac{GW} injection is present with a $t_c$ outside the $[11.0, 11.2]$ s range. The behaviour of the \ac{ML} model for these windows is not guaranteed to resemble that of a pure noise sample or a noisy signal sample, and thus the false positive label is not accurate. Since we cannot confidently assign the false positive label to these samples, it is not possible to include all foreground dataset triggers for the computation of the \ac{ROC} curve.

    All tests were run on 16 CPU cores and an NVIDIA Quadro RTX 8000 GPU. Each epoch of training takes $\approx2$ hours, excluding validation. Running \texttt{Sage} on evaluation mode, it takes $\approx16$ hours to generate triggers on 1 month of testing data. 
    
    \subsection{\label{subsec:Comparison with matched-filtering}Comparison with Matched Filtering}

        We compare \texttt{Sage} to the \texttt{PyCBC} submission in MLGWSC-1~\cite{DP5_MLGWSC1_2023} for the realistic D4 injection study. The submission is based on the standard configuration of \texttt{PyCBC} used for archival searches of compact binary mergers, as seen in \cite{4OGC_PyCBC_configuration} (particularly the ``Broad" configuration \textit{without} an extra tuning term in the ranking statistic for a specific signal population). The template bank they used to evaluate this injection study included templates that account for spins aligned with the orbital angular momentum, did not include precession or higher-order modes, and had mass parameters that were bound by the testing distributions provided in table \ref{tab:testing_dataset_priors} and were placed in a stochastic manner. They account for the presence of glitches~\cite{glitches_1, glitches_2, glitches_3} using waveform consistency checks \cite{blip_glitch_veto, chi_squared_test} and ensure that the amplitude, phases, and time of arrival in each detector are consistent with the astrophysical population \cite{population_consistency_check}. \texttt{PyCBC} was run separately on the ``background" and ``foreground" datasets, and a ranking statistic was assigned to potential candidates based on an empirically measured noise distribution and the consistency with the expected \ac{GW}~\cite{population_consistency_check} (without attempting to optimise sensitivity for a particular signal population).

        Although \texttt{PyCBC} uses a template bank that does not account for precession or higher-order modes, we don't expect the detection efficiency to be too different from a template bank that does, since neither of those effects is detectable in this injection study (refer to section \ref{fig:testing_dataset_priors}). Although there would exist slight differences in the signal morphology between injections generated using \verb|IMRPhenomXPHM| and the relatively simpler templates used by \texttt{PyCBC}, the network \ac{SNR} of each detected injection should not vary significantly due to the simpler templates. Thus, although we use \verb|IMRPhenomPv2| to train \texttt{Sage}, there is no precedent to expect a better detection efficiency from \texttt{Sage} solely due to the usage of a waveform model that is relatively more similar to the injections.

        \subsubsection{\label{subsubsec:Injection Study}Injection Study}

            In this section, we describe each of the runs in table \ref{tab:summary_of_runs}, compare them and describe how \ac{ML} biases influence their detection performance. All aspects of the methodology are consistent with section \ref{sec:Methods}, and any changes or choices made are described within each run description. Some choices are common for all runs and are described here:

            For all runs, we rescale the optimal network \ac{SNR} of all samples during training, as mentioned in section \ref{subsec:Training Dataset Priors}. From experimentation, we found that using a half-normal distribution allowed for a higher detection efficiency than using a uniform distribution in the same range. The half-normal is described by the probability density function,
            \begin{equation}
                f(x)=\sqrt{\frac{2}{\pi}}\exp{\frac{-x^2}{2}},
            \end{equation}
            where $x\geq0$. The samples generated using this density are scaled by a factor of 4 and shifted by a factor of 5. This places the lower bound at an \ac{SNR} of 5. From a sample size of $10^6$, we find that $99\%$ of the distribution lies below an \ac{SNR} of $\approx15.3$ and the maximum \ac{SNR} is $\approx24.3$.

\begin{figure}[htbp]
                \centering
                \includegraphics[width=1.0\linewidth]{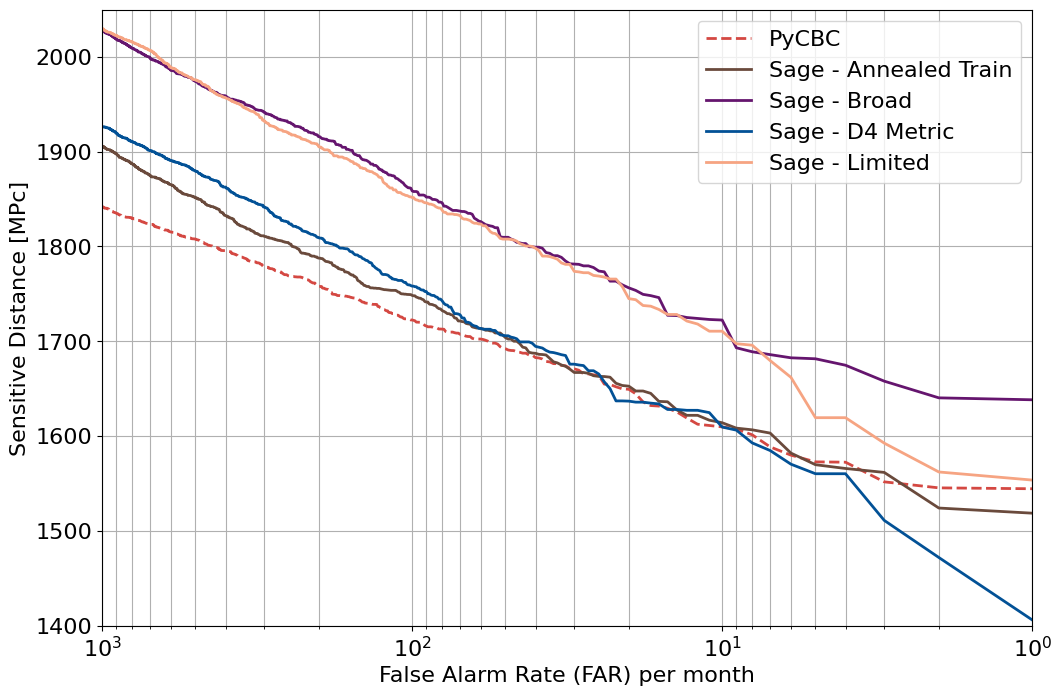}
                \caption{The sensitive distance metric as a function of false alarm rate for \texttt{PyCBC} and a set of different \texttt{Sage} analyses of the testing dataset.}
                \label{fig:compare_sensitive_distance_sage_runs}
            \end{figure}

\begin{figure*}[htbp]
                \centering
                \includegraphics[width=1.0\linewidth]{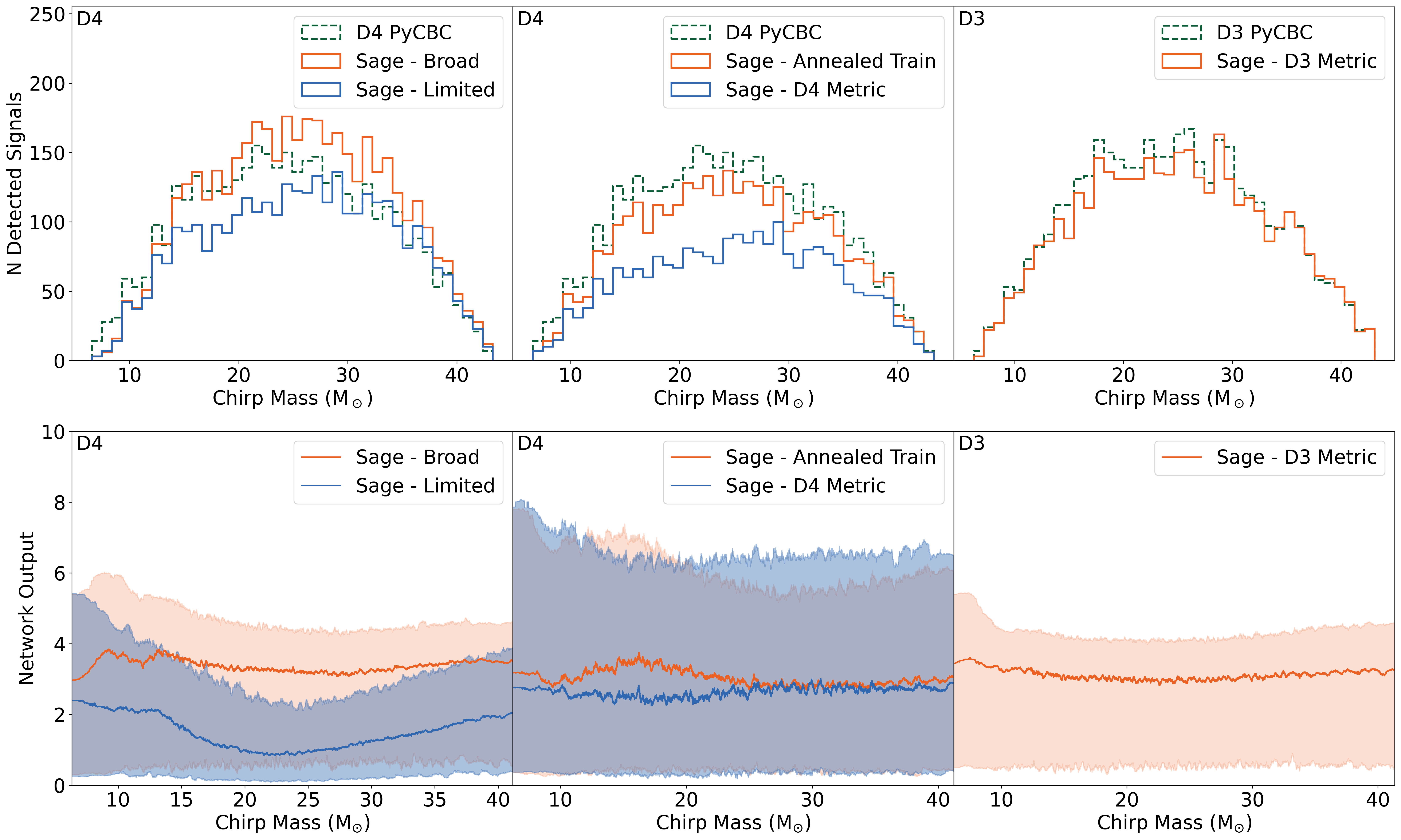}
                \caption{\textit{(top)} 1D histograms of detected signals along the chirp mass dimension of different \texttt{Sage} runs with \texttt{PyCBC} at an \ac{FAR} of 1 per month. The \textit{(bottom)} row of Fig. \ref{fig:compare_main_runs} shows the ranking statistic output by the network compared to the chirp mass $\mathcal{M}_c$ of 250,000 signals. The ranking statistic is put through a running mean across $\tau$ where we use a window size of 500 samples and a step size of 1 sample. The error bar on each plot indicates the region between the 5$^{\text{th}}$ and 95$^{\text{th}}$ percentile region of the network output at a given chirp mass.}
                \label{fig:compare_main_runs}
            \end{figure*}

            \textit{Sage - Limited}: We use detector noise from O3a(Day 36-51) and O3b(Day 1-113) for training. This is then recoloured using precomputed PSDs from the same data with a probability of 0.4 (value chosen via fine-tuning). Figure \ref{fig:compare_sensitive_distance_sage_runs} shows a comparison of the sensitive distances of \texttt{Sage - Limited} and \texttt{PyCBC} as a function of \ac{FAR}. We see that our run achieves a higher sensitive distance than \texttt{PyCBC} at all \ac{FAR}s from 1000/month to 1/month. However, Fig. \ref{fig:compare_main_runs} (top left) shows that the run does not find as many signals as \texttt{PyCBC}. The higher sensitive distance is due to finding a few more high chirp mass signals than \texttt{PyCBC}. As described in section \ref{subsec:Issues}, this is a consequence of a bias against low chirp mass signals (incurred from biases due to insufficient information, limited sample representation, sample difficulty and spectral bias) and a parameter-dependent weighting of the sensitive distance metric. Although our bias mitigation strategies improved the detection efficiency at lower chirp masses compared to runs without them, it is still not sufficient to match \texttt{PyCBC}'s detection efficiency.

            \textit{Sage - Broad}: The presence of \ac{OOD} \ac{PSD}s in the 30-day D4 testing dataset, as seen in Fig. \ref{fig:o3a_psds}, might lead to reduced detection efficiency due to the model's inability to extrapolate to unseen \ac{PSD}s. In this experiment, we determine whether closing this gap would affect the detection efficiency. The only change we introduce in this experiment is to use O3a(Day 1-30) to precompute the \ac{PSD}s used to recolour the training data O3a(Day 36-51) and O3b(Day 1-113), instead of using the training data itself. This intentionally introduces some mild overlap between the training and testing datasets, although the recoloured noise realisations are strictly speaking different from the testing dataset. 
            
\begin{figure}[htbp]
                \centering
                \includegraphics[width=1.0\linewidth]{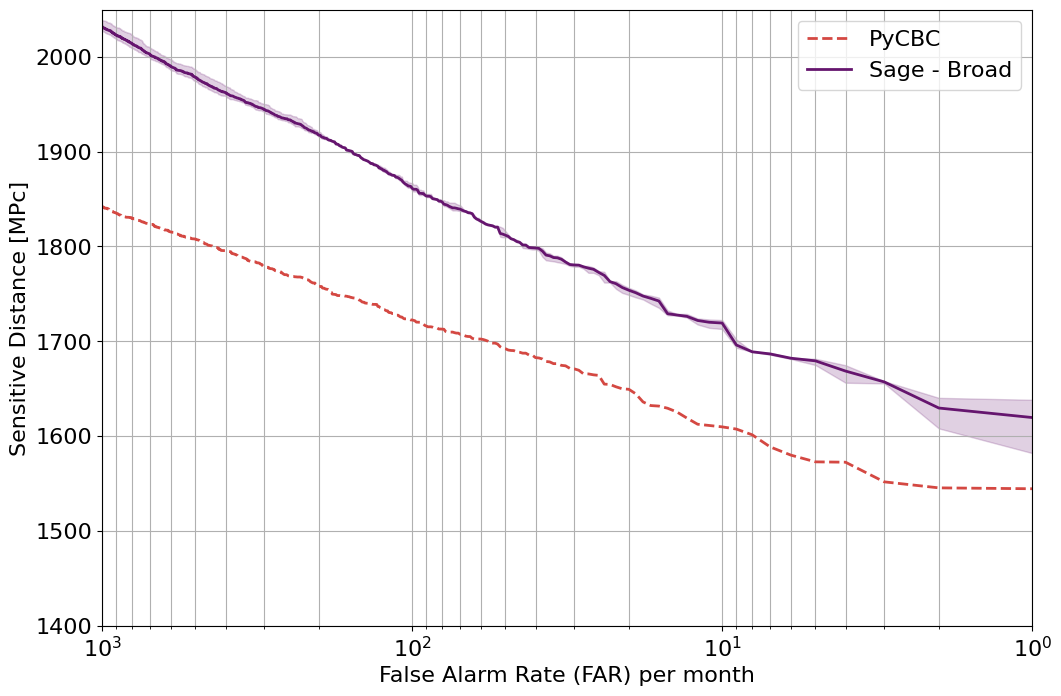}
                \caption{The sensitive distance metric as a function of \ac{FAR} for PyCBC and the \texttt{Sage - Broad} run on the testing dataset. The solid violet line is the mean sensitive distance of 3 runs made with the ``Broad" configuration, and the error bar represents the max and min values obtained from the runs.}
                \label{fig:compare_sensitive_distance_sage_to_pycbc}
            \end{figure}

\begin{figure}[htbp]
                \centering
                \includegraphics[width=1.0\linewidth]{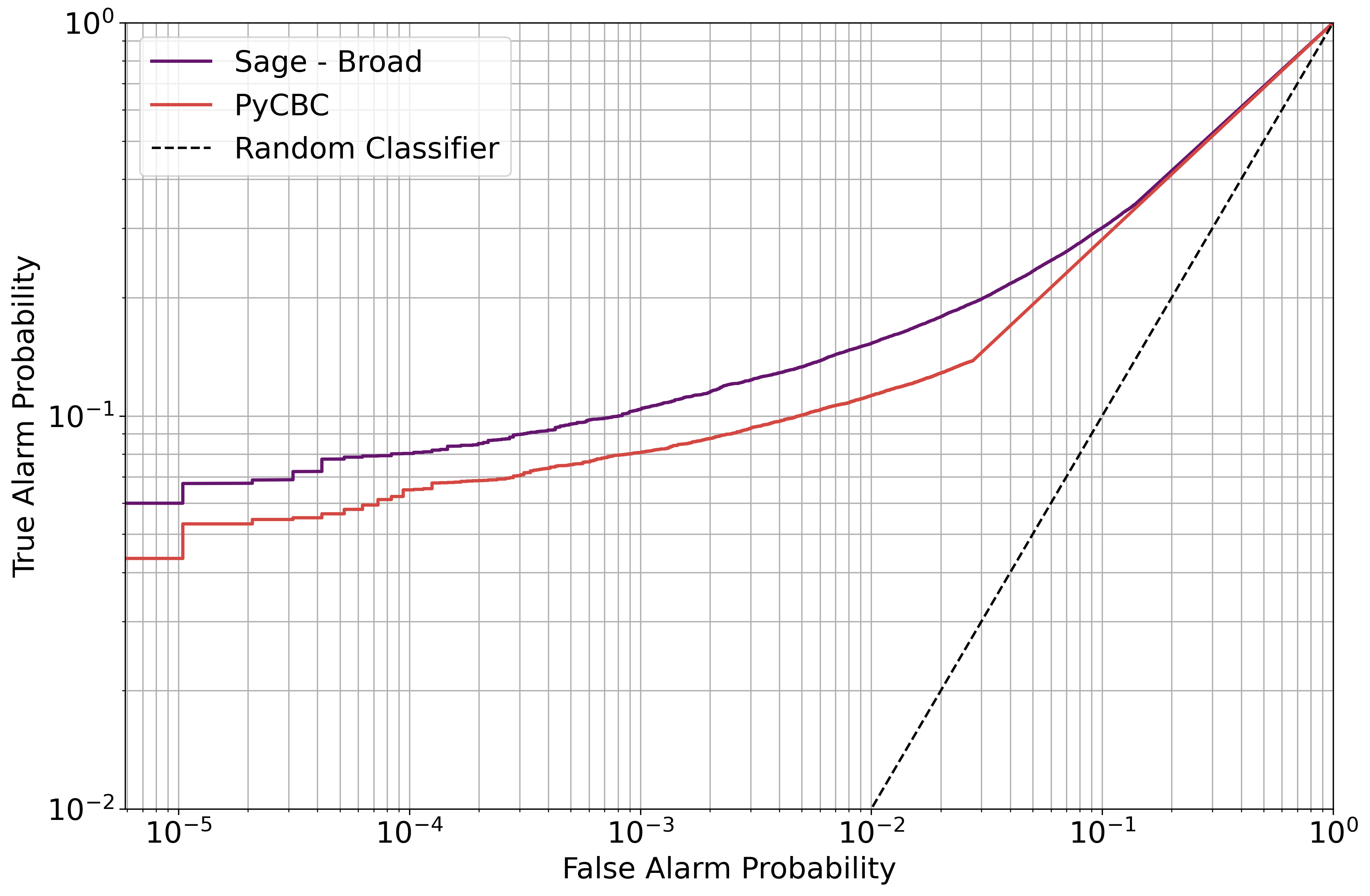}
                \caption{The true alarm probability as a function of false alarm probability (ROC curves) for PyCBC and the \texttt{Sage - Broad} run that achieved the highest sensitive distance on the testing dataset.}
                \label{fig:compare_roc_sage_to_pycbc}
            \end{figure}

            From Fig. \ref{fig:compare_sensitive_distance_sage_runs}, it is immediately evident that we see a significant boost in the sensitive distance, from an \ac{FAR} of 10/month to 1/month. We see an increase of $\approx5.5\%$ and $\approx6.5\%$ in sensitive distance when compared to \texttt{Sage - Limited} and \texttt{PyCBC} respectively at an \ac{FAR} of 1 per month. Figure~\ref{fig:compare_sensitive_distance_sage_to_pycbc} shows the median sensitive distance for this run with an error bar made by repeating this run 3 times with different seeds. The seeds were set such that both signal and noise realisations for all epochs during training are different but within the same distribution and that the networks have different random initialisations for their trainable parameters. We also provide the parameter-independent \ac{ROC} curve that compares this result to \texttt{PyCBC} in Fig. \ref{fig:compare_roc_sage_to_pycbc}. At a false alarm probability of $10^{-5}$ we see an $\approx38\%$ increase in the true alarm probability. Figure \ref{fig:compare_main_runs} (top left) also shows the 1D chirp mass histograms of the detected signals, comparing \texttt{Sage - Limited} and \texttt{Sage - Broad}. The two runs perform similarly at chirp masses below $\approx10\,\text{M}_{\odot}$, but the latter outperforms the former significantly everywhere else. 

            The increase in detection performance when compared to \texttt{Sage - Limited} points to two possible causes: [i] an increase in the variance of training noise realisations (via their \ac{PSD}), or [ii] all testing noise \ac{PSD}s are in-distribution for \texttt{Sage - Broad}, but the same is not true for \texttt{Sage - Limited}. We will investigate the root cause in the following runs and in section \ref{subsubsec:Effects of PSD}.

            \textit{Caveats of the evaluation metrics}: We generally attribute an increase in the true alarm probability of an \ac{ROC} curve at a given false alarm probability, and the sensitive distance metric at a given \ac{FAR} to better detection performance. However, this is not entirely correct, and both evaluation metrics have their disadvantages. Assuming Gaussian noise, a model that is biased against low chirp mass signals will operate on a narrower parameter space and thus detect more high chirp mass signals at any given \ac{FAR}. This bias would result in a significant increase in the sensitive distance metric, given the chirp mass dependence of the sensitive volume (see equation \ref{eqn:sensitive_volume}). A parameter-independent metric, like the \ac{ROC} curve, would still only account for the number of detections made and not the distribution of the detections, thus masking any biases. The use of only a single figure of merit based on a pre-determined set of signals may be of very limited application to the real-world problem of detecting signals with an unknown parameter distribution. For instance, an algorithm may achieve a high figure of merit while being almost completely insensitive to signals in some region of parameter space, which may be much more prevalent in reality than in the evaluation set. In other words, optimising detection for a known target population may imply a greater risk of losing sensitivity for the unknown real population.
            
\begin{figure*}[htbp]
                \centering
                \includegraphics[width=1.0\linewidth]{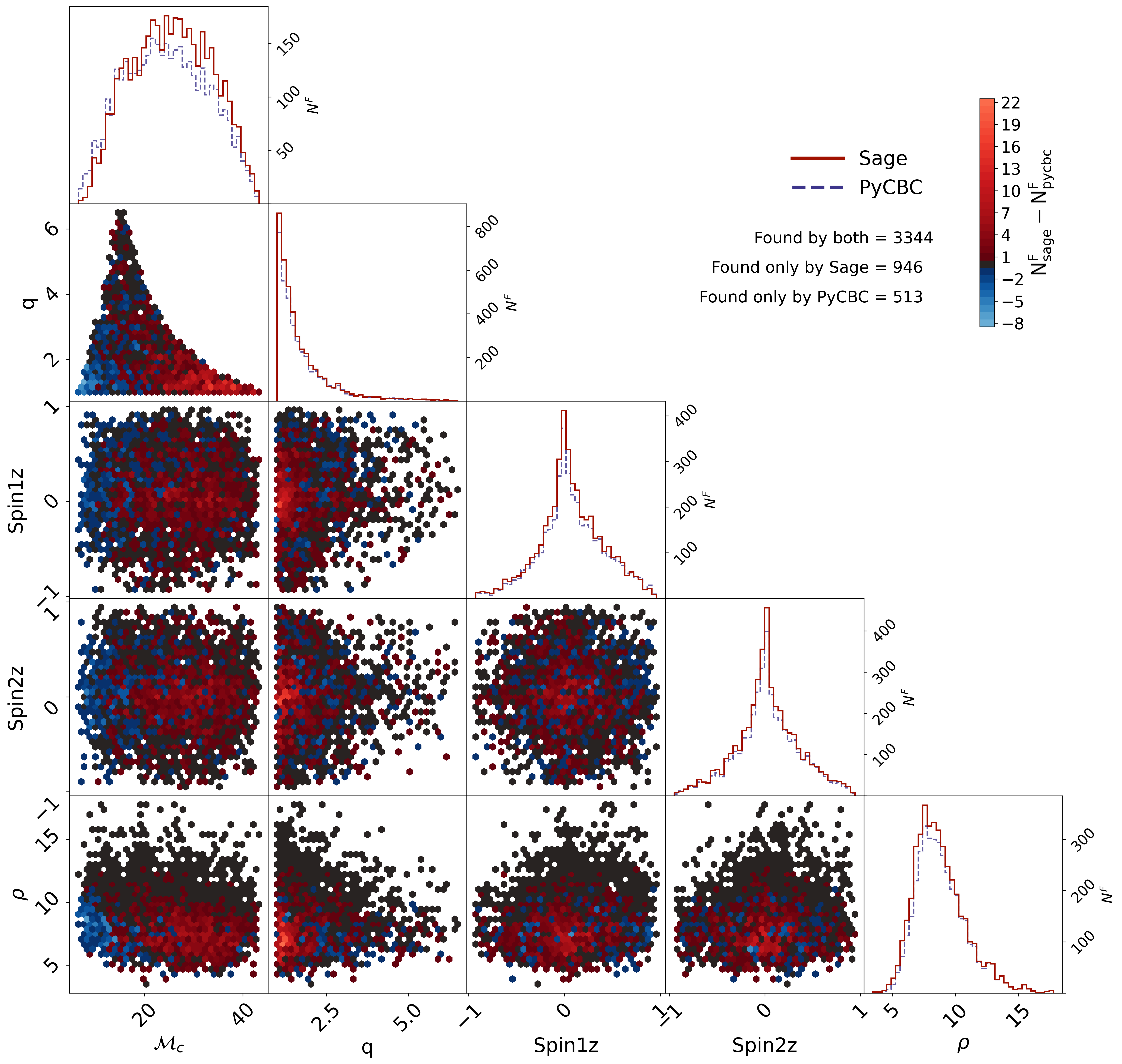}
                \caption{1D and 2D histograms of detected signals for different \ac{GW} parameters at an \ac{FAR} of 1 per month, comparing the best \texttt{Sage - Broad} run and \texttt{PyCBC}. We chose \ac{GW} parameters where we expected to see learning bias. The diagonal subplots contain 1D histograms of the detected signals for a given parameter. The 2D histograms are hexagonal binned plots, where each bin is coloured using $N^F_{\text{sage}}-N^F_{\text{pycbc}}$ where $N^F_{\text{sage}}$ and $N^F_{\text{pycbc}}$ is the number of signals detected by \texttt{Sage} and \texttt{PyCBC} respectively. We use a divergent colour map that is set to be black when $N^F_{\text{sage}}-N^F_{\text{pycbc}}$ is zero, is progressively brighter in red along the positive axis, and is progressively brighter in blue along the negative axis. So, red and blue patches can be attributed to regions where \texttt{Sage} and \texttt{PyCBC} detect more signals, respectively.}
                \label{fig:compare_sage_to_pycbc}
            \end{figure*}

            We visualise the 1D and 2D distributions of the detected signals as histograms in Fig. \ref{fig:compare_sage_to_pycbc}, to compare and visualise any differences between \texttt{Sage - Broad} and \texttt{PyCBC}. We find that 3344 signals are detected by both pipelines, 513 signals were found only by \texttt{PyCBC} and 946 signals were found only by \texttt{Sage} at an \ac{FAR} of 1/month. This is equivalent to an increase of $\approx11.2\%$ in the number of signals detected compared to \texttt{PyCBC} in the MLGWSC-1 realistic injection study. We find no anomalous biases present across any parameter that is not directly related to chirp mass, mass ratio or \ac{SNR}. A full set of 1D histograms for all relevant intrinsic and extrinsic parameters are provided as 1D histograms in Fig. \ref{fig:compare_histograms_sage_pycbc_all_params} in appendix \ref{app:sage-broad detection sensitivity}.

            \textit{Important note}: In Fig. \ref{fig:compare_sage_to_pycbc}, \texttt{Sage - Broad} and \texttt{PyCBC} operate on two different mass distributions - the former on U$(m_1, m_2)$ and the latter on U($\tau_0$, $\tau_3$). For a fair comparison of their detection sensitivities, the search statistic for \texttt{PyCBC} must be fine-tuned and optimised for the U$(m_1, m_2)$ mass distribution (see \cite{bbh_tuning_mf_1, bbh_tuning_mf_2} for relevant information). This fine-tuning will likely lead to an increase in the number of signals detected by \texttt{PyCBC} in the high chirp mass region. However, given that the effect of blip glitch noise is much higher in this region, it is not straightforward to speculate on the extent of improvement that might be seen in a fine-tuned \texttt{PyCBC}~\cite{blip_glitch_veto, blip-glitches-O3}. This study is left for future work. Furthermore, the small number of low \ac{SNR}, low chirp mass signals missed by \texttt{Sage} in Fig. \ref{fig:compare_sage_to_pycbc} could be attributed to this difference in mass distribution.

            \textit{Sage - D3 Metric}: We found many training training distribution configurations that reduced the bias across the chirp mass and mass ratio dimensions, but for this run, we investigate using $U(\tau_0, \tau_3)$ () to set the training distribution for the component masses. A detailed description for using this distribution for training is defined in section \ref{subsec:Training Dataset Priors}. The key difference between $U(m_1, m_2)$ (which we used in the previous runs) and $U(\tau_0, \tau_3)$ is that the latter has an underdensity of training data samples in the high chirp mass region (see Fig. \ref{fig:training_dataset_priors}). For this run, we train and test on the D3 dataset setting with simulated coloured Gaussian noise instead of real noise. This setup is ideal for studying the effects of varying the training distributions of \ac{GW} parameters, as it avoids introducing learning biases from real detector noise.
            
            Figure \ref{fig:compare_main_runs} (top right) compares the 1D histograms of the detected signals at an \ac{FAR} of 1 per month of \texttt{Sage - D3 Metric} to \texttt{PyCBC}. We see that using $U(\tau_0, \tau_3)$ as the training mass distribution alongside simulated Gaussian noise produces a detection efficiency very similar to \texttt{PyCBC} at all chirp masses. This justifies the use of the parameter distribution on real detector noise as well. 
            
            \textit{Sage - D4 Metric}: This run also uses $U(\tau_0, \tau_3)$ to set the training distribution for component masses, but uses real detector noise (similar to all other runs on D4) instead of simulated coloured Gaussian noise. Compared to \texttt{Sage - D3 Metric} we see that \texttt{Sage - D4 Metric} has significantly lower detection efficiency, in Fig. \ref{fig:compare_main_runs} \textit{(top middle)}. This reduction is solely due to the attributes of real noise; it is either due to the inability to properly reject non-Gaussian artefacts or due to the difference between the training and testing noise \ac{PSD}s. 
            
            \textit{Sage - Annealed Train}: An underdensity of training data samples in the high chirp mass region results in worse noise rejection capability, as seen in \texttt{Sage - D4 Metric}, and an overdensity results in a bias against long duration signals, as seen in \texttt{Sage - Broad}. We tackle this issue by viewing the \ac{GW} detection problem as the learning of two independent tasks: noise rejection capability and unbiased learning of the \ac{GW} parameter space. Since the learning of these tasks simultaneously is not trivial, we introduce the concept of annealed training. For the \texttt{Sage - Annealed Train} run, we train a network similar to \texttt{Sage - Broad} for 20 epochs (with a mass distribution given by U$(m_1, m_2)$) and then transition the mass distribution to U($\tau_0$, $\tau_3$) over the course of the next 20 epochs. During the transition phase, the ratio between the number of samples generated using U($\tau_0$, $\tau_3$) to U$(m_1, m_2)$ varies as a function of epoch.
            
            The intent of annealing is two-fold: taking advantage of catastrophic forgetting~\cite{overcoming_catastrophic_forgetting} and the concept of ``unforgettable" examples~\cite{example_forgetting, unforgettable_examples}. In the context of continual learning, catastrophic forgetting is the tendency of a neural network to abruptly lose knowledge relevant to a previous task as information about the current task is incorporated~\cite{overcoming_catastrophic_forgetting}. This is undesirable in continual learning, but beneficial to our problem for shifting the training distribution from U$(m_1, m_2)$ to U($\tau_0$, $\tau_3$). Unforgettable examples (or more precisely, difficult-to-forget examples) are input data samples that are learnt early during training and typically have low losses during the bulk of the training \cite{unforgettable_examples}. This allows for all high \ac{SNR} signals to be remembered well throughout training and not be forgotten during the annealing process. This should allow for any noise rejection capability learnt via high \ac{SNR} signal features to also be retained.

            Figure \ref{fig:compare_main_runs} (top middle) compares the 1D chirp mass histograms of the detected signals at an \ac{FAR} of 1 per month between the \texttt{Sage - Annealed Train} and \texttt{Sage - D4 Metric} runs. We see a significant increase in the detection efficiency of the former run compared to the latter, proving the efficacy of the annealing training strategy.

            \textit{Network output behaviour}: We evaluate all the above runs on a validation dataset of 500,000 samples (1:1 ratio between signal and noise classes) comprised of signals and noise from the testing dataset distribution. Figure \ref{fig:compare_main_runs} (bottom) shows the network output obtained from this validation dataset for the signal class as a function of its corresponding chirp mass. The larger the network output, the more confident the model is to place that signal in the signal class. Notable observations have been described below.
            
            \textit{Observation 1}: There is a significant increase in the confidence of \texttt{Sage - Broad} when classifying signals across the whole range of chirp masses when compared to \texttt{Sage - Limited}. This is especially evident above a chirp mass of $\approx15\,\text{M}_{\odot}$. \texttt{Sage - Broad} has the least classification confidence for chirp masses below $\approx10\,\text{M}_{\odot}$ which contributes to the lower number of detections in this chirp mass regime compared to \texttt{PyCBC}.

            \textit{Observation 2}: The annealing procedure for \texttt{Sage - Annealed Train} improves the classification confidence from a chirp mass of $\approx10\,\text{M}_{\odot}$ till $\approx25\,\text{M}_{\odot}$. This indicates that the increase in detection efficiency seen in Fig. \ref{fig:compare_main_runs} (top middle) for \texttt{Sage - Annealed Train} over \texttt{Sage - D4 Metric} is primarily due to retaining information in this chirp mass region when annealed from the U$(m_1, m_2)$ mass distribution. Surprisingly, signals with a chirp mass above $\approx25\,\text{M}_{\odot}$ do not contribute much to the improved noise rejection capability.
            
            \textit{Observation 3}: On average the \texttt{Sage - D3 Metric} network is a lot more confident with classification than \texttt{Sage - D4 Metric} throughout the chirp mass range. Although the latter, on average, has lower confidence, it detects many signals across the whole range of chirp mass with higher confidence than \texttt{Sage - D3 Metric} and even \texttt{Sage - Broad}. Using the metric as the training distribution leads to a lower \textit{bias due to limited sample representation}, and in turn, the network is provided with more \ac{GW} domain knowledge. This allows it to be more confident with some signals. However, \texttt{Sage - Annealed Train} and \texttt{Sage - D4 Metric} do not seem to perform well at lower \ac{SNR}s leading to the worse average performance (there are more low \ac{SNR} than high \ac{SNR} signals; see the \ac{SNR} distribution in Fig. \ref{fig:testing_mock_datset_all_priors} in appendix \ref{app:testing dataset priors}). 

            \textit{Observation 4}: Interestingly, the upper error bar on all runs shows that the networks find longer signals with the highest ranking-statistics. However, other than \texttt{Sage - D3 Metric}, none of the other runs are good at low \ac{SNR} low chirp mass signals, leading to a lower average network output at low chirp masses. 

        \subsubsection{\label{subsubsec:learning_dynamics}Learning Dynamics}

\begin{figure}[htbp]
                \centering
                \includegraphics[width=1.0\linewidth]{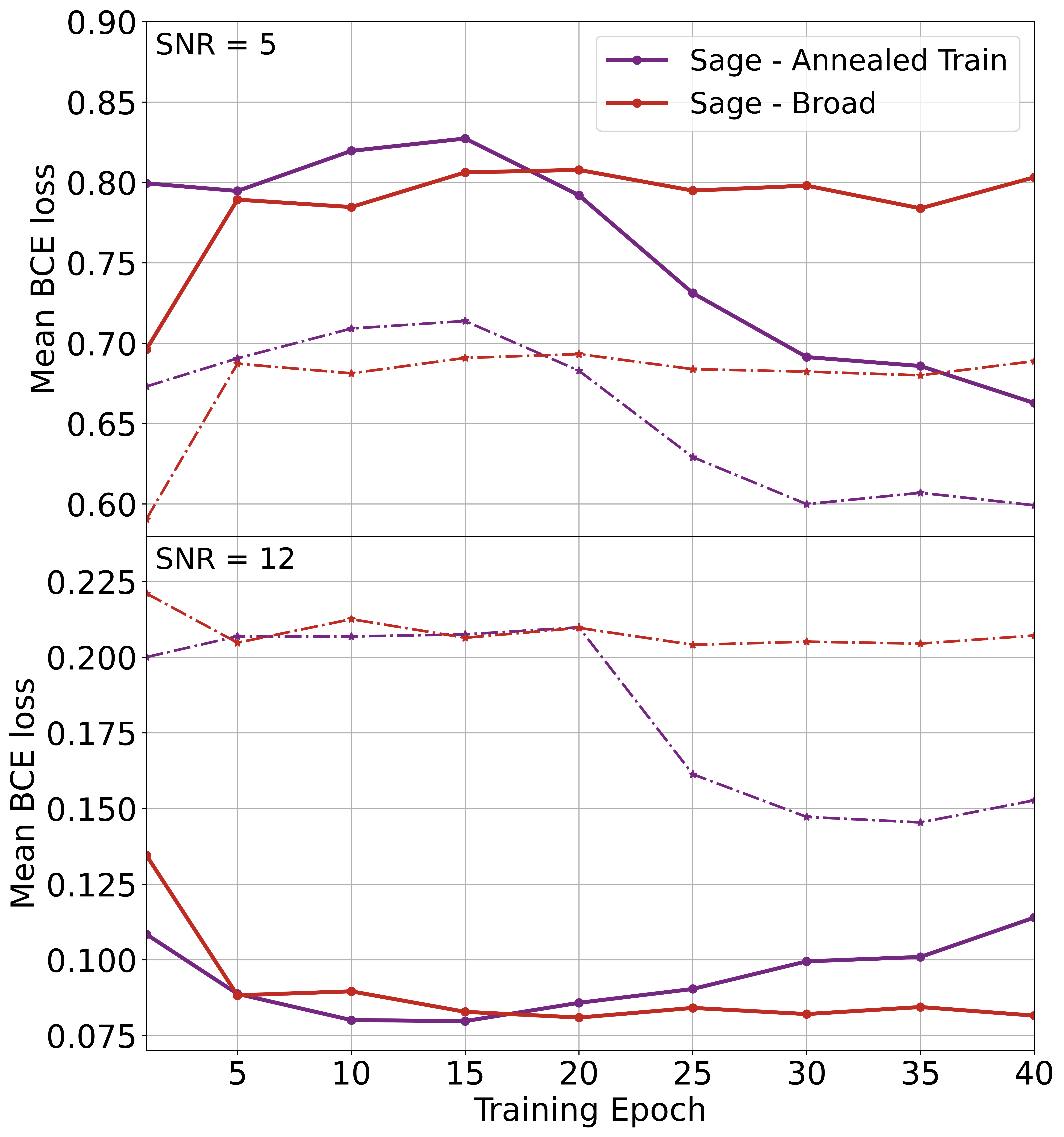}
                \caption{Mean BCE loss as a function of training epochs (only for the signal class). The plots compare the auxiliary dataset validation loss curves for the \texttt{Sage - Broad} and \texttt{Sage - Annealed Train} runs. The solid lines correspond to the low chirp mass dataset U[$\mathcal{M}_c^{\text{low}}$, $\mathcal{M}_c^{\text{median}}$] and the dashed lines correspond to the high chirp mass dataset U[$\mathcal{M}_c^{\text{median}}$, $\mathcal{M}_c^{\text{high}}$]. As the plot text states, the top plot is in the low \ac{SNR} regime ($=5$), and the bottom row is in the high \ac{SNR} regime ($=12$).}
                \label{fig:compare_learning_dynamics}
            \end{figure}

            Here, we investigate the difference between \texttt{Sage - Broad} and \texttt{Sage - Annealed Train} to potentially bridge the gap between the two runs. We create 4 auxiliary validation sets to analyse the learning dynamics in different regions of the parameter space. We chose all unique combinations without duplicates, between the two sets of distributions: \{\mbox{U[$\mathcal{M}_c^{\text{low}}$, $\mathcal{M}_c^{\text{median}}$]}, \mbox{U[$\mathcal{M}_c^{\text{median}}$, $\mathcal{M}_c^{\text{high}}$]}\} and \{\mbox{$\delta$(\ac{SNR}=5)}, \mbox{$\delta$(\ac{SNR}=12)}\}, where $\delta$ is the Dirac delta functions. This creates 4 datasets with different chirp mass and \ac{SNR} regimes. At the end of each training epoch, we freeze the weights and run the network on each of these auxiliary validation datasets with 500,000 samples (1:1 ratio between signal and noise classes). This provides us with the learning dynamics of the network in each regime. Figure \ref{fig:compare_learning_dynamics} depicts these 4 losses (obtained only for the signal class) over the course of 40 epochs of training for the runs \texttt{Sage - Broad} and \texttt{Sage - Annealed Train}. 

            \begin{figure*}[htbp]
                \centering
                \includegraphics[width=1.0\linewidth]{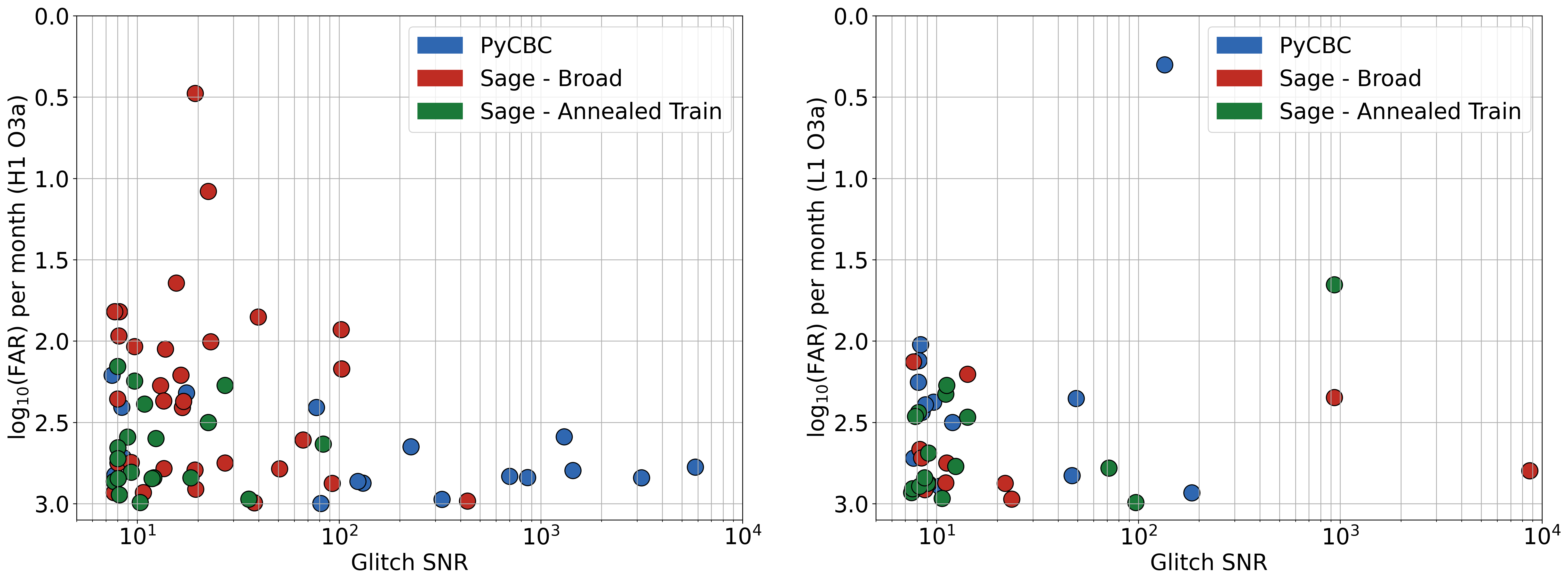}
                \caption{Log \ac{FAR} per month as a function of glitch \ac{SNR} (obtained from GravitySpy \cite{gravity_spy_o3ab}) for the H1 and L1 detectors. The scatter plot shows the glitch triggers produced by different \texttt{Sage} runs and \texttt{PyCBC} from the two detectors in the O3a testing dataset noise.}
                \label{fig:glitch_triggers_sage_and_pycbc}
            \end{figure*}
            
            \textit{Low \ac{SNR} regime} (see Fig. \ref{fig:compare_learning_dynamics} (top)): For the \texttt{Sage - Broad} run, we see that low chirp mass signals are more difficult to learn compared to high chirp mass signals, given that they have higher losses. It is interesting that both losses increase with more training, contradictory to typical loss curve dynamics. This is likely due to the need to reject glitches which could be similar in morphology compared to these signals. 
            
            \textit{High \ac{SNR} regime} (see Fig. \ref{fig:compare_learning_dynamics} (bottom)): We see a surprising trend where low chirp mass signals have much lower losses compared to high chirp mass signals, despite the expected biases against low chirp mass signals. Both losses have a decreasing trend, which is consistent with expectations since they should be easier to distinguish from non-Gaussian noise given their \ac{SNR}. 

            \textit{Effect of annealing}: The transition from $U(m_1, m_2)$ to $U(\tau_0,\tau_3)$ happens at epoch 20. Here, we see a mild increase in the loss for high \ac{SNR} low chirp mass signals despite having an overdensity of low chirp mass signals, which is indicative of a contribution to noise rejection capability. We also observe that the losses pertaining to high chirp mass signals decrease steadily during the annealing procedure at both \ac{SNR} regimes. However, despite having lower losses than \texttt{Sage - Broad} at high chirp masses, \texttt{Sage - Annealed Train} is not as performant. This is because we don't show the total loss function, which includes the noise class as well. The total loss of \texttt{Sage - Annealed Train} is higher than \texttt{Sage - Broad}, which explains the discrepancy.

            Given the overall lower losses for the signal class in \texttt{Sage - Annealed Train} compared to \texttt{Sage - Broad}, the key to improving the detection efficiency of \texttt{Sage - Annealed Train} must lie in improving the noise rejection capability. This requires an in-depth investigation of the noise behaviour and its effects in the learning of different regions of the \ac{GW} parameter space. We leave this study for the future. 

        \subsubsection{\label{subsubsec:Glitch rejection capability}Glitch Rejection Capability}

            In this section, we investigate the effect of the presence of glitches in the testing dataset. If the highest noise triggers are caused by glitches, the model would be extremely brittle to the testing dataset noise realisations and its detection efficiency at lower \ac{FAR}s has the potential to vary significantly with different datasets.
            
            We get the start time and duration of all glitches reported in the \texttt{Omicron} trigger pipeline \cite{omicron_trigger_pipeline}, with a glitch \ac{SNR} above 7.5 and a peak frequency of the noise event between $10$ Hz and $2048$ Hz, for the O3a testing dataset noise segments in the H1 and L1 detectors. The triggers that lie within the start GPS time and end GPS time of each glitch are obtained from \texttt{Sage} for different experiments and \texttt{PyCBC} for the D4 submission to MLGWSC-1. We then maximise these triggers to obtain the ranking statistic of a single trigger that represents each glitch. \texttt{Omicron} produces a glitch \ac{SNR} from the time-frequency representation of the strain and cannot be compared to the integrated signal \ac{SNR}. We plot the maximised ranking statistic of triggers below an \ac{FAR} of 1000/month against its corresponding glitch \ac{SNR} in Fig. \ref{fig:glitch_triggers_sage_and_pycbc} for the runs \texttt{Sage - Annealed Train} and \texttt{Sage - Broad} and compare it with \texttt{PyCBC}.
            
            We observe that all triggers produced by \texttt{Sage - Annealed Train} in H1 and L1 are above an \ac{FAR} of 100/month and 1/day, respectively. \texttt{Sage - Broad} consistently seemed to produce a lot more triggers comparatively and at \ac{FAR}s below the lowest \ac{FAR} observed in the \texttt{Sage - Annealed Train} for the H1 detector. However, both runs seem to perform similarly well in L1, with the annealed run producing the trigger with the lowest \ac{FAR}. Although we expect the annealed run to have better glitch rejection capability (due to lower \textit{bias due to limited sample representation} and, in turn, higher \ac{GW} domain knowledge), we cannot conclude much from this comparison since this is a relatively small testing dataset. However, comparing our results with \texttt{PyCBC}, we can conclude that the performance is similar, except for \texttt{PyCBC} producing a trigger at an \ac{FAR} of 2/month in L1. 
            
            We investigate a little further into the comparison of generated triggers for different types of glitches in the appendix (see Fig. \ref{fig:glitch_swarmplot_sage_and_pycbc} in appendix \ref{app:network behaviour for glitch types}).

        \subsubsection{\label{subsubsec:Effects of PSD}Effects of Non-stationary Noise}

            Since \texttt{Sage} and \texttt{PyCBC} seem to have similar glitch rejection capability, the primary cause for performance degradation in some \texttt{Sage} runs must be due to the differences in noise \ac{PSD}s between the training and testing dataset. We investigate this possibility by comparing the results between three runs: \texttt{Sage - D3 Metric}, \texttt{Sage - Limited} and \texttt{Sage - Broad}. With this investigation, we test three hypotheses: [i] if there is no difference between the training and testing noise \ac{PSD} distribution, the model should perform identically in both settings, [ii] a model trained with a broader noise \ac{PSD} distribution will be more confident in predicting the presence of a \ac{GW} signal compared to a model with a narrower distribution, and [iii] if the testing noise \ac{PSD} distribution is out-of-distribution compared to the training noise \ac{PSD} distribution, the model might not be able to extrapolate to the unknown \ac{PSD}s.

            \begin{table}[!htbp]
            \setlength{\tabcolsep}{12pt}
            \begin{tabular}{c c} 
                \toprule \toprule
                {Parameter} & {Value} \\ \midrule
                Approximant & IMRPhenomPv2 \\
                Mass 1 & 34.79 \\
                Mass 2 & 26.21 \\
                Right Ascension & 3.11 \\
                Declination & -0.62 \\
                Inclination & 0.83 \\
                Coalescence Phase & 1.75 \\
                Polarisation & 2.88 \\
                Time of Coalescence & 11.09 \\
                Spin1x & -0.30 \\
                Spin1y & 0.11 \\
                Spin1z & 0.01 \\
                Spin2x & 0.28 \\
                Spin2y & -0.25 \\
                Spin2z & -0.09 \\
                Chirp Mass & 26.24 \\
                Mass Ratio & 1.33 \\
                \bottomrule \bottomrule
            \end{tabular}
            \caption{Parameters used to produce the \ac{GW} signal used in section \ref{subsubsec:Effects of PSD}. The \ac{SNR} is rescaled to the required value, which dictates the rescaled distance parameter.}
            \label{tab:single_signal_params}
            \end{table}
            
            We create four different testing datasets (also known as a \textit{testset} within the plots), each with 500,000 samples of noise for studying the effects of time-varying \ac{PSD}s on \texttt{Sage}. These datasets are described below:
            
            \textit{Datasets 1 \& 2}: The first and second testing datasets contain simulated coloured Gaussian noise with \ac{PSD}s from the same distribution as that described for the \texttt{Sage - D3 Metric} run. The seed for the training and testing datasets for \texttt{Sage - D3 Metric} runs determines the random choice of \ac{PSD}s from the precomputed set when generating the noise samples. The first testing dataset uses the same seed as the training dataset and the second does not. 

            \textit{Datasets 3 \& 4}: The third and fourth datasets contain noise samples retrieved from O3a(Day 36 - 51) + O3b(Day 1 - 113) and O3a(Day 1 - 30) segments respectively. The inclusion or omission of O3a(Day 1 - 30) from the training or testing datasets significantly affects the variance of the noise \ac{PSD}s as seen in Fig. \ref{fig:o3a_psds}, where the yellow (in H1) and purple (in L1) \ac{PSD}s lie within this segment. The seed for these two datasets determines the random start time for each noise sample. The start time is uniformly distributed along the segments provided. The testing seed is not equal to the training seed for either of the datasets.
            
\begin{figure*}[htbp]
                \centering
                \includegraphics[width=1.0\linewidth]{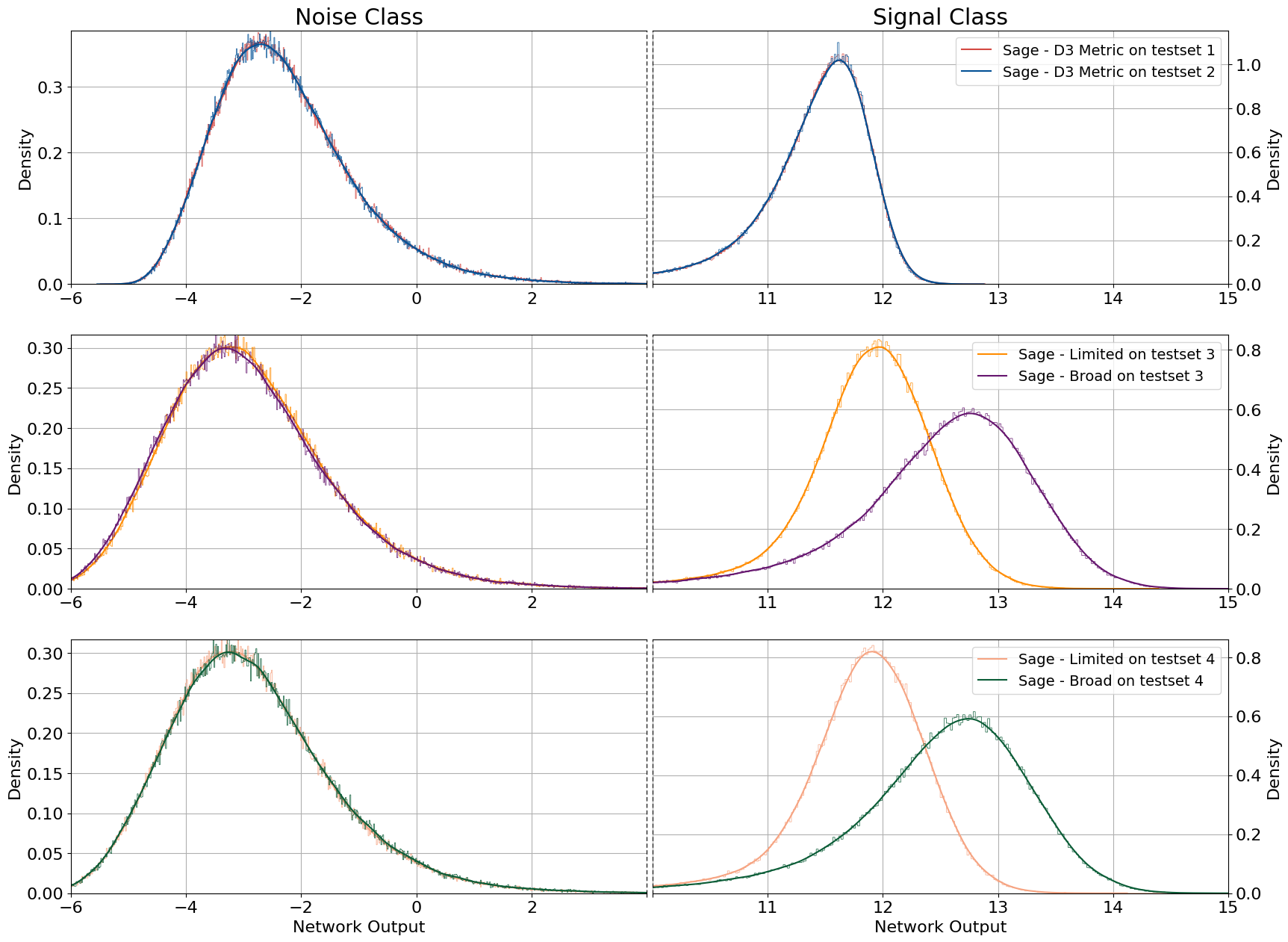}
                \caption{Normalised histograms of ranking statistics generated by \texttt{Sage} runs for samples from the signal and noise class on a particular testing dataset of 500,000 samples (1:1 ratio between the signal and noise class). A kernel density estimate of each histogram is plotted over each histogram. The left column corresponds to the noise class and the right column corresponds to the signal class. We truncate both histograms in the region of overlap between the two classes for plotting purposes.}
                \label{fig:psd_robustness}
            \end{figure*}

            We then generate a single arbitrarily chosen \ac{GW} signal with parameters provided in table \ref{tab:single_signal_params} and rescale its network optimal \ac{SNR} across H1 and L1 to be a fixed value of 10. This signal is injected at the provided time of coalescence into half of the noise samples in each dataset, producing a 1:1 ratio between the signal and noise classes. A summary of the different runs made is provided in table \ref{tab:summary_psd_robustness_runs}. The rest of this section is dedicated to the analysis of Fig. \ref{fig:psd_robustness}.

            \begin{table*}[htbp]
            \setlength{\tabcolsep}{12pt}
            \begin{tabular}{l l l c} 
                \toprule \toprule
                {Pointer} & {Model} & {Train PSD Distribution} & {Test PSD Distribution} \\ \midrule
                Row 1 Red & Sage - D3 Metric & fixed set of O3 PSDs & \multirow{2}{*}{same as train} \\
                Row 1 Blue & Sage - D3 Metric & fixed set of O3 PSDs & \\ \midrule
                Row 2 Orange & Sage - Limited & O3a(Day 36 - 51) + O3b(Day 1 - 113) & \multirow{2}{*}{same as Sage - Limited} \\
                Row 2 Purple & Sage - Broad & O3a(Day 1 - 51) + O3b(Day 1 - 113) & \\ \midrule
                Row 3 Pink & Sage - Limited & O3a(Day 36 - 51) + O3b(Day 1 - 113) & \multirow{2}{*}{O3a(Day 1 - 30)} \\
                Row 3 Green & Sage - Broad & O3a(Day 1 - 51) + O3b(Day 1 - 113) &\\
                \bottomrule \bottomrule
            \end{tabular}
            \caption{Summary of run configurations used to make Fig. \ref{fig:psd_robustness}. The first column points to a histogram in a particular row of Fig. \ref{fig:psd_robustness}. The second column provides the name of the \texttt{Sage} run used to generate the histogram (details about the runs can be obtained from section \ref{subsubsec:Injection Study}). The third column describes the PSD distribution used to generate the training dataset samples. \texttt{Sage - D3 Metric} uses a fixed set of O3 PSDs as described in section \ref{subsec:Data Generation}. \texttt{Sage - Broad} and \texttt{Sage - Limited} both use noise samples from O3a(Day 36 - 51) + O3b(Day 1 - 113) for training, but the former uses \ac{PSD}s from O3a(Day 1 - 30) to recolour the training samples, thus broadening the training \ac{PSD} distribution. The last column describes the PSD distribution used to generate the testing dataset samples.}
            \label{tab:summary_psd_robustness_runs}
            \end{table*}

            \textit{General description of Fig. \ref{fig:psd_robustness}}: After evaluating a particular testing dataset with a \texttt{Sage} run, we histogram the network output (or ranking statistic) produced by the model for the signal and noise class. The plots on the left column correspond to the noise class and the ones on the right to the signal class. The histograms should resemble a noncentral chi-squared distribution for both classes given the behaviour of real noise \ac{PSD}s. This is true for the signal class as they all contain the same signal with a fixed network optimal \ac{SNR} and \ac{GW} parameters. The intersection of the tails of the two classes contributes to type-I and type-II errors in the detection problem. The classification confidence of a model for a given dataset can be associated with the interclass distance (distance between average network outputs of the classes) and the reduction in type-I and type-II errors.

            \textit{Row 1 in Fig. \ref{fig:psd_robustness}}: The \texttt{Sage - D3 Metric} is evaluated on the first two testing datasets to produce the first row of results in Fig. \ref{fig:psd_robustness}. We see that the model produces similar results for both testing datasets (1 \& 2). This corroborates the hypothesis that the model should perform identically in the training and testing settings, given that the \ac{PSD} distribution is identical (ignoring non-Gaussian noise). 

            \textit{Row 2 in Fig. \ref{fig:psd_robustness}}: Here, we test \texttt{Sage - Limited} and \texttt{Sage - Broad} on the third dataset, with noise retrieved from O3a(Day 36 - 51) + O3b(Day 1 - 113). Our naive expectation might be that \texttt{Sage - Limited}, which is trained on the same noise segment, should perform better than \texttt{Sage - Broad} as it is trained on precisely the required \ac{PSD} realisations. However, we observe that \texttt{Sage - Broad} is more confident in classifying a signal injected into the same noise realisations than \texttt{Sage - Limited}, corroborated by the larger interclass distance. Moreover, \texttt{Sage - Broad} also seems to perform slightly better when classifying pure noise realisations from this dataset as seen on the left plot. This corroborates the hypothesis that a model trained with a broader noise \ac{PSD} distribution will be more confident in predicting the presence of a \ac{GW} signal compared to a model with a narrower distribution. 
            
            \textit{Reason for classification confidence}: This might not be simply due to a broader \ac{PSD} distribution, but rather the inclusion of more challenging noise \ac{PSD} realisations. Resolving the sources of type-I and type-II errors might lead to more confident models. The outcome of better generalisation due to the inclusion of harder training samples has been observed in many previous works \cite{hard_sample_mining_1, hard_sample_mining_2, hard_sample_mining_3}, and is thus a reasonable conclusion. However, we also found that simply decreasing the lower bound of the training \ac{SNR} distribution did not yield the same results. Thus, the improvement in confidence is likely due to better generalisation on noise \ac{PSD}s due to addressing the \textit{bias due to lack of variation}.

            \textit{Row 3 in Fig. \ref{fig:psd_robustness}}: We repeat the same procedure as Row 2, but using the fourth dataset, with noise retrieved from O3a(Day 1 - 30). The behaviour of the two networks is similar to that seen in Row 2, with the exception that both networks are now equally confident about the noise class outputs. Although \texttt{Sage - Broad} performed better than \texttt{Sage - Limited} in testing dataset 3 for the noise class, this behaviour did not transfer to \ac{PSD}s in testing dataset 4. The similar performance of the two models on the noise class indicates that \texttt{Sage - Limited} is able to successfully extrapolate to unknown \ac{PSD}s with higher power. The unchanged behaviour with the signal class in Row 2 and Row 3 signifies that neither network is overfitting.

            \textit{Is real noise better than Gaussian noise for training?}: Comparing Row 1 to Rows 2 \& 3, we see that models trained on real noise (\texttt{Sage - Limited \& Broad}) have higher interclass distance than the model trained on simulated coloured Gaussian noise (\texttt{Sage - D3 Metric}). This is due to two reasons: [i] the precomputed \ac{PSD}s used for \texttt{Sage - D3 Metric} does not vary significantly, thus leading to nearly white noise after whitening with the median \ac{PSD}, [ii] absence of non-Gaussian noise artefacts. Previous works \cite{closer_look_at_memorisation, learning_and_memorisation} have established that deep networks learn patterns easily and require less network capacity to do so, but white Gaussian noise (or the lack of patterns) is not learnt but memorised and takes a large amount of network capacity. Moreover, trying to learn Gaussian noise was shown to lead to complex decision boundaries and required longer training times. Thus, the lack of patterns in noise samples hinders learning.

            \textit{Implications}: Data pre-processing procedures that effectively whiten the noise samples (by using an estimated PSD or glitch removal), counter-intuitively, might hinder learning for the \ac{GW} detection problem. Using a broader training distribution for noise \ac{PSD} might allow the model to understand and interpolate the \ac{PSD} space rather than memorise it. Interpolation, in turn, leads to better generalisation and reduces the need to use larger training datasets (data efficiency).

            \begin{figure*}[htbp]
                \centering
                \includegraphics[width=1.0\linewidth]{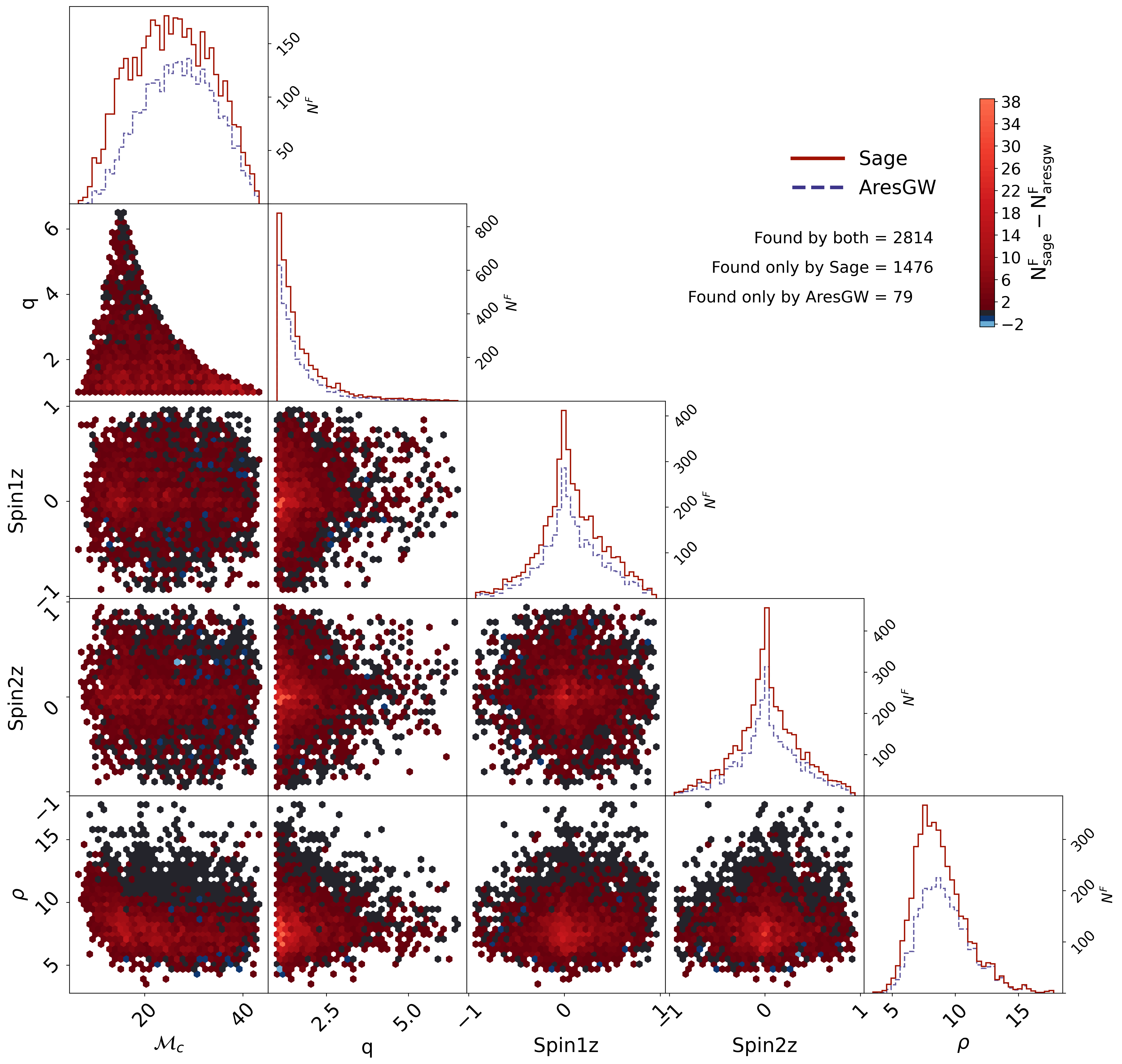}
                \caption{1D and 2D histograms of detected signals for different \ac{GW} parameters at an \ac{FAR} of 1 per month, comparing the best \texttt{Sage - Broad} run and \texttt{AresGW}. We chose \ac{GW} parameters where we expected to see learning bias. The diagonal subplots contain 1D histograms of the detected signals for a given parameter. The 2D histograms are hexagonal binned plots, where each bin is coloured using $N^F_{\text{sage}}-N^F_{\text{aresgw}}$ where $N^F_{\text{sage}}$ and $N^F_{\text{aresgw}}$ is the number of signals detected by \texttt{Sage} and \texttt{AresGW} respectively. We use a divergent colour map that is set to be black when $N^F_{\text{sage}}-N^F_{\text{aresgw}}$ is zero, is progressively brighter in red along the positive axis, and is progressively brighter in blue along the negative axis. So, red and blue patches can be attributed to regions where \texttt{Sage} and \texttt{AresGW} detect more signals, respectively.}
                \label{fig:compare_sage_to_aresgw}
            \end{figure*}

    \subsection{\label{subsec:Comparison with ML}Comparison with Other ML Pipelines} 

        In this section, we compare our results with recent \ac{ML} pipelines that have evaluated their work on the datasets provided in MLGWSC-1 \cite{DP5_MLGWSC1_2023}. We do a more in-depth comparison with \texttt{AresGW}~\cite{DP6_Nousi_2023} and \texttt{Aframe}~\cite{aframe} given that they exhibit the highest detection sensitivity among the pipelines considered. 

        \subsubsection{Comparison with \texttt{AresGW}}

        \textit{Data/architecture differences between \texttt{AresGW} and \texttt{Sage}}: The \texttt{AresGW} architecture is a 1D ResNet-54 trained using the curriculum learning strategy \cite{DP6_Nousi_2023, curriculum_learning_survey}. For training, they use a $1$-second sample length, 12 days of real O3a noise data and 38,000 different waveforms (each injected into 19 different background segments, leading to 740,000 different training samples). They use the same training mass distribution to generate waveforms as \texttt{Sage - Broad}.
        
        Figure \ref{fig:compare_sage_to_aresgw} shows the distribution of detections across different parameters and compares the number of detections between \texttt{Sage - Broad} and \texttt{AresGW} at an \ac{FAR} of 1/month. We see that 2814 signals were found by both pipelines, 1476 signals were found only by \texttt{Sage} and 79 signals were found only by \texttt{AresGW}. This is equivalent to an increase of $\approx48.29\%$ in the number of signals detected at that \ac{FAR}. \texttt{Sage - Broad} particularly outperforms its counterpart in the lower chirp mass regime.

        The \texttt{AresGW} pipeline was enhanced in \cite{Koloniari_gw_events} and included the following major changes: using 35 days of real O3a noise instead of 12, a low-pass filter at $350$ Hz for all samples, and changes to how the triggers are generated and aggregated during testing. The relative difference in detection sensitivity between \texttt{Sage} and enhanced-\texttt{AresGW} should not change by much due to the additional noise data, since the latter does not mitigate the biases in section \ref{subsec:Issues}. This statement holds even if enhanced-\texttt{AresGW} were trained with 160 days of real O3 noise, 80 million training samples with a length of $12$ seconds and the same training strategies as \texttt{Sage}. Justification for these statements is provided with sufficient detail in ablation study II and III (see section \ref{subsec:Ablation study}).

        \subsubsection{Comparison with \texttt{Aframe}}
        \textit{Data/architecture differences between \texttt{Aframe} and \texttt{Sage}}: The \texttt{Aframe} architecture is a 1D ResNet-54 trained using the curriculum learning strategy~\cite{aframe, curriculum_learning_survey}. For training, they use a $1.5$-second sample length, $\approx4.5$ days of real O3 noise data and 75,000 different waveforms (each injected into a random noise segment obtained from the $\approx4.5$ days of real noise). A key area of difference that sets apart \texttt{Aframe} from the other pipelines we compare to is the training mass distribution. They use $p(m_1)\propto m_1^{-2.35}$ and $p(m_2|m_1)\propto m_2$ as per GWTC-3 \cite{GWTC-3}, which places a higher emphasis on lower chirp mass or longer duration signals. Moreover, they train on a mass range between 5 and 100 $M_{\odot}$. 

\begin{figure*}[htbp]
            \centering
            \includegraphics[width=1.0\linewidth]{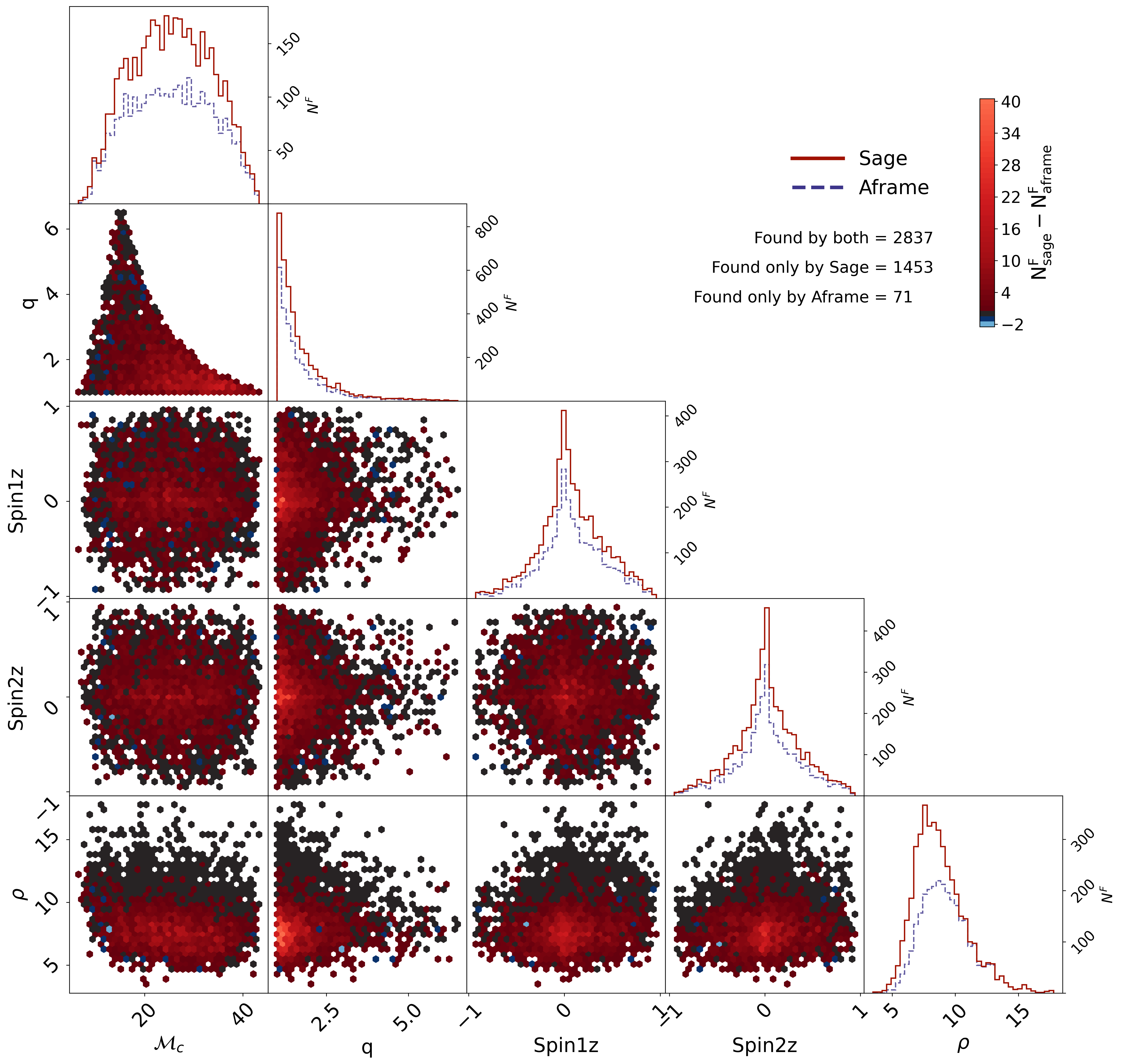}
            \caption{1D and 2D histograms of detected signals for different \ac{GW} parameters at an \ac{FAR} of 1 per month, comparing the best \texttt{Sage - Broad} run and \texttt{Aframe}. We chose \ac{GW} parameters where we expected to see learning bias. The diagonal subplots contain 1D histograms of the detected signals for a given parameter. The 2D histograms are hexagonal binned plots, where each bin is coloured using $N^F_{\text{sage}}-N^F_{\text{aframe}}$ where $N^F_{\text{sage}}$ and $N^F_{\text{aframe}}$ is the number of signals detected by \texttt{Sage} and \texttt{Aframe} respectively. We use a divergent colour map that is set to be black when $N^F_{\text{sage}}-N^F_{\text{aframe}}$ is zero, is progressively brighter in red along the positive axis, and is progressively brighter in blue along the negative axis. So, red and blue patches can be attributed to regions where \texttt{Sage} and \texttt{Aframe} detect more signals, respectively.}
            \label{fig:compare_sage_to_aframe}
        \end{figure*}

        Figure \ref{fig:compare_sage_to_aframe} shows the distribution of detections across different parameters and compares the number of detections between \texttt{Sage - Broad} and \texttt{Aframe}\footnote{Results obtained via private communication.} at an \ac{FAR} of 1/month. We see that 2837 signals were found by both pipelines, 1453 signals were found only by \texttt{Sage} and 71 signals were found only by \texttt{Aframe}. The number of signals detection is similar to \texttt{AresGW}, but the distribution is expectedly different given the different training mass distribution (see Fig. \ref{fig:compare_sage_to_aresgw} for reference). The 1D histogram of chirp mass in Fig. \ref{fig:compare_sage_to_aframe} shows that \texttt{Aframe} addresses the \textit{bias due to limited sample representation} of low chirp mass signals via the training mass distribution, thus detecting more longer duration signals than \texttt{AresGW}. The 1D histogram of optimal \ac{SNR} is similar to that of \texttt{AresGW}, indicating a limited capacity to effectively detect low \ac{SNR} signals.

        For a more meaningful comparison, we must train \texttt{Sage} on the $p(m_1)\propto m_1^{-2.35}$ distribution, which will likely detect more long duration signals than \texttt{Sage - Broad}, as indicated by the \texttt{Sage - Annealed Train} run in figure \ref{fig:compare_main_runs}. As \texttt{Aframe} uses a $1.5$-second sample length, they do not address the \textit{bias due to insufficient information}, and a testing dataset that contains significantly more long-duration signals will enunciate this bias. We leave this analysis to the future.

\begin{figure}[htbp]
            \centering
            \includegraphics[width=1.0\linewidth]{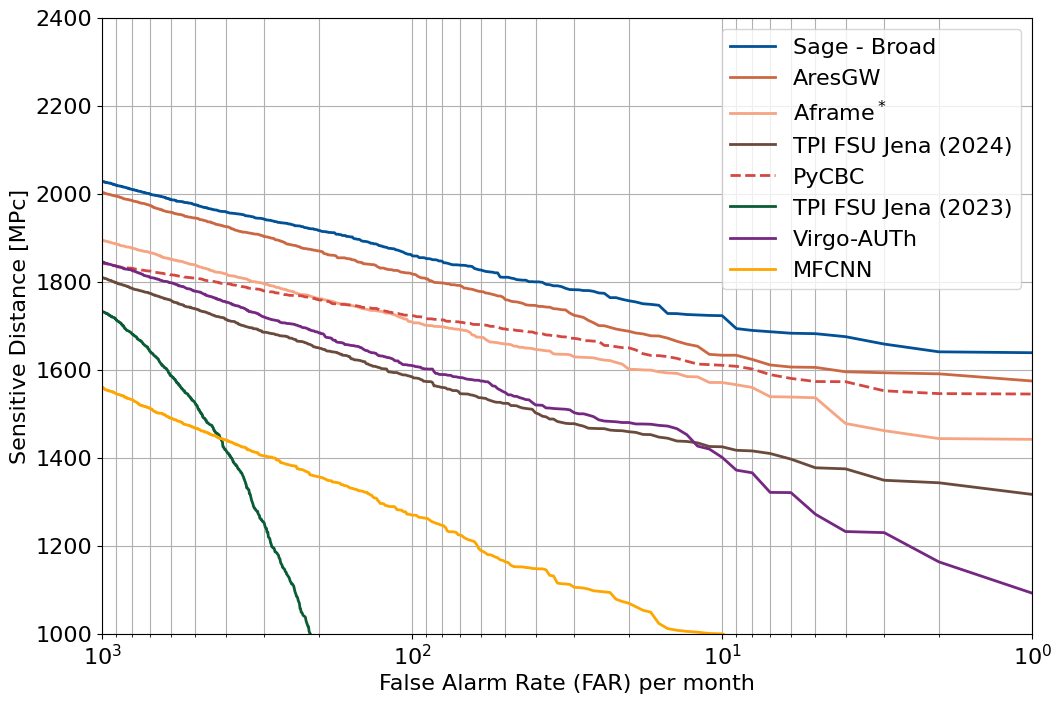}
            \caption{The sensitive distance metric as a function of \ac{FAR} per month for \texttt{Sage - Broad}, other \ac{ML} pipelines and \texttt{PyCBC} on the D4 testing dataset. The \texttt{AresGW} results were obtained from \cite{DP6_Nousi_2023}, \texttt{Aframe} and \texttt{TPI FSU Jena} (2024) results via private communication, and all other results were obtained from \cite{DP5_MLGWSC1_2023}. The \texttt{Aframe} results are marked with an asterisk ($^*$) as their pipeline is not optimised for the D4 mass distribution in MLGWSC-1~\cite{DP5_MLGWSC1_2023}.}
            \label{fig:compare_ml_pipelines}
        \end{figure}

        \textit{Consolidated comparison}: We consolidate the comparison of the detection performance for different pipelines in table \ref{tab:summary_sensitive_distance} and shows the sensitive distance metric of multiple \ac{ML} pipelines and \texttt{PyCBC} at different \ac{FAR}s in Fig. \ref{fig:compare_ml_pipelines}. We see an increase of $\approx4.1\%$ in sensitive distance compared to \texttt{AresGW} at an \ac{FAR} of 1 per month. For the sake of completeness, we provide the sensitive distance curve of \texttt{Aframe} in Fig. \ref{fig:compare_ml_pipelines}. The disadvantage of using a parameter-dependent metric is clearly evident when comparing our \texttt{Sage - Annealed Train} run with \texttt{AresGW}. Although our run finds $\approx15.5\%$ more signals at an \ac{FAR} of 1 per month, it scores $\approx3.5\%$ less on the sensitive distance metric, due to the chirp mass dependence. 

        \begin{table*}[htbp]
        \setlength{\tabcolsep}{12pt}
        \begin{tabular}{c c c c c c} 
            \toprule \toprule
            {$\text{Pipeline}$} & {$\text{FAR}\;100/\text{month}$} & {$\text{FAR}\;10/\text{month}$} & {$\text{FAR}\;1/\text{month}$} & {$N^{F}_{\text{10/month}}$} & {$N^{F}_{\text{1/month}}$}\\ \midrule
            PyCBC~\cite{DP5_MLGWSC1_2023} & 1722.008 & 1607.307 & 1538.768 & 4364 & 3857  \\
            CWB~\cite{DP5_MLGWSC1_2023} & 1364.195 & 1440.848 & 1460.617 & -- & -- \\ \midrule
            Aframe$^*$~\cite{aframe} & 1708.237$^*$ & 1570.450$^*$ & 1441.291$^*$ & 4252$^*$ & 2908$^*$  \\
            AresGW~\cite{DP6_Nousi_2023} & 1817.074 & 1632.455 & 1573.914 & 3878 & 2893 \\
            TPI-FSU Jena (2024)~\cite{DP7_Zelenka_2024} & 1583.625 & 1424.370 & 1316.161 & 2932 & 1609 \\
            MFCNN~\cite{DP5_MLGWSC1_2023} & 1269.034 & 984.995 & 587.930 & 444 & 89  \\ \midrule
            Sage - Broad & \textbf{1858.554} & \textbf{1722.352} & \textbf{1638.204} & \textbf{4760} & \textbf{4290} \\
            Sage - Limited & 1851.586 & 1710.573 & 1553.479 & 4731 & 3306 \\
            Sage - D4 Metric & 1758.124 & 1609.436 & 1405.947 & 4179 & 2316 \\
            Sage - Annealed Train & 1748.338 & 1613.982 & 1518.557 & 4280 & 3341 \\
            \bottomrule \bottomrule
        \end{tabular}
        \caption{Summary of sensitive distance measures and number of signals detected in the realistic testing dataset for different pipelines and \texttt{Sage} runs, for different \ac{FAR}s. $N^{F}_{\text{1/month}}$ and $N^{F}_{\text{10/month}}$ are the number of found injections at an \ac{FAR} of 1 per month and 10 per month respectively. The \texttt{Aframe} results are marked with an asterisk ($^*$) as their pipeline is not optimised for the D4 mass distribution in MLGWSC-1~\cite{DP5_MLGWSC1_2023}.}
        \label{tab:summary_sensitive_distance}
        \end{table*}

    \subsection{\label{subsec:Ablation study}Ablation Study on \texttt{Sage}} 

        Although a common practice for pipelines in the \ac{ML} literature \cite{ablation_studies}, ablation studies are generally not performed in the \ac{ML} for \ac{GW} literature. Given the diversity of methods that can be used in \ac{ML}, it is imperative to prove the necessity of important parts of our methodology.
        
        Although we have shown proof and explanations for the performance gains seen in \texttt{Sage} compared to previous \ac{ML} pipelines, the source of this performance is not apparent. We try to understand the contribution of some of the major components described in the methodology (see section \ref{sec:Methods}) by systematically removing them and evaluating \texttt{Sage}'s performance on the same realistic testing dataset. The methodology used throughout the ablation study is the same as that seen in section \ref{sec:Methods}, and any changes made are described in each study. The training seeds are fixed for all studies, so they use the same signal and noise realisations for all epochs. The studies that use a limited dataset are trained on the same pre-made dataset. All comparison plots are made at an \ac{FAR} of 1/month. This ablation study is not exhaustive and only targets some of the major differences that led to significant performance gains.
        
        \textit{Study I - Decreased Noise and Signal Realisations}: For the first study, we removed several major components of \texttt{Sage} that were meant to increase the variance in the signal and noise classes during training. For the signal class, we generate the plus and cross polarisations of $10^6$ \ac{GW} signals with the same generation settings as \texttt{Sage - Broad}. Each iteration of every epoch is given a unique seed, which is used to set the right ascension, declination, and polarisation for the antenna pattern before calculating the strain of the waveform as measured by the detector. A time shift is applied for the given detector relative to the assumed geocentric frame to obtain the strain for each of the considered detectors. We then rescale the network \ac{SNR} of the system by drawing \ac{SNR} values from a uniform distribution within the bounds [5, 15]. This procedure limits the \ac{GW} parameter augmentation to the four parameters mentioned above. All other parameters have a fixed $10^6$ realisations throughout the training procedure. For the noise class, we retrieve $10^6$ samples from O3a(Day 31 - 51) noise segment (with sample overlap) and do not apply any noise recolouring.

\begin{figure*}[htbp]
            \centering
            \includegraphics[width=1.0\linewidth]{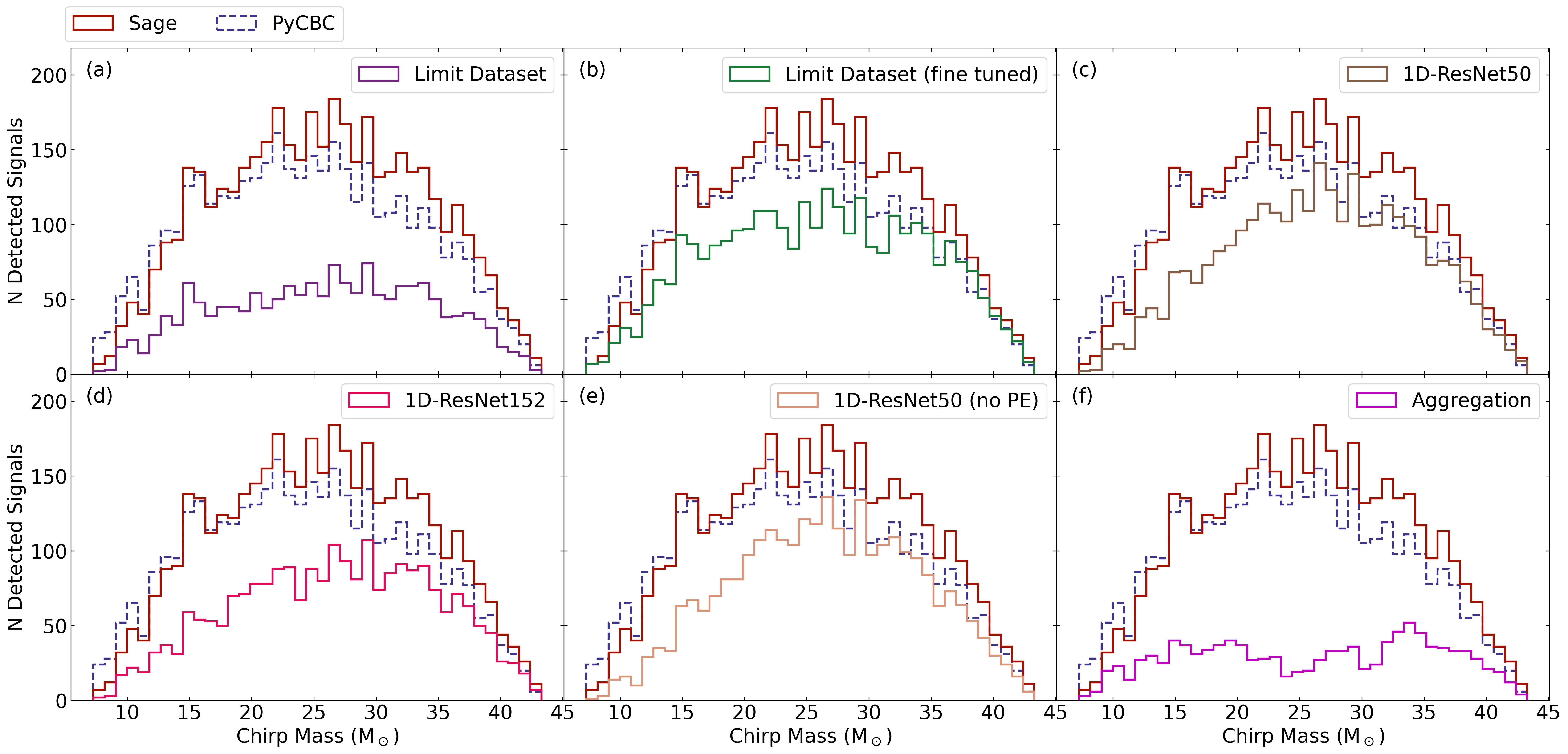}
            \caption{1D chirp mass histograms of detected signals for all ablation studies compared with \texttt{Sage - Broad} and \texttt{PyCBC} at an \ac{FAR} of 1 per month. Panels (a) and (b) correspond to ablation study I (Decreased Noise and Signal Realisations), panel (c) shows study II (Using a Simpler Network Architecture), panel (d) shows study III (Using a Simple (but deeper) Network Architecture), panel (e) shows study IV (Abandon Regularisation), and panel (f) shows study V (Effects of Aggregation).}
            \label{fig:all_ablation_studies}
        \end{figure*}

        Figure \ref{fig:all_ablation_studies}(a) shows the chirp mass histogram of the number of found injections compared to \texttt{PyCBC} at an \ac{FAR} of 1/month. It also shows reference histograms for the \texttt{PyCBC} and best \texttt{Sage - Broad} results on this dataset, taken from Fig. \ref{fig:compare_sage_to_pycbc}. We only use chirp mass here since it is the parameter along which we observe the most bias. Although we use the same network architecture and fix all other settings, the detection performance is significantly reduced. We detected signs of overfitting during training, likely because many of the training hyperparameters were tuned for the original setting and higher variance in the noise and signal classes.

        We fine-tune the hyperparameters of the network to work on the reduced dataset for a fairer comparison. Although the \texttt{Sage} architecture reduces some of the biases, the distribution of the parameters, the amount of variation and the training strategy are paramount to achieving a good detection performance. Figure \ref{fig:all_ablation_studies}(b) shows an increase in performance due to the fine-tuning but does not compare well with our best results. All ablation studies henceforth have been fine-tuned to account for the corresponding changes made.

        \textit{Study II - Using a Simpler Network Architecture}: In this study, we use a 1D ResNet-50 \cite{resnet}, which is comparable in size to the networks used in \cite{aframe, DP6_Nousi_2023, Koloniari_gw_events}. Due to the lack of relevant inductive biases (a network architecture or training strategy particularly designed for the problem) in the standard 1D ResNet, we observe in Fig. \ref{fig:all_ablation_studies}(c) that the detection sensitivity has significantly degraded compared to \texttt{Sage}. This study indicates that the architectures used in \texttt{AresGW}~\cite{DP6_Nousi_2023}, enhanced-\texttt{AresGW}~\cite{Koloniari_gw_events} and \texttt{Aframe}~\cite{aframe} would not experience a significant increase in detection sensitivity even if they were trained on 160 days of real O3 noise, 80 million training samples of length $12$-seconds and the same training strategies as \texttt{Sage}. This empirically justifies the requirement of appropriate inductive biases.
                
        \textit{Study III - Using a Simple (but deeper) Network Architecture}: For the third study, we use a 1D ResNet-152 \cite{resnet}. We chose this architecture for two reasons: [i] This is a fairer comparison to \texttt{Sage} as the 1D ResNet-50 is significantly smaller in terms of the number of trainable parameters, and [ii] deeper networks are known to learn long term dependencies better \cite{effective_receptive_field} which also offers an advantage over the 1D ResNet-50. However, we see in Fig. \ref{fig:all_ablation_studies}(d) that the 1D ResNet-152 performs worse than the much shallower 1D ResNet-50. This is because we fixed the number of training iterations to be the same for all ablation studies. Bigger networks are typically less data efficient than shallower networks~\cite{big_network_waste_capacity}, which explains the slightly worse performance in comparison to a 1D ResNet-50. However, this is \textit{only} true when the proper inductive biases are not used. The 1D ResNet-152 and \texttt{Sage} have nearly the same number of trainable parameters, with the former being slightly larger. The reduction in performance seen in Fig. \ref{fig:all_ablation_studies}(d) is thus purely due to the lack of proper inductive bias. The \texttt{Sage} architecture allows for significantly higher data efficiency.
        
        \textit{Study IV - Abandon Regularisation}: All previous ablation studies have predicted a point estimate of the chirp mass and time of coalescence of the supposed \ac{GW} signal in each sample. Section \ref{subsec:Architecture} of the methodology describes this point estimate as a regularisation method that provides an additional inductive bias to distinguish between the signal and noise classes. We repeat study II but remove this regularisation and see in Fig. \ref{fig:all_ablation_studies}(e) that it leads to a comparative reduction in detection performance with Fig. \ref{fig:all_ablation_studies}(c). Since the training seeds were exactly the same between studies II and IV, we can safely conclude that the difference is purely due to the lack of this regularisation feature.
        
        \textit{Study V - Effects of Aggregation}: For the final study, we emphasise the aggregating effects of supervised learning by using an architecture that is not well suited to handling glitches in the data. Instead of our CBAM ResNet-50 backend classifier, we use a Res2Net \cite{res2net}, a multiscale backbone to the standard ResNet architecture. We see that in Fig. \ref{fig:all_ablation_studies}(f), this relatively minor change led to a drastic reduction in detection performance. We found that certain architectures led to a completely invariant ranking statistic for all signals above a certain threshold in \ac{SNR} (let us name this $T_{\text{snr}}$). This threshold was typically present close to the signal-noise class decision boundary. Say we have 1 month of unseen testing data. The presence of a single unseen difficult-to-classify glitch in this dataset could generate a trigger with a ranking statistic close to $T_{\text{snr}}$ and instantly reduce the detection sensitivity at an \ac{FAR} of 1 per month. This sort of \textit{aggregation} of triggers close to the decision boundary can be mildly alleviated when using larger networks, and significantly mitigated when using networks with proper inductive biases.

\section{\label{sec:Running Sage on Live Data}Running Sage on Live Data}
Sections \ref{subsubsec:Injection Study}, \ref{subsubsec:Glitch rejection capability} and \ref{subsubsec:Effects of PSD} prove that \texttt{Sage} is able to achieve highly sensitive detection performance with good glitch rejection capability and limited extrapolation capability with noise \ac{PSD}s. Thus, we believe that \texttt{Sage} can handle real-time detection of \ac{GW} alongside other existing production-level pipelines. We dedicate this section to detail minor modifications to \texttt{Sage} that will allow for its usage on an unknown live data feed.

Say we want to run \texttt{Sage} on the fifth observing run (O5). Before running a search on live data, we must first train the model with noise realisations that resemble the new data, to bridge the gap due to different noise \ac{PSD}s. We can obtain a small amount of noise \ac{PSD}s from the new data and recolour previously known noise realisations from O3 and O4, giving us a large dataset. This data can then be used to fine-tune \texttt{Sage} over 1 or 2 epochs, which allows it to adapt to the new noise \ac{PSD}s. To achieve a low enough \ac{FAR}, we can modify \texttt{Sage} in one of the following ways: 

\textit{Method 1}: Our runs in the injection study assume a known background dataset to estimate the \ac{FAR} of an event. In the case of live data, we can retain the current multi-detector architecture of \texttt{Sage} and perform a time-slide analysis \cite{time_slides} of the strain data with limited data to achieve low \ac{FAR}s. To perform a time-sliding technique, the model must first be trained on a dataset that includes a signal in one detector but not the other. This will incentivise the model to produce appropriate null triggers when operating on time-slid data. However, this will use unnecessary learning capacity (since time-slid data is not of interest astrophysically) and will likely increase training times. For each time-slid data sample, the model must run in evaluation mode; to achieve an \ac{FAR} of 1 per year, we must evaluate the model on 1 year of time-slid data. The primary disadvantage of this method is that going to very low \ac{FAR}s might become infeasible given evaluation times.

\textit{Method 2}: We can run \texttt{Sage} on a single-detector coincident analysis setting and perform time-sliding. Background estimation can be performed similar to \texttt{PyCBC Live}, by storing a rolling buffer of detector triggers and performing the time-sliding technique to estimate the \ac{FAR}. Shifting the triggers beyond the light travel time between detectors will negate some background triggers via non-physical coincidences. Contrary to the first method, this will lead to highly efficient time sliding of \textit{triggers} instead of the strain itself. However, a single-detector \ac{ML} model will not be able to take advantage of some of the architectural features introduced in section \ref{subsec:Architecture}, like channel-wise attention in the backend classifier. This might reduce detection efficiency. 

\textit{Running \texttt{Sage} on O3}: Say we want to conduct an offline search on O3 data with \texttt{Sage}. Since \texttt{Sage} is trained on the entirety of O3 noise data (as labelled by current matched-filtering searches), there is the natural concern that our training dataset possibly includes as-of-yet undetected signals in the noise class. However, given that the ratio between the volume of undetected signals to genuine noise is negligible, the following biases will prevent the model from learning undetected signals as noise: [i] the \textit{bias due to limited sample representation} and [ii] the significant imbalance between noise and signal features in the noise class (\textit{bias due to limited feature representation}).


\section{\label{sec:Limitations}Limitations of \texttt{Sage}}
\textit{Persistent biases}: Although we tried to address the biases mentioned in section \ref{sec:Introduction}, our best results in Fig. \ref{fig:compare_sage_to_pycbc} are still marginally biased against lower chirp masses. An attempt to alleviate all mentioned biases, as seen for the \texttt{Sage - Annealed Train} run in Fig. \ref{fig:compare_main_runs}, leads to an $\approx13.3\%$ decrease in the number of signals detected at an \ac{FAR} of 1 per month when compared to \texttt{PyCBC}.

\textit{Evaluation speed}: A larger testing dataset would allow for performance evaluations at lower \ac{FAR}s. However, even at the high evaluation speed of the analysis, the current coherent strategy is limited due to the issues raised in section \ref{sec:Running Sage on Live Data} \textit{method 1}. Since evaluating on 1 month of testing data takes $\approx16$ hours, achieving an \ac{FAR} of 1 in 10 years, for example, would require $\approx80$ days. Even if a model had an evaluation speed of 30 minutes for 1 month of testing data, it would still take 2.5 days. Thus, section \ref{sec:Running Sage on Live Data} \textit{method 1} becomes quickly impractical.

\textit{Mild train-test overlap}: The best result we present uses noise \ac{PSD}s obtained from the testing dataset to recolour the training noise realisations. The resulting recoloured noise realisation is technically new since the testing data \ac{PSD} is only an estimate and whitened noise realisation (before being recoloured) is different from testing data. Although the network does not exhibit any signs of overfitting, this method still introduces mild train-test overlap. It was shown in section \ref{subsubsec:Effects of PSD} that the performance gain was \textit{not} due to this overlap, but the result of a broader distribution of noise \ac{PSD}s in the training dataset. We can thus alleviate the mild train-test overlap by generating artificial \ac{PSD}s using a normalising flow \cite{normalising_flows} and retraining \texttt{Sage} once before an online search with some representative \acp{PSD}. It is possible to retrain an \ac{ML} model such that it does not need further retraining throughout an observing run, as evidenced by \cite{adapting-to-noise-ml}.

\textit{Small testing dataset}: The number of glitches, although substantial ($\mathcal{O}(10^4)$) within the month of O3a testing noise, is not sufficient to fully analyse the glitch rejection capability of \texttt{Sage}. Moreover, there is no guarantee that the glitch profiles of \textit{all} glitch types were fully learnt given the imbalance of data for different glitch types (see Fig. \ref{fig:glitch_swarmplot_sage_and_pycbc} in appendix \ref{app:network behaviour for glitch types}). 

\textit{Need for large training dataset}: Our current work requires a large training dataset in order to reduce biases. Providing sufficient variation to all \ac{GW} parameters and including all known noise realisations will prove difficult in the future. Thus, we have to work toward creating an information-efficient pipeline that is able to adapt to unforeseen changes in noise characteristics without the need for retraining.

\section{\label{sec:Conclusion}Conclusion}
We introduce \texttt{Sage}, an \ac{ML}-based \ac{BBH} detection pipeline, that is shown to be capable of detecting $\approx11.2\%$ and $\approx48.29\%$ more signals than the benchmark \texttt{PyCBC} submission in \cite{DP5_MLGWSC1_2023} and the \texttt{AresGW} results in \cite{DP6_Nousi_2023} for the Machine Learning Gravitational Wave Search Challenge (MLGWSC-1) injection study, respectively, at an \ac{FAR} of one per month in O3a noise. We achieve this performance by identifying a non-exhaustive set of 11 interconnected biases that are present in the usage of \ac{ML} for the \ac{GW} detection problem and create mitigation tactics and training strategies to address them concurrently. We prove that \texttt{Sage} is robust to \ac{OOD} variation in noise power spectral densities and can reject non-Gaussian transient noise artefacts effectively. Explanations and rationale have been provided for all method choices, and an ablation study was conducted to provide evidence for our claims. Our work addresses the need for interpretability in the application of \ac{ML} to \ac{GW} detection.

The biases, mitigation tactics and training strategies that we provide are general, and the salient ideas are transferable to the detection of other sources of \ac{GW}s and to other domains of \ac{GW} data analysis. We re-iterate some of our primary empirical contributions for the application of \ac{ML} to the \ac{GW} detection problem: 

\noindent \textbf{Negating the need for \ac{PSD} estimation}: Contrary to previous \ac{ML}-based detection pipelines operating on real detector noise \cite{DP6_Nousi_2023, aframe, DP7_Zelenka_2024, Koloniari_gw_events}, we show that \texttt{Sage} does not need to use a \ac{PSD} estimate for whitening any input sample (see section \ref{sec:Methods}, \textit{Data Transformation}). We use a single approximate \ac{PSD} instead for all samples.

\noindent \textbf{Mitigating biases might aid in glitch rejection}: In section \ref{subsubsec:Glitch rejection capability}, we obtain all the triggers generated by \texttt{Sage} and \texttt{PyCBC} during glitch events in one month of O3a noise and prove that \texttt{Sage} is capable of effectively rejecting glitches. We also argue about the potential benefits of mitigating biases for glitch rejection.

\noindent \textbf{\ac{OOD} \ac{PSD}s are good for classification}: In section \ref{subsubsec:Effects of PSD}, we show that using a broader range of noise \ac{PSD}s for the training data, which could even be \ac{OOD} to the testing dataset, greatly aids in detection performance.

\noindent \textbf{Real noise is better than Gaussian noise for training}: We show in section \ref{subsubsec:Effects of PSD} that models trained with real detector noise produce more confident classifiers than models trained with simulated coloured Gaussian noise.

\noindent \textbf{Parameter efficiency via inductive biases}: Via an ablation study (see section \ref{subsec:Ablation study}) we show that a model that is larger than \texttt{Sage} in terms of number of trainable parameters is not able to achieve nearly the same detection sensitivity as \texttt{Sage}, \textit{if} it does not have the proper inductive biases.

\begin{acknowledgments}
NN would like to extend special thanks to Will Benoit and Ond\v rej Zelenka for running their respective \ac{ML}-based search pipelines, \cite{aframe} and \cite{DP7_Zelenka_2024}, on the D4 testing dataset provided in MLGWSC-1~\cite{DP5_MLGWSC1_2023}. We appreciate the useful comments from Thomas Dent and Nikolaos Stergioulas on our paper. NN wishes to acknowledge and appreciate the support of Joseph Bayley, Michael Williams and Christian Chapman-Bird. We would also like to extend our sincere gratitude to the PHAS-ML group members from the University of Glasgow, for their fruitful weekly meetings. NN is supported by the College Scholarship offered by the School of Physics and Astronomy (2021-2025), University of Glasgow. CM is supported by STFC grant ST/Y004256/1. This material is based upon work supported by NSF's LIGO Laboratory, a major facility fully funded by the National Science Foundation.
\end{acknowledgments}

\appendix
\counterwithin{figure}{section}

\section{\label{app:testing dataset priors}Testing Dataset Parameter Distributions}
\begin{figure*}[htbp]
    \centering
    \includegraphics[width=1.0\linewidth]{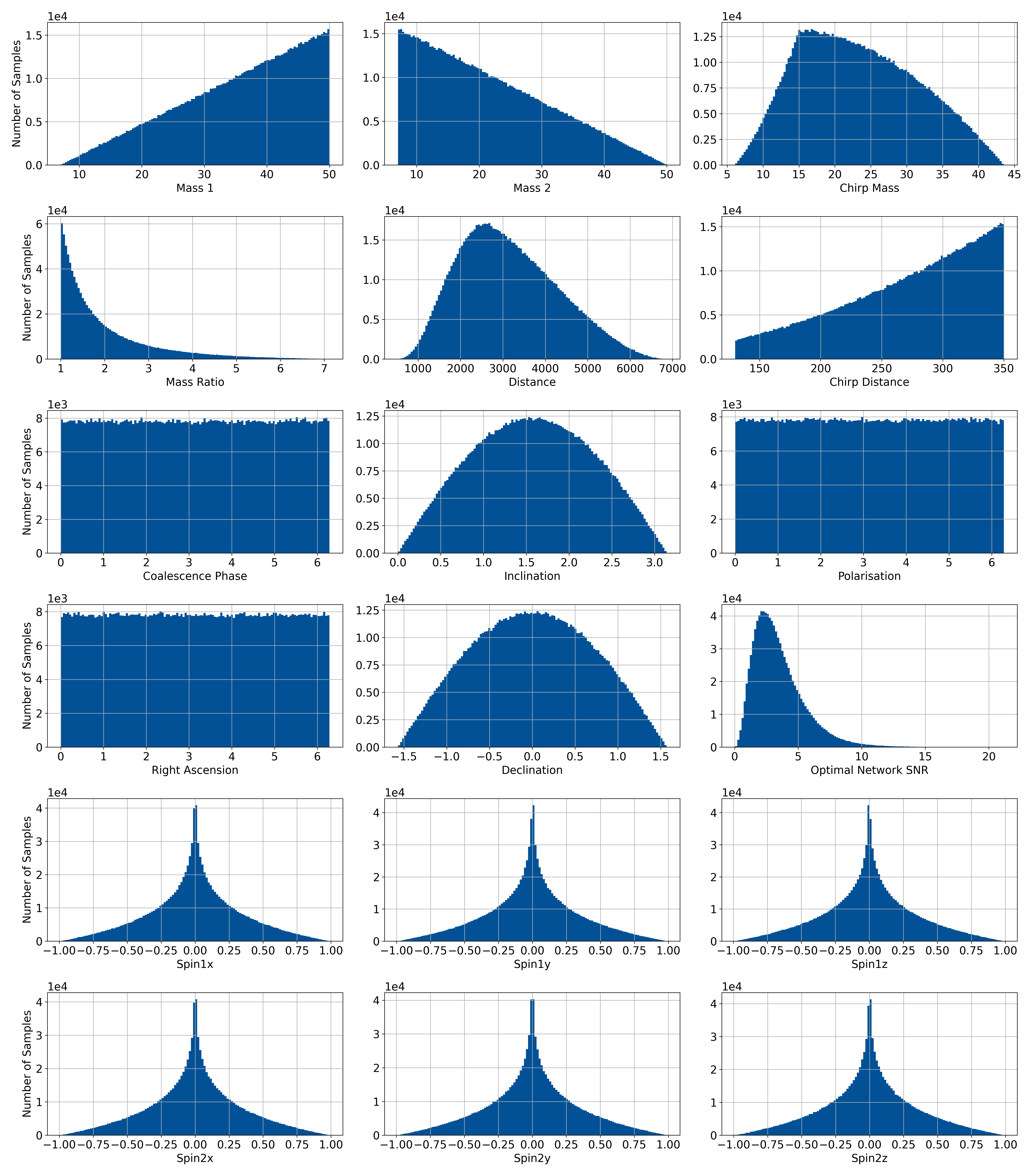}
    \caption{1D histograms of \ac{GW} parameters showing the parameter distributions based on the D4 testing dataset generated from $10^6$ samples. The distributions are fully defined in table \ref{tab:testing_dataset_priors}.}
    \label{fig:testing_mock_datset_all_priors}
\end{figure*}

Figure \ref{fig:testing_mock_datset_all_priors} shows the gravitational wave parameter distributions for the D4 testing dataset introduced in the MLGWSC-1~\cite{DP5_MLGWSC1_2023}. This was generated using the data generation \texttt{ini} file provided in the MLGWSC-1 GitHub repository for use with \texttt{PyCBC} (see \href{https://github.com/gwastro/ml-mock-data-challenge-1/blob/master/ds4.ini}{this https url}).

\section{\label{app:sage-broad detection sensitivity}Sage-Broad Detection Sensitivity}
\begin{figure*}[htbp]
    \centering
    \includegraphics[width=1.0\linewidth]{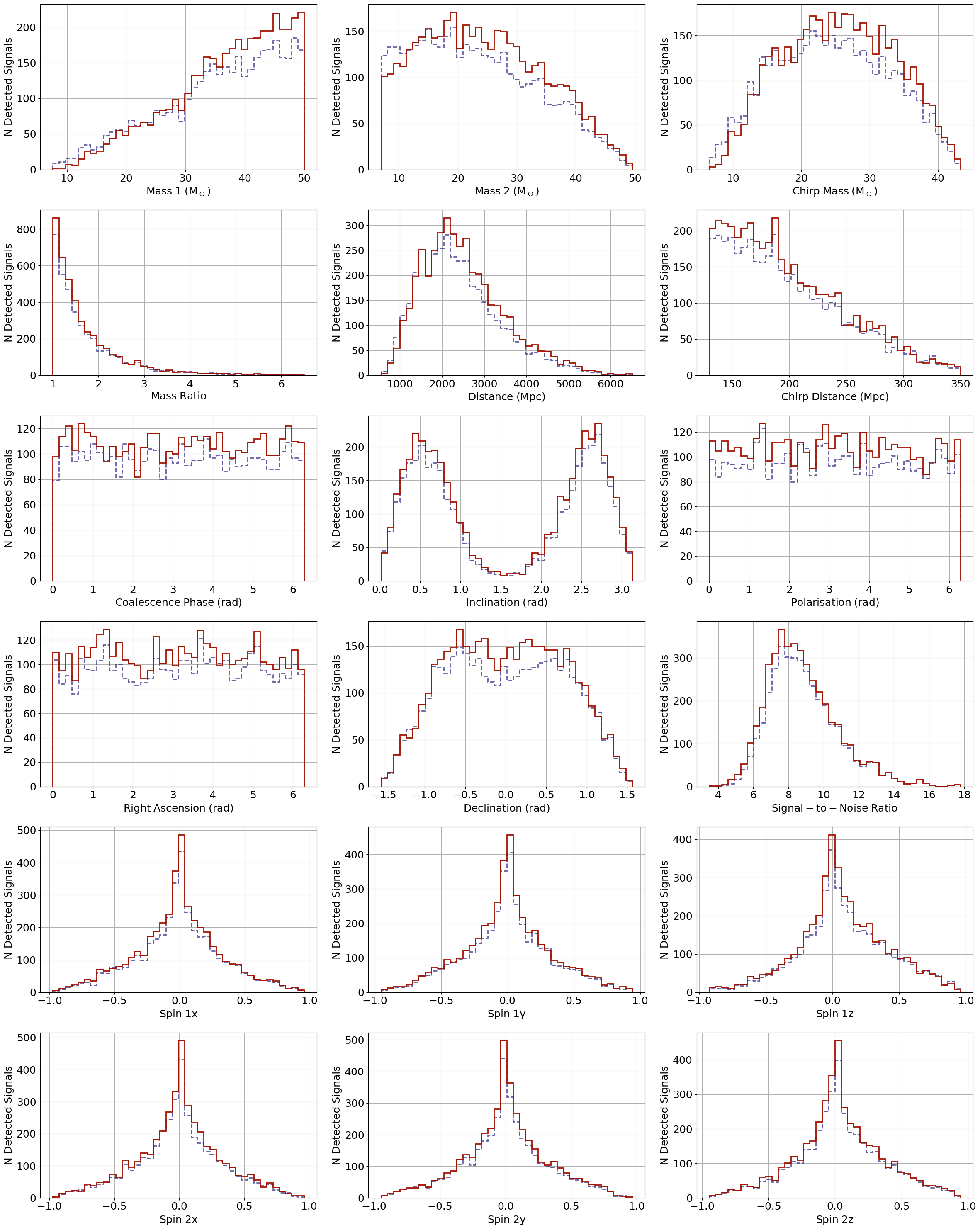}
    \caption{1D chirp mass histograms of different \ac{GW} parameters showing the number of detections made by \texttt{Sage - Broad} and \texttt{PyCBC} at an \ac{FAR} of 1 per month for the D4 testing dataset.}
    \label{fig:compare_histograms_sage_pycbc_all_params}
\end{figure*}

Figure \ref{fig:compare_histograms_sage_pycbc_all_params} shows the 1D histograms of detected events along different gravitational wave parameters at an \ac{FAR} of 1 per month for \texttt{Sage - Broad}. Figures \ref{fig:compare_sage_to_pycbc}, \ref{fig:compare_sage_to_aresgw}, and \ref{fig:compare_sage_to_aframe} show the same for a limited number of parameters - chirp mass, mass ratio, spin-1z, spin-2z and optimal network SNR. Here, we provide a full comparison of \texttt{Sage - Broad} and \texttt{PyCBC}.

\section{\label{app:network behaviour for glitch types}Network Behaviour for Different Glitch Types}
\begin{figure*}[htbp]
    \centering
    \includegraphics[width=1.0\linewidth]{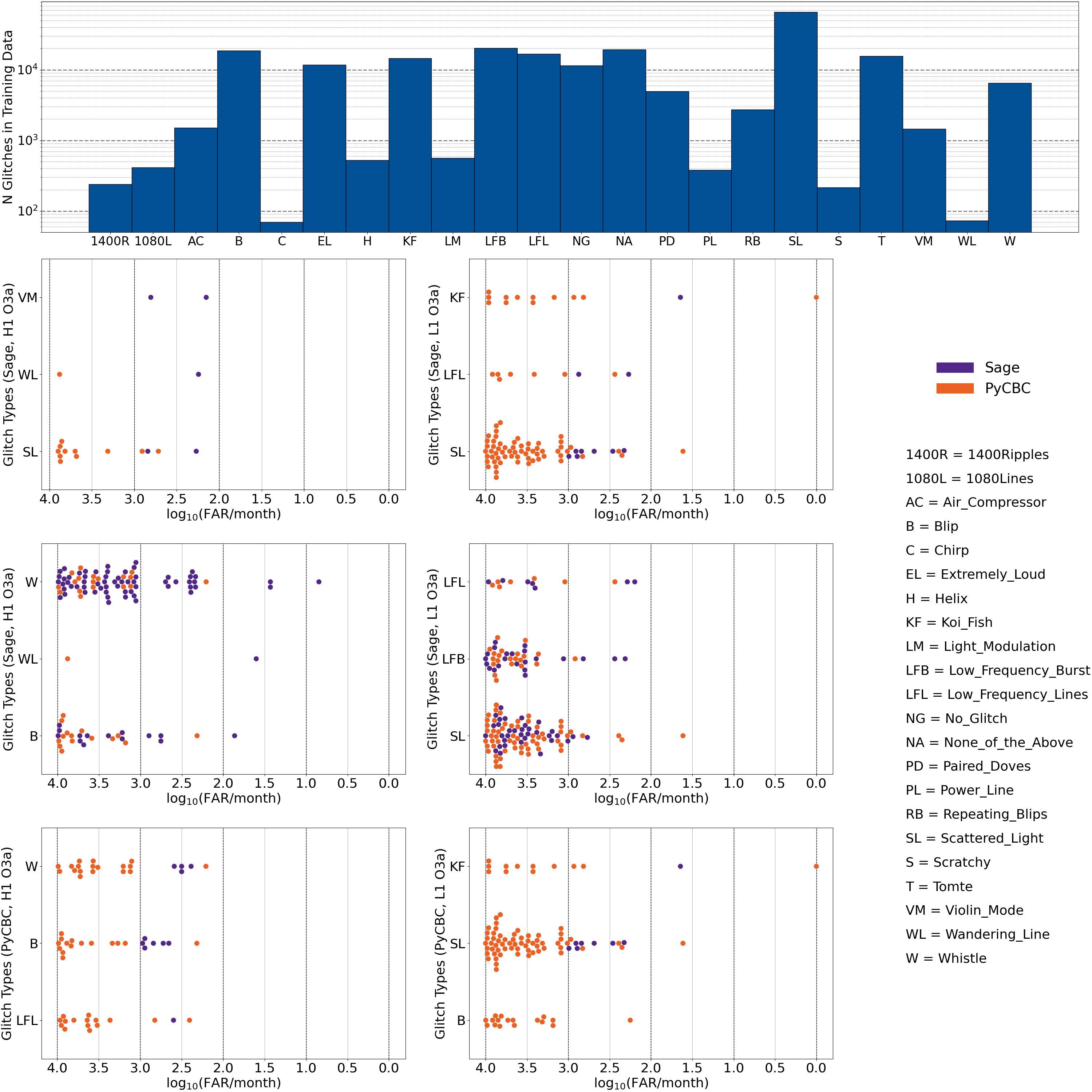}
    \caption{Glitch triggers from different glitch types at \ac{FAR}s from $10^4$ per month to 1 per month for different \texttt{Sage} runs and \texttt{PyCBC}. The left and right columns correspond to the H1 and L1 detectors, respectively. The top row corresponds to the \texttt{Sage - Annealed Train} run, the middle row to the \texttt{Sage - Broad} run and the bottom row to \texttt{PyCBC}. The glitch types chosen for each panel produced the top three triggers for that configuration (``top" here refers to triggers with the lowest observed \ac{FAR}s).}
    \label{fig:glitch_swarmplot_sage_and_pycbc}
\end{figure*}

We investigated a little further into the comparison of generated triggers for different types of glitches (see Fig. \ref{fig:glitch_swarmplot_sage_and_pycbc}). We arrived at three major observations:
            
\textit{Observation 1}: The training component mass distribution and its related learning biases determine the glitch types that the \ac{ML} model finds most difficult to classify.

\textit{Observation 2}: The \texttt{Sage - Annealed Train} run typically produced fewer glitch triggers between an \ac{FAR} of $10^4$ per month to 1 per month, when compared to \texttt{Sage - Broad}.

\textit{Observation 3}: We did not observe any noticeable correlation between the difficult-to-classify glitch types and their number of occurrences in the training data. In particular, glitch types that were not significantly present in the training dataset were not any more difficult to classify than the glitch types that were present in abundance. This might be an artefact of a limited testing dataset.

\section{\label{app:parameter point estimate}Parameter Point Estimate}
\begin{figure*}[htbp]
    \centering
    \includegraphics[width=1.0\linewidth]{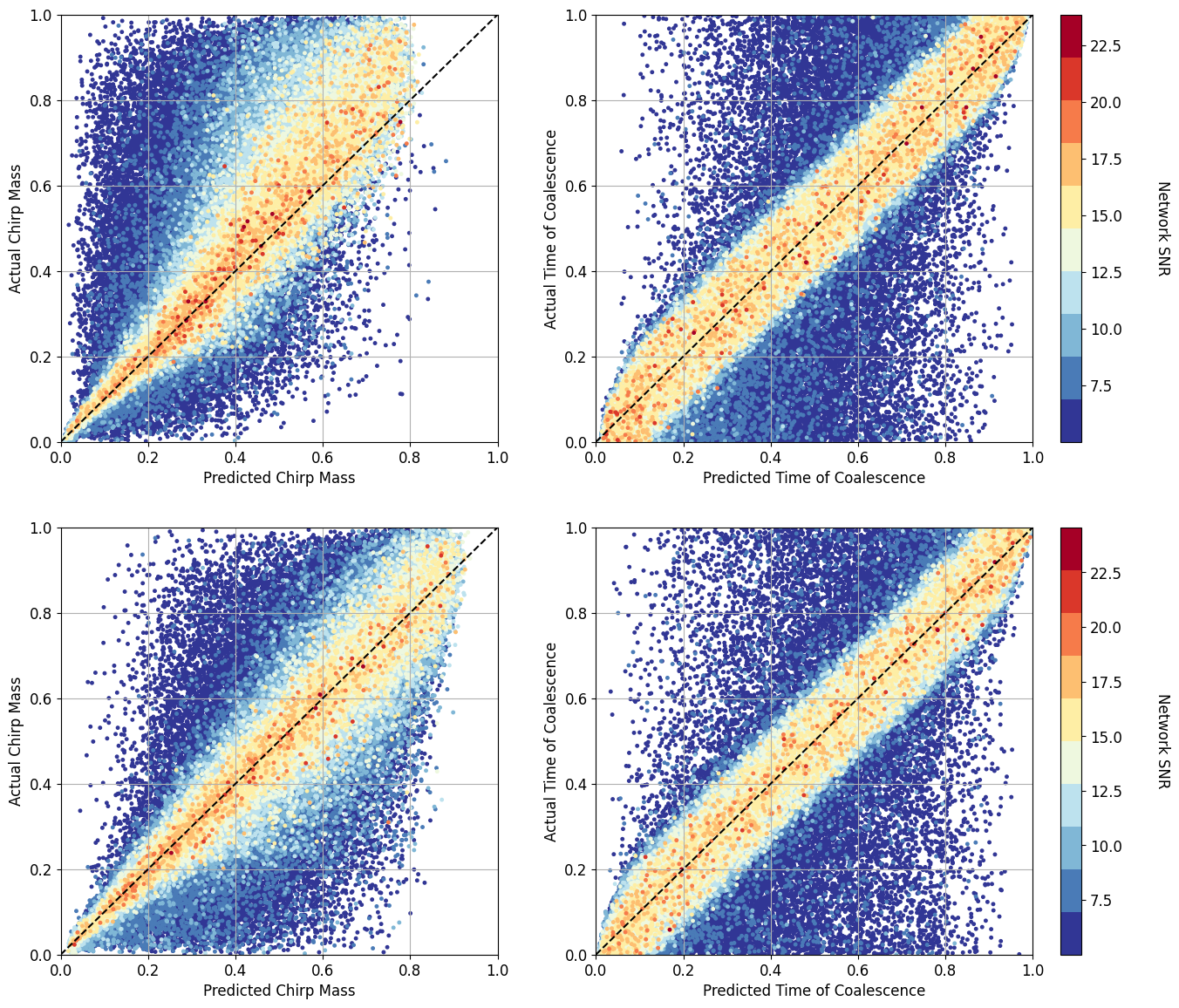}
    \caption{Actual value of \ac{GW} parameters: chirp mass $\mathcal{M_c}$ and time of coalescence $t_c$, plotted against the network predicted value for a validation epoch in different \texttt{Sage} runs. The points are coloured based on the network optimal \ac{SNR} of their corresponding gravitational wave signal. The top row corresponds to the \texttt{Sage - Annealed Train} run and the bottom row to the \texttt{Sage - Broad} run.}
    \label{fig:parameter_point_estimate}
\end{figure*}

Figure~\ref{fig:parameter_point_estimate} shows parity plots of true vs. predicted chirp mass and coalescence time for the validation dataset from \texttt{Sage - Annealed Train} and \texttt{Sage - Broad}, evaluated at the epoch with the lowest validation loss. These point estimates are included primarily to provide an auxiliary inductive bias during training, while also serving to passively diagnose parameter-specific training biases.

\textit{Diagnosing Biases}: For both \texttt{Sage - Annealed Train} (Fig.~\ref{fig:parameter_point_estimate}, top) and \texttt{Sage - Broad} (Fig.~\ref{fig:parameter_point_estimate}, bottom), the chirp mass is typically underestimated near the upper end of its range, even at high network \ac{SNR}. Two possible causes are: [i] insufficient representation of high chirp mass signals in the training data, and [ii] model limitations in capturing short-duration waveform morphology. The second is unlikely, as it would produce random errors rather than systematic bias. Supporting the first explanation, \texttt{Sage - Broad}, which trains on $U(m_1, m_2)$, shows reduced bias compared to \texttt{Sage - Annealed Train}, which trains on $U(\tau_0, \tau_3)$ and is more underrepresented at high chirp masses.

\clearpage
\bibliography{sage}

\end{document}